\documentclass[12pt,preprint2]{aastex}
\newcommand{\kms}{km~s$^{-1}$}
\newcommand{\lam}{$\lambda$}
\newcommand{\hi}{H~{\sc i}}
\newcommand{\fuse}{{\it FUSE}}
\newcommand{\stis}{STIS}

\begin{document}

\title{Intrinsic Absorption in the Spectrum of Mrk~279:
Simultaneous {\it Chandra},
\fuse\, and \stis\ Observations}

\author{Jennifer E.\ Scott\altaffilmark{1}, 
Gerard A.\ Kriss\altaffilmark{1,2},
Julia C.\ Lee\altaffilmark{3,4},
Nahum Arav\altaffilmark{5},
Patrick Ogle\altaffilmark{6}, 
Kenneth Roraback\altaffilmark{3},
Kimberly Weaver\altaffilmark{7,2},
Tal Alexander\altaffilmark{8}, 
Michael Brotherton\altaffilmark{9},
Richard F.\ Green\altaffilmark{10}, 
John Hutchings\altaffilmark{11},
Mary Elizabeth Kaiser\altaffilmark{2},
Herman Marshall\altaffilmark{3}, 
William Oegerle\altaffilmark{12},
\& Wei Zheng\altaffilmark{2}}

\altaffiltext{1}{Space Telescope Science Institute, 3700 San Martin Drive,
Baltimore, MD  21218 USA;
[jescott,gak]@stsci.edu}

\altaffiltext{2}{Center for Astrophysical Sciences, Department of Physics and Astronomy,
The Johns Hopkins University, Baltimore, MD 21218 USA; [kaiser,zheng]@pha.jhu.edu}

\altaffiltext{3}{Department of Physics and Center for Space Research,
Massachusetts Institute of Technology, 77 Massachusetts Avenue, NE80, Cambridge, MA 02139;
[jlee,roraback,hermanm]@space.mit.edu}

\altaffiltext{4}{Chandra Fellow}

\altaffiltext{5}{Center for Astrophysics and Space Astronomy, University of Colorado,
Boulder, CO  80309 USA;
arav@colorado.edu}

\altaffiltext{6}{
Mail Code 238-332, 
Jet Propulsion Lab,
4800 Oak Grove Drive,
Pasadena, CA 91109;
pmo@sgra.jpl.nasa.gov}

\altaffiltext{7}{Laboratory for High Energy Astrophysics,
National Aeronautics and Space Administration,
Goddard Space Flight Center,
Greenbelt, MD 20771; kweaver@cleo.gsfc.nasa.gov}

\altaffiltext{8}{Weizmann Institute of Science, P.O. Box 26,
Rehovot 76100, Israel; tal.alexander@weizmann.ac.il}

\altaffiltext{9}{Department of Physics and Astronomy, University of Wyoming,
Laramie, WY, 82071; mbrother@uwyo.edu}

\altaffiltext{10}{Kitt Peak National Observatory, National Optical Astronomy Observatories, 
P.O. Box 26732, 950 North Cherry Avenue, Tucson, AZ 85726  USA; rgreen@noao.edu}

\altaffiltext{11}{Herzberg Institute of Astrophysics, National
Research Council of Canada, Victoria, BC V9E 2E7, Canada; john.hutchings@hia.nrc.ca}

\altaffiltext{12}{Laboratory for Astronomy and Solar Physics, 
National Aeronautics and Space Administration,
Goddard Space Flight Center,
Greenbelt, MD 20771; oegerle@uvo.gsfc.nasa.gov}

\addtocounter{footnote}{-12}

\begin{abstract}
We present a study of the intrinsic X-ray and far-ultraviolet absorption in the
Seyfert 1.5 galaxy Markarian 279 using simultaneous observations 
from the {\it Chandra X-ray Observatory},
the 
Space Telescope Imaging Spectrograph aboard the 
{\it Hubble Space Telescope},
and the {\it Far Ultraviolet Spectroscopic Explorer (FUSE)}.
We also present \fuse\ 
observations made at three additional epochs. 
We detect the Fe K$\alpha$ emission line in the
{\it Chandra} spectrum, and its flux is consistent with the
low X-ray continuum flux level of Mrk~279 at the time of the observation. 
Because of low signal-to-noise ratios (S/N) in the {\it Chandra} spectrum, 
no O~{\sc vii} or O~{\sc viii} absorption features
are observable in the {\it Chandra} data, but the UV spectra reveal 
strong and complex absorption
from \hi\ and high-ionization species such as O~{\sc vi}, N~{\sc v}, and C~{\sc iv}, as
well as from low-ionization species such as C~{\sc iii}, N~{\sc iii}, 
C~{\sc ii}, and N~{\sc ii} in some velocity components.
The far-UV spectral coverage of the \fuse\ data provides information on 
high-order Lyman series
absorption, which we use to calculate the optical 
depths and line and continuum 
covering fractions in the intrinsic \hi\ absorbing gas in a self-consistent fashion.  
The UV continuum flux of Mrk~279 decreases
by a factor of $\sim$7.5 over the time spanning these observations and we discuss the
implications of the response of the absorption features to this change.
From arguments based on the velocities, profile shapes, covering fractions
and variability of the UV absorption, we conclude
that some of the absorption components, 
particularly those showing prominent low-ionization lines, 
are likely associated with the host galaxy of Mrk~279, and possibly 
with its interaction with a close companion galaxy,
while the remainder arises
in a nuclear outflow.
\end{abstract}

\keywords{galaxies: active --- galaxies: individual (Mrk~279) --- 
galaxies: Seyfert --- quasars: absorption lines --- ultraviolet: galaxies ---
X-ray: galaxies}

\section{Introduction}
Approximately one-half of all low-redshift active galactic nuclei (AGNs) show absorption edges
due to highly ionized gas, ``warm absorbers'', in their X-ray spectra 
(Reynolds 1997; George et al.\ 1998).  
Among these objects, there is a one-to-one correspondence
with those that show high-ionization UV absorption lines in their spectra
\cite{cren1999}. 
This correlation indicates that the two phenomena are
closely related, but the exact relationship remains unclear. 

This problem has been addressed by the use of 
high-resolution spectroscopy in the UV and/or X-ray,
which has provided a variety of detailed information on
the physical state of the intrinsic gas in
several AGNs:
Mrk~509  (Kriss et al.\ 2000a; Kraemer et al.\ 2003; Yaqoob et al.\ 2003);
NGC~3516 (Kraemer et al.\ 2002);
NGC~3783 (Kraemer, Crenshaw, \& Gabel 2001; Kaspi et al.\ 2001;
 Blustin et al.\ 2002; Gabel et al.\ 2003a,b);
NGC~4151 (Crenshaw et al.\ 2000; Kraemer et al.\ 2001);
NGC~5548 (Mathur et al.\ 1999; Crenshaw \& Kraemer 1999;
 Brotherton et al.\ 2002; Kaastra et al.\ 2002; Steenbrugge et al.\ 2003); and
NGC~7469 (Kriss et al.\ 2000b; Kriss et al.\ 2003; Blustin et al.\ 2003).
In general, it is difficult to associate a particular UV absorption component
with the X-ray absorber unambiguously, in part because the ionization modeling
is sensitive to the shape of the UV-to-X-ray continuum (Kaspi et al.\ 2001), 
and observations to
date have not always been performed simultaneously in the UV and X-ray.

The intrinsic absorption lines are typically blueshifted with respect to the
systemic redshift of the AGN, indicating radial outflow.
The UV absorption is often highly variable, due either
to changes in the ionization state of the absorbing gas in response to changes in 
the continuum flux of the AGN 
(Krolik \& Kriss 1995; Shields \& Hamann 1997; Crenshaw et al.\ 2000; Kraemer et al.\ 2002) 
or to bulk motion of the gas in the direction transverse to the line of sight
(Crenshaw \& Kraemer 1999; Kraemer et al.\ 2001).  
These properties, along with measurements of partial coverage
of the continuum and broad emission line region (BELR) for some absorbers, 
indicate that this gas is indeed intrinsic to the AGN. 

Measurements of the covering fractions of the absorbers are important for
an accurate determination of the column density of the gas, for
building a picture of the geometry of the absorbing structures, and for
testing physical models.  
Covering fractions 
less than 1
indicate partial covering of the continuum source or BELR 
(Barlow \& Sargent 1997; Hamann et al.\ 1997) by the absorbing gas.
It is generally not possible to measure covering fractions
for the continuum and BELR separately along with absorption optical depths,
though doublets and saturated regions in Lyman series lines have been
used as tools for disentangling these quantities (Hamann et al.\ 1997;
Ganguly et al.\ 1999; Gabel et al.\ 2003a).  Additionally, covering fractions vary
along the absorption profiles and these velocity-dependent
covering fractions play a key role in determining the profile shapes 
(Arav, Korista, \& de Kool 2002), 
though some caution that 
finite instrumental resolution limits the ability to determine these
quantities accurately in the profile wings (Ganguly et al.\ 1999).

In models of intrinsic absorption in AGNs,
the gas responsible for the UV and X-ray absorption
arises either in a wind emanating from the accretion disk
(K\"{o}nigl \& Kartje 1994; Murray et al.\ 1995),
or in a wind driven off the obscuring torus
(Krolik \& Kriss 1995, 2001).
In each of these scenarios, the absorbing material lies in different orientations
with respect to the BELR and at different distances
from the AGN central engine.
Observations of variability in the intrinsic absorption in response
to changes in the AGN continuum flux provide information on  
the distance of the absorbers from the continuum-emitting region,
allowing one to distinguish between these two models.

Mrk~279 ($z=0.0305$, see \S~\ref{sec-intabs})
has been studied at wavelengths ranging from the X-ray to the radio
and is among the brightest Seyfert galaxies in the UV and X-rays.
As part of an X-ray and far-UV monitoring campaign of several low-redshift AGNs,
we have obtained spectra of Mrk~279 using the  
{\it High Energy Transmission
Grating Spectrometer} on the {\it Chandra} X-ray satellite,
the Space Telescope Imaging Spectrograph (STIS) onboard
the {\it Hubble Space Telescope},  and the 
{\it Far Ultraviolet Spectroscopic Explorer (FUSE)}.
We observed Mrk~279 at four epochs over a period
of 2.4 yr with \fuse, simultaneously with \stis\ and with
{\it Chandra} 
for the most recent of the \fuse\ observations.
The {\it Chandra} data cover spectral regions that may be expected to show
absorption features due to 
highly ionized silicon, magnesium, neon, iron, oxygen, and nitrogen.  
In particular, the absorption lines and edges from O~{\sc vii}
and O~{\sc viii} are useful diagnostics of the ionization state of the
absorbing gas, since the \fuse\ spectra cover the O~{\sc vi}~\lam\lam1032,1038
doublet.
The spectral coverage of
\fuse\ also provides information on \hi\ absorption in Ly~$\beta$
and higher order
Lyman transitions, and C~{\sc iii}~\lam977,
and the \stis\ observations cover Ly$\alpha$ and the N~{\sc v}~\lam\lam1238, 1242,
Si~{\sc iv}~\lam\lam1393, 1402, and C~{\sc iv}~\lam\lam1548, 1550 doublets. 
This data set is unique in several respects:
the information on high-order
Lyman series absorption from the \fuse\ spectra allows us to calculate optical depths and line and
continuum covering fractions in the absorbers; 
the UV observations made at different epochs permit us to investigate the
effects of UV continuum variability on the intrinsic absorption; and
the simultaneous UV and X-ray observations we have obtained give us the
ability, in principle, to perform photoionization modeling of the 
warm absorber 
with a self-consistent UV-to-X-ray spectral energy distribution. 

\section{Observations}
\subsection{X-ray: {\it Chandra}} 
\label{sec-cxobs}

We observed Mrk~279 on 2002 May 18-19
with the High Energy Transmission 
Grating Spectrometer (HETGS) on the {\it Chandra X-Ray Observatory}.
The HETGS
\footnote{For further details on
{\it Chandra} instruments, see http://cxc.harvard.edu/proposer/POG.}
provides high resolving power, up to $\sim$1000,
from 0.4~keV to 10.0~keV through the use of a Rowland design and two gratings:
(1) the Medium Energy Grating (MEG), which 
intercepts X-rays from the two outer mirror shells of the 
High Resolution Mirror Assembly (HRMA) and gives good spectral coverage from
0.4~keV to 5.0~keV (2.5-31~\AA);
and (2) the High Energy Grating (HEG), which
intercepts
X-rays from the two inner shells of the HRMA and is
optimized for the
0.8-10.0~keV (1.2-15~\AA) spectral region. 
We reduced the X-ray spectra using
CIAO version 2.3, filtering 
on energy ($E<10$~keV), hot columns, and grade$=$(0, 2, 3, 4, 6).  We
removed streak events from CCD 8 using {\it destreak}.
We extracted the $m=\pm1$ order spectra from a spatial region $3\farcs6$ wide, centered
on the peak of the zeroth-order image.
We flux-calibrated the spectra 
using the aspect histogram and standard grating response matrix functions.
See Table~\ref{table-obs} for a summary of the {\it Chandra}
observation details.

\begin{figure}
\epsscale{0.67}
\plotone{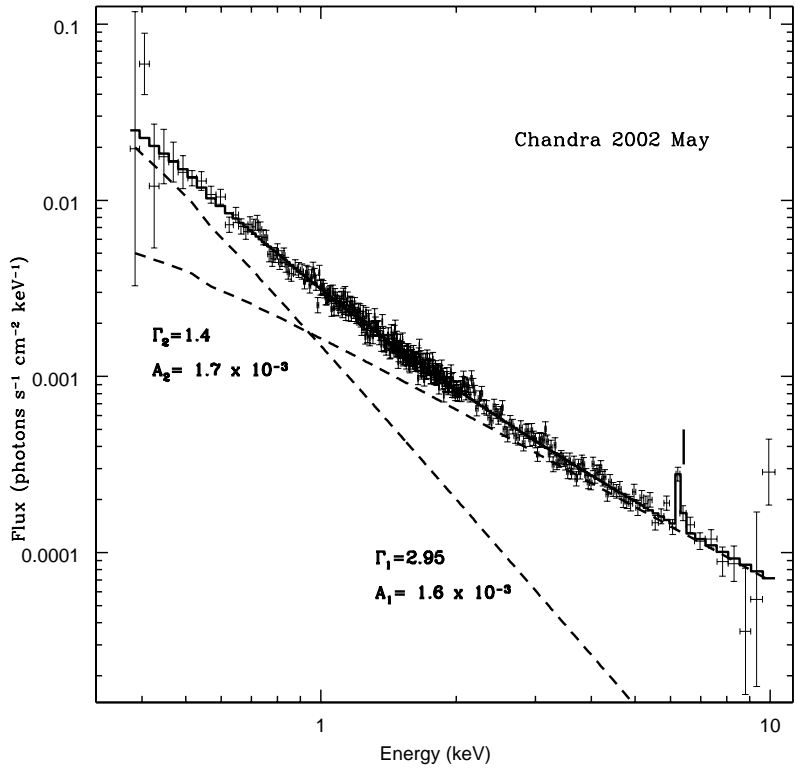}
\caption{\small Combined {\it Chandra}  MEG and HEG spectrum of Mrk~279 
with fits to continuum and Fe K$\alpha$ line.
The individual power laws that constitute the continuum fit are
shown by dashed lines.
The tick mark denotes the position of the Fe K$\alpha$ line.
\label{fig:chand}}
\epsscale{1.0}
\end{figure}

Mrk~279 was in a low flux state during the {\it Chandra} observation, and
the resulting S/N in the individual MEG and HEG spectra is low.
Therefore, we combined the MEG and HEG spectra at MEG resolution and 
rebinned adaptively to fit the continuum model. 
We fixed the bin 
width at energy $E$ so that $S/N > 10$,
but we restricted the width to be no larger than 5\%$E$ wide. 
Many bins with $E < 0.7$ keV and $E > 7$ keV have SNR $< 10$.
We show the merged spectrum in Figure~\ref{fig:chand}.

\subsection{Ultraviolet: {\it FUSE} and  STIS}
\fuse\ (Moos et al.\ 2000; Sahnow et al.\ 2000) consists of
four primary mirrors feeding four Rowland circle spectrographs
and two, two-dimensional, photon-counting microchannel plate
detectors. Two of the mirror/spectrograph pairs employ optics
coated with Al+LiF for efficient spectral coverage from $\sim$990 to 1187~\AA,
while two use optics coated with SiC for coverage down to $\sim$905~\AA. 
The data were recorded in photon-address mode, which delivers a time-tagged 
list of event positions in the downlinked data stream.

We observed Mrk~279 at four
different epochs with \fuse\, 
through the $30\arcsec \times 30\arcsec$ low-resolution aperture.
See Table~\ref{table-obs} for the details of the observations.
Because the two 2002 January observations were made within several hours of 
each other, we combined the two spectra.
As a result of mirror alignment problems during the 2002 May observation, 
we are only able to analyze data from the 
LiF1 channel. 
We used the standard \fuse\ calibration pipelines (see
Sahnow et al.\ 2000), to extract the spectra, to subtract dark current, and 
to perform wavelength and flux calibrations.
We used CALFUSE version 2.0.5 for the 1999 and 2000 data,
and we used CALFUSE version 2.1.6 for the 2002 data.  
However, Mrk~279 was in a low state during the 2002 observations,
resulting in a low signal-to-noise ratio (S/N) in the 2002 January
spectra in particular, because their exposure times were relatively short. 
Therefore, because
improved background subtraction algorithms were implemented in later
versions of CALFUSE, we re-reduced the 2002 January data using CALFUSE version 2.2.3.
We combined the spectra from each detector
and binned by 3-8 pixels, improving
the S/N while preserving the full
spectral resolution, $\sim$20~\kms.

To correct the wavelength scale of the \fuse\ spectra for zero-point offsets
due to placement of the target in the $30\arcsec$ aperture, we measure the
positions of low-ionization Galactic absorption lines from species such as
Ar~{\sc i}, Fe~{\sc ii}, {\sc O~i}, and $\rm H_2$ and compare these with
the Galactic 21 cm \hi\ spectrum along the line of sight to Mrk 279
(Wakker et al.\ 2001).
The \hi\ emission has a complex profile with
multiple peaks. The brightness-temperature-weighted mean heliocentric
velocity is $-55$~\kms.  Correcting the spectra from the individual
observation epochs to this value required shifts of 0.05 to 0.10 \AA.
We estimate that the residual systematic errors in the \fuse\ wavelength scale
are on the order of 10 \kms, and that the flux scale is accurate to
$\sim$10\%.

We also obtained a spectrum of Mrk~279 with the \stis\, FUV-MAMA
contemporaneously with the 2002 May 18 \fuse\ observation
in a five-orbit exposure through the $0.2\arcsec \times 0.2\arcsec$ aperture
using the medium-resolution E140M echelle grating covering 1150-1730~\AA. 
See Table~\ref{table-obs} for other details of the \stis\ observation.
We used CALSTIS version 2.12a to process the spectra.
This version automatically corrects for echelle
scattered light using the Lindler \& Bowers (2000) two-dimensional 
algorithm as implemented in
the CALSTIS pipeline.  Several ISM absorption features from our
own Galaxy (Si~{\sc iii}~\lam1206, 
Si~{\sc ii}~\lam1260, C~{\sc ii}~\lam1334)  
are saturated in the line troughs, with residual fluxes 
consistent with zero to 
within 1.5\% of the 
adjacent continuum levels.  This indicates a reasonable 
scattered light correction given that the one-dimensional
algorithm used previously
had been found to result in residual saturated-line core fluxes
of up to 10\% of the adjacent continuum levels (Lindler \& Bowers 2000;
Crenshaw et al.\ 2000).
We extracted the individual echelle orders from each exposure
and summed them using a set of IRAF
scripts to produce the merged spectrum.
As described above, we measure the positions of lines from low-ionization
species, S~{\sc ii}, Si~{\sc ii}, O~{\sc i}, C~{\sc ii}, and Fe~{\sc ii}
in the \stis\ spectrum, and we compare these with the mean heliocentric
velocity of the \hi\ 21 cm emission.  We find that no correction to the wavelength
scale derived by the CALSTIS pipeline is necessary.
We obtained the \stis\ spectrum and the 2002 May \fuse\
spectrum simultaneously with the {\it Chandra}
data described in Section~\ref{sec-cxobs}.

We fit the \fuse\ and \stis\  spectra with power-law continua and Gaussian
emission lines.  The full spectra are shown in Figures~\ref{fig:fusespec} and
\ref{fig:stisspec}.
The emission features we identify in the \fuse\ spectra include 
C~{\sc iii}~\lam977, N~{\sc iii}~\lam989, Ly$\beta$,
O~{\sc vi}~\lam\lam1032,1038, S~{\sc iv}\lam\lam1063,1073,
and He~{\sc ii}~\lam1085. 
In the \stis\ spectrum, we identify emission lines from Ly$\alpha$,
N~{\sc v}~\lam\lam1238, 1242, Si~{\sc iv}~\lam\lam1393, 1402, 
C~{\sc iv}~\lam\lam1548, 1550, and others. We discuss the
continuum and emission line fits to the spectra in more 
detail in Section~\ref{sec-fusespec} below.

\begin{figure}
\epsscale{0.57}
\plotone{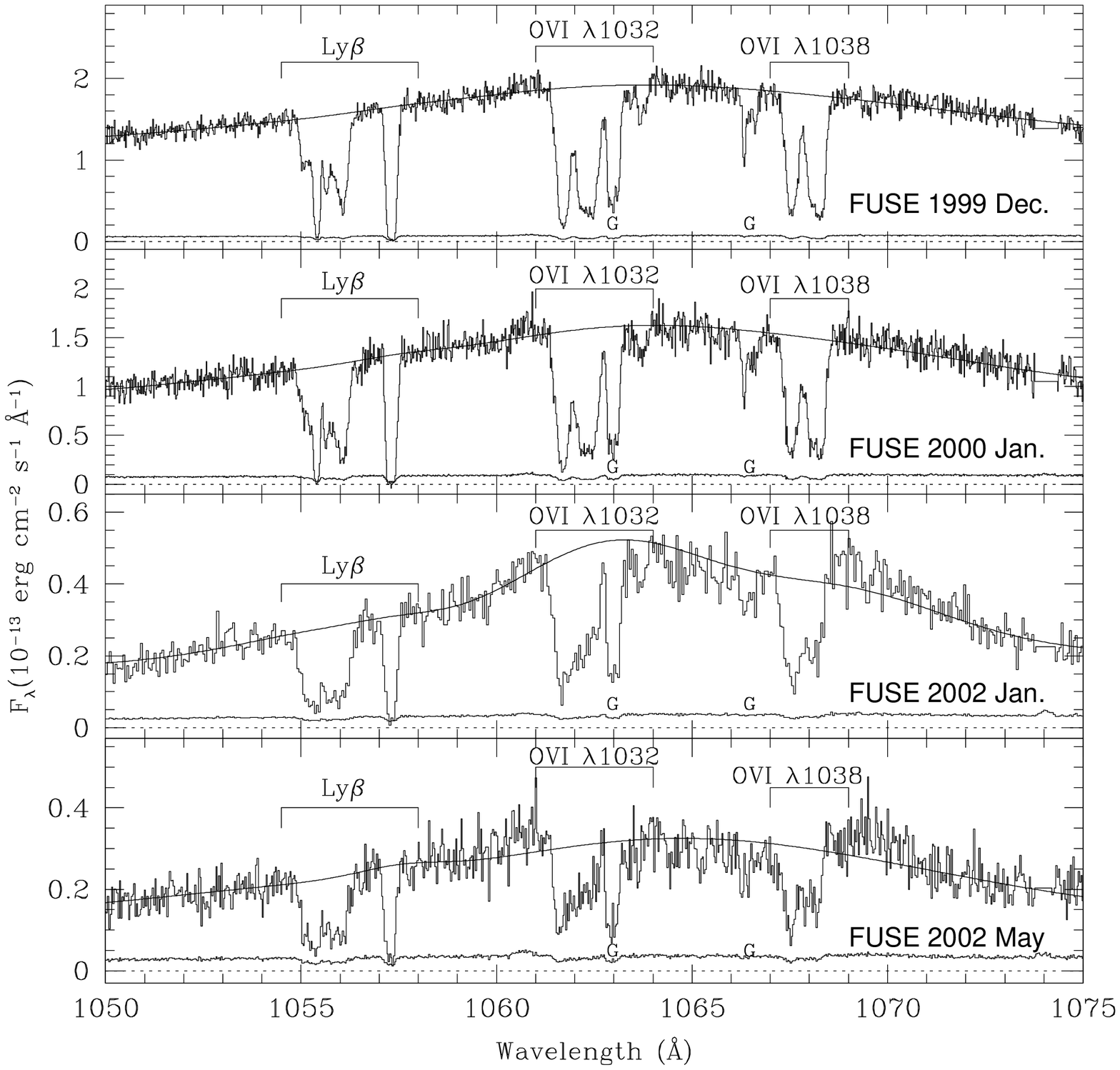}
\caption{\small Ly$\beta$ and O~{\sc vi} intrinsic absorption in \fuse\ spectra of Mrk~279.
Galactic ISM absorption features marked with ``G'', see also 
Figures~\ref{fig:ly99}-\ref{fig:met502}.
\label{fig:fusespec2}}
\plotone{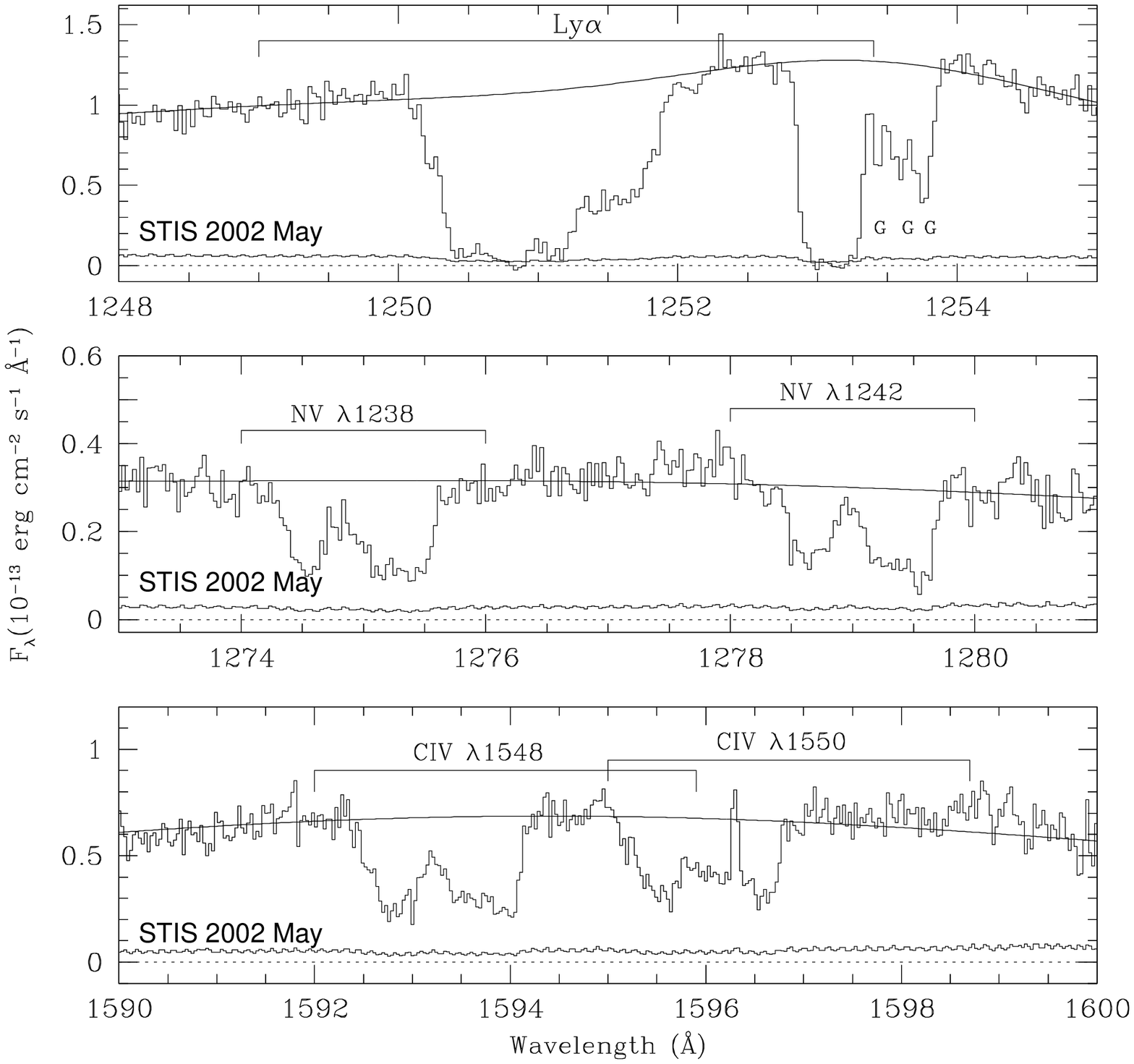}
\caption{\small Ly$\alpha$ and N~{\sc v} and C~{\sc iv} doublets in the
\stis\ spectrum of Mrk~279.
Galactic ISM absorption features marked with ``G'', see also Figures~\ref{fig:ly502} 
and \ref{fig:metstis}.
\label{fig:stisspec2}}
\epsscale{1.0}
\end{figure}

We find intrinsic absorption in these UV spectra extending from $-$580 to +160 \kms. 
Figure~\ref{fig:fusespec2} shows a close-up view
of the Ly$\beta$ and O~{\sc vi} absorption profiles in each \fuse\ spectrum,
and Figure~\ref{fig:stisspec2} shows close-up views of
the Ly$\alpha$, N~{\sc v}, and C~{\sc iv} absorption profiles in the \stis\ spectrum. 
In Ly$\beta$, we find five distinct components of absorption, 
named 1-5 from the strongest and reddest component blueward. 
We also identify two subcomponents of Component 2, which we label 
2a and 2b, and one subcomponent
of Component 4, which we label 4a.  These components and subcomponents are 
discerned most clearly in the Ly$\beta$ and C~{\sc iii}~\lam977 profiles
in the 1999 December spectrum, which are shown with these components labeled  
in the first panel of Figure~\ref{fig:ly99}
and the second panel of Figure~\ref{fig:met99}, respectively.
A slight excess in the flux is visible in the 2002 January and May \fuse\ spectra
at $\sim$1069-1070 \AA, in the bottom two panels of Figure~\ref{fig:fusespec2}.
The lack of a corresponding feature at $\sim$1063.5 rules out a component
of narrow O~{\sc vi} emission, so perhaps this is due to a complex of
N~{\sc i} airglow lines in this region appearing more prominently in the low flux state 
spectra.  In any case, the unabsorbed continuum
is fitted well otherwise, so we contend that this has no significant impact on
our results.
The intrinsic UV absorption is 
discussed further in Section~\ref{sec-intabs}.

\section{Analysis of the {\it Chandra} Spectrum}
\subsection{Continuum Fit and Spectral Features}
\label{sec-xray}

During the {\it Chandra} observation, the 2-10~keV flux was 
$1.2 \times 10^{-11}$~ergs~cm$^{-2}$~s$^{-1}$.  This is a low flux  
state for Mrk~279, as the flux in this band 
varies between $\sim1 \times 10^{-11}$ and $5 \times 
10^{-11}$ (Weaver, Gelbord \& Yaqoob 2001).  
Previous low flux levels occurred in 1979 (HEAO-1) and 1991 (BBXRT).
Dividing the {\it Chandra} dispersed MEG and HEG data into 1000 s bins,
we find a 10-15\% decrease in the count rate starting 75 ks into
the observation and lasting 20 ks.

The best-fit continuum model consists of two power laws
modified by absorption due to a Galactic hydrogen column
of $1.8 \times 10^{20}$ cm$^{-2}$ (Dickey \& Lockman 1990), giving
$\Delta \chi^2 / dof$ = 306/327.  
The two power laws have photon indices 
$\Gamma_1 = 2.95 \pm 0.25$ and $\Gamma_2 = 1.4 \pm 0.1$
and normalizations  
$\rm A_1 = (1.6 \pm 0.6) \times 10^{-3}$ and
$\rm A_2 = (1.7 \pm 0.6) \times 10^{-3}$ 
photons cm$^{-2}$ s$^{-1}$ keV$^{-1}$ at 1~keV in  the emitted frame. 
We included additional multiplicative model components 
to account for calibration effects.  We use:
(1) an ad hoc correction for ACIS pileup effects, which also
reduces the effect of the Ir-M edge (Marshall et al.\ 2004a);
(2) a correction to bring the front-side-illuminated chip quantum efficiencies 
into agreement with the back-side-illuminated chip quantum efficiencies 
(Marshall et al.\ 2004a); and
(3) an ACIS contamination correction (Marshall et al.\ 2004b).

The most significant emission feature we identify in the
combined MEG and HEG spectrum is the
Fe K$\alpha$ emission line redshifted to 1.99 \AA,
which we discuss in further detail below in Section~\ref{sec-kalpha}.
We do not detect O~{\sc viii} Ly~$\alpha$ absorption at 
18.98 \AA\ rest, or the
strongest O~{\sc vii} line at 21.60 \AA.  We use the 
flux error array and the continuum fit
discussed above to derive 3$\sigma$ equivalent width limits at 
the expected locations of these features, 0.06 and  0.14~\AA, respectively.  
Assuming absorption on the linear part of the curve of growth,
we derive upper limits on the column densities
of O~{\sc vii} and O~{\sc viii}, which are both $\sim$5 $\times 
10^{16}$~cm$^{-2}$.  
However, the column density of either of these species
could be substantially larger
if either line is below the equivalent width threshold and saturated, which
is the case for the 3$\sigma$ limits quoted above if the X-ray absorbers
have Doppler parameters typical of intrinsic UV absorbers, 100-200 \kms.
(See Fig.\ 2 of Gabel et al.\ 2003a.) For $b=100$ \kms,
and unity covering fraction, the equivalent width limits
for O~{\sc vii}  and O~{\sc viii} imply $N$(O~{\sc vii}) $< 6 \times 10^{19}$~cm$^{-2}$
and $N$(O~{\sc viii}) $< 3 \times 10^{19}$~cm$^{-2}$.

\subsection{Fe K$\alpha$ Line}
\label{sec-kalpha}
The most significant line that appears in the X-ray
spectrum is the fluorescent Fe K$\alpha$ emission line at 
a rest-frame energy of $6.42 \pm 0.02$~keV
($1.93 \pm 0.01$~\AA). 
This feature is marked in Figure~\ref{fig:chand}.
The best Gaussian fit to the Fe K$\alpha$ line in the HEG spectrum
has an intrinsic velocity width of $4200^{+3350}_{-2950}$ \kms\ FWHM,
and the width of the instrumental line spread function at the position
of the line is $\sim$1500 \kms.
The corresponding line flux and equivalent width are
$2.4\pm 0.8 \times 10^{-5}$ photons s$^{-1}$~cm$^{-2}$ and
$180 \pm 60$~eV, respectively.

Spectra of Mrk~279 from the {\it Advanced Satellite for Cosmology and
Astrophysics (ASCA)} revealed variability of the Fe~{\sc i} K$\alpha$
profile on $\sim$10 ks timescales as the 2-10~keV continuum flux
increased by 20\% (Weaver et al.\ 2001). The X-ray continuum source of
Mrk~279 was at least twice as bright during the 1995 {\it ASCA} observations
compared with the 2002 {\it Chandra} observation.  
Because the best-fit line width from the {\it Chandra} spectrum is comparable to that
obtained with {\it ASCA}, we can make direct comparisons. 
The line flux is approximately one-half the strength it was in the {\it
ASCA} observation.  Since the continuum is also one-half as strong,
this indicates that the line flux has scaled with the continuum flux.
With {\it ASCA}, the peak energy of the line changed from $\sim$6.3~keV
in the low state to $\sim$6.5~keV in the high state. The peak
energy of the line observed with {\it Chandra} is located at the
average of these two energies. At 90\% confidence
the low-state energy measured by {\it ASCA}, $6.30^{+0.10}_{-0.12}$~keV,
is consistent with the energy measured here with {\it Chandra}/HETGS.
There appears to be  a
blueward asymmetry to the line profile, but the current
data do not warrant
detailed modeling with accretion disk line profiles.
The S/N in this region is such that we cannot place meaningful constraints on
a Compton reflection component. 

\section{Analysis of the \fuse\ and \stis\ Spectra}
\subsection{Continuum and Emission Lines}
\label{sec-fusespec}

\begin{figure}
\epsscale{0.6}
\plotone{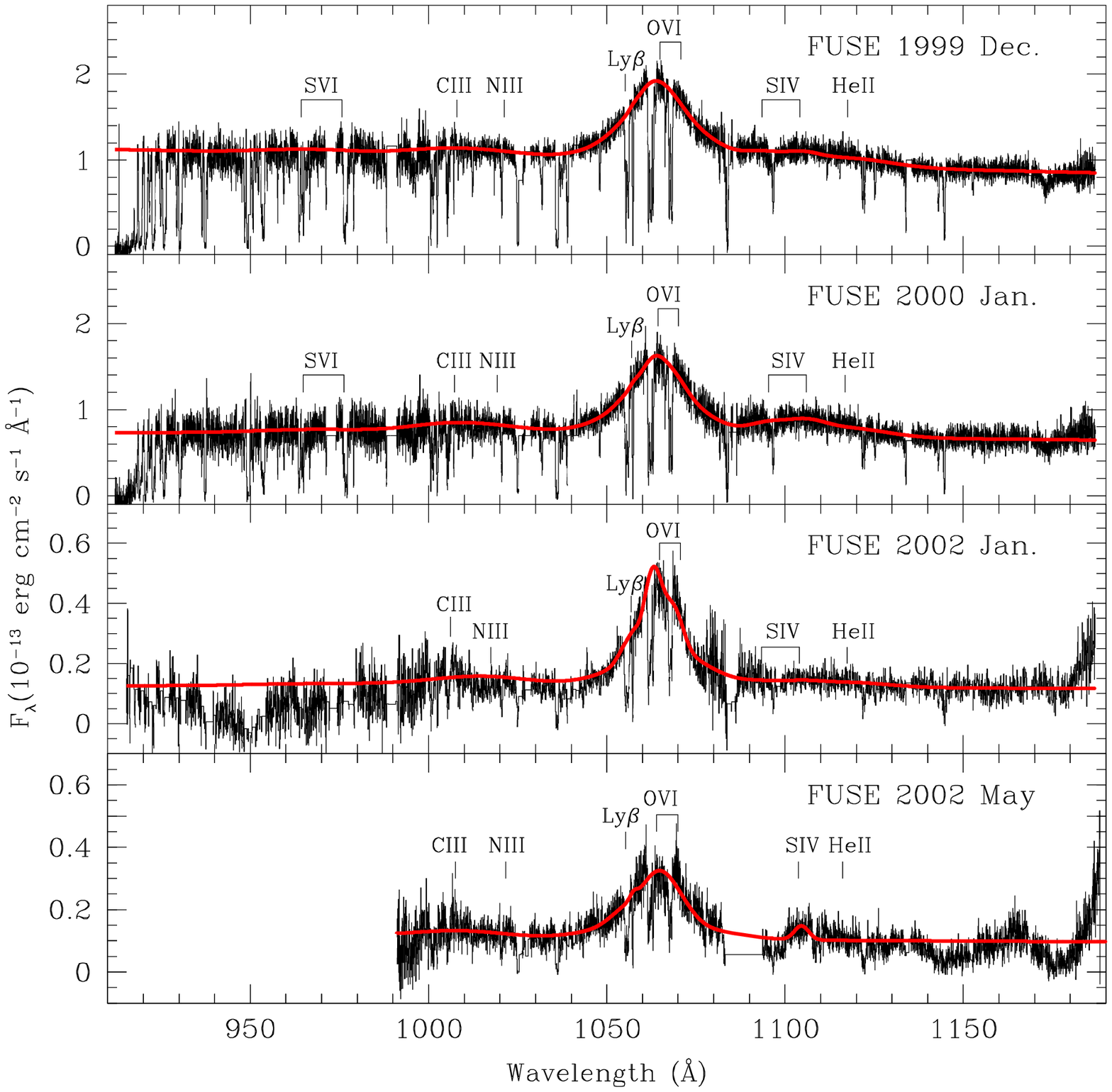}
\caption{\small \fuse\ spectra of Mrk~279 with continuum and emission line
fits (red).
\label{fig:fusespec}}
\plotone{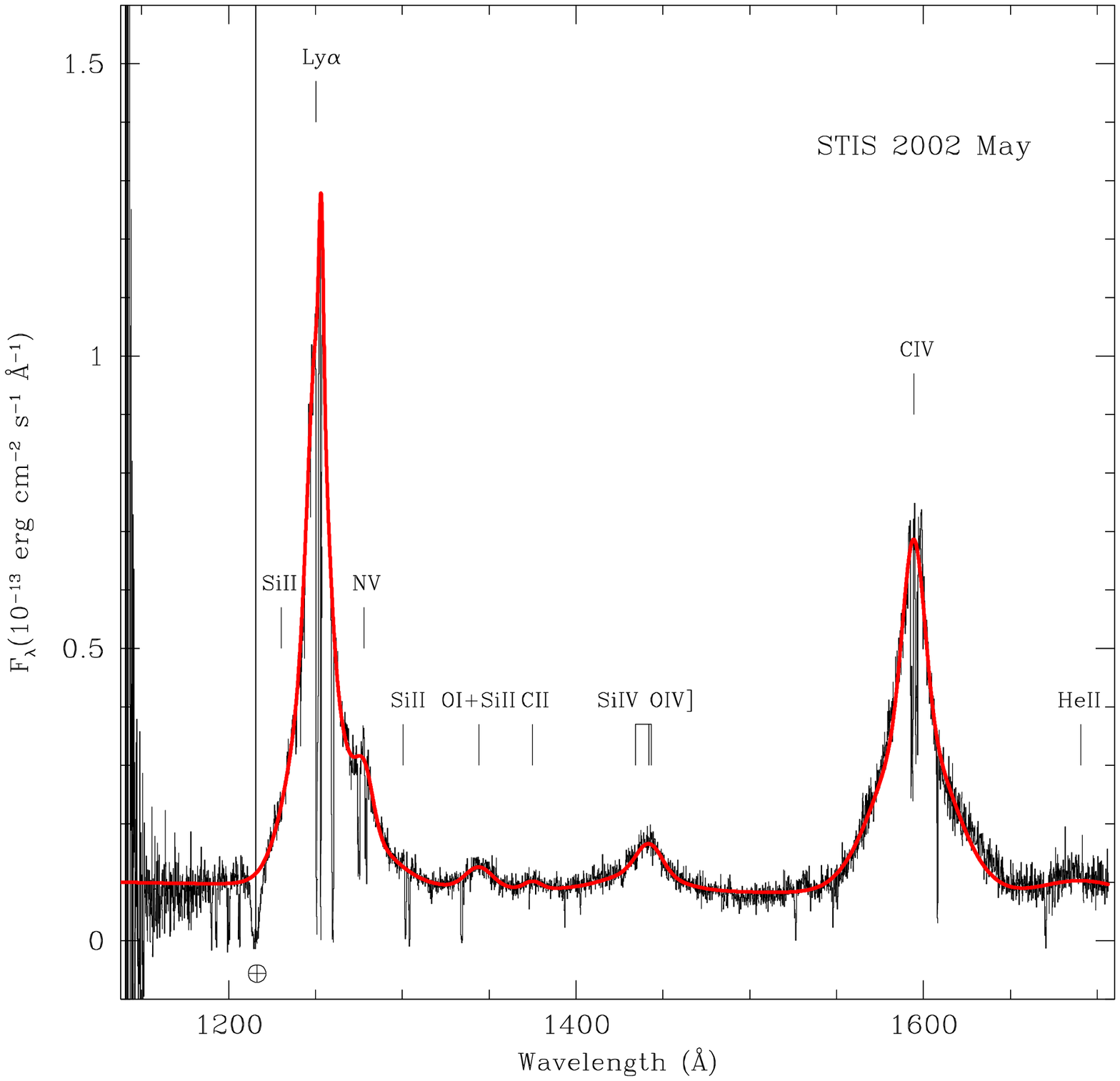}
\caption{\small \stis\ spectrum of Mrk~279 with continuum and emission line
fits (red).
\label{fig:stisspec}}
\epsscale{1.0}
\end{figure}

We use the IRAF\footnote{IRAF is distributed by the National Optical 
Astronomy Observatories,
which are operated by the Association of Universities for Research
in Astronomy, Inc., under cooperative agreement with the National
Science Foundation.}
task {\it specfit}  (Kriss 1994)
and spectral regions unaffected by absorption features
to fit the continuum and emission lines in the \fuse\ and \stis\ spectra
of Mrk~279.  For the continuum, we fit 
a power law 
of the form $f_{\lambda} \propto \lambda^{-\alpha}$, and for the 
emission lines, we fit Gaussian profiles. 
The continuum and emission line fits at each of the four observation epochs
are plotted in Figures~\ref{fig:fusespec} and \ref{fig:stisspec}
and the fit parameters are tabulated in
Tables~\ref{table-contspec} and 
\ref{table-emspec}. 

We include 
the Galactic extinction
law of Cardelli, Clayton \& Mathis (1989) with R$_{V}=3.1$ and 
E(B-V)=0.016 (Schlegel, Finkbeiner, \& Davis 1998) in the continuum fits.
The flux at 1000~\AA\ faded by 21\% between the 1999 December 
and 2000 January observations.  
The fitted spectral index to the 927-1187~\AA\ range 
is bluer for the brighter spectrum by a factor of 1.5.  
By 2002, the continuum flux of Mrk~279 at 1000~\AA\ was $\sim$7.5 times fainter than 
observed in 1999 December.

Since the 2002 May \fuse\ and \stis\ spectra 
were taken simultaneously, we
combine them into a single
spectrum for the continuum and emission line fits.
The fits to the \stis\ spectrum are shown in Figure~\ref{fig:stisspec}.
The emission features fitted in each \fuse\ spectrum  
include broad (FWHM $\approx$ 7000-10,000~\kms)
Ly$\beta$, O~{\sc vi}~\lam\lam1032,1038, S~{\sc vi}~\lam\lam933,944, 
C~{\sc iii}~\lam977,
N~{\sc iii}~\lam991, S~{\sc iv}~\lam\lam~1063, 1073, and He~{\sc ii}~\lam1085. 
(See, e.g.
Zheng et al.\ 2001, Telfer et al.\ 2002.)
These are labeled in Figure~\ref{fig:fusespec}.
The Ly$\beta$ contributions to the O~{\sc vi} emission complexes were
required to achieve a good fit to the profiles, consistent with the
results of Laor et al.\ (1994, 1995).
Also included in the fit are Ly$\beta$ and 
O~{\sc vi}~\lam\lam1032,1038 components of intermediate velocity width 
(FWHM $\approx$
1200-2000~\kms). We include a narrow
component of
Ly$\beta$, primarily because the emission line fit to the \stis\ 
spectrum described below requires this 
narrow component in Ly$\alpha$ to achieve a reasonable fit.
We allow its flux to vary in all fits but keep the centroid and FWHM
fixed to values found from the fit to the 2002 May \fuse\ and \stis\
spectra, in which the width of the narrow component 
is restricted to a range that is consistent with 
widths found for narrow, forbidden [O{\sc iii}]\lam5007, $\sim$600-700 \kms\
(Heckman et al.\ 1981; Feldman et al.\ 1982).
We fix 
the relative intensities of 
all doublets to the optically thin ratio of 2:1. 

In the \stis\ portion of the spectrum 
we fit
broad emission features due to Si~{\sc ii}~\lam1192, Ly$\alpha$, 
N~{\sc v}~\lam\lam1238, 1242, 
Si~{\sc ii}~\lam1260, O~{\sc i} + Si~{\sc ii}~\lam1304, C~{\sc ii}~\lam1335,
Si~{\sc iv}~\lam\lam1393, 1402, O~{\sc iv]}~\lam1402,
C~{\sc iv}~\lam1549, and
He~{\sc ii}~\lam1640. 
We also include
intermediate velocity width (FWHM $\approx$ 2000~\kms) components of
Ly$\alpha$, N~{\sc v}, 
and C~{\sc iv}, and a narrow (FWHM $\approx$ 700~\kms)
component of Ly$\alpha$ in the fit, as discussed above.  
These features are labeled in Figure~\ref{fig:stisspec}.

\subsection{Interstellar and Intergalactic Absorption Features}
\label{sec-ism}

\begin{figure}
\epsscale{0.83} 
\plotone{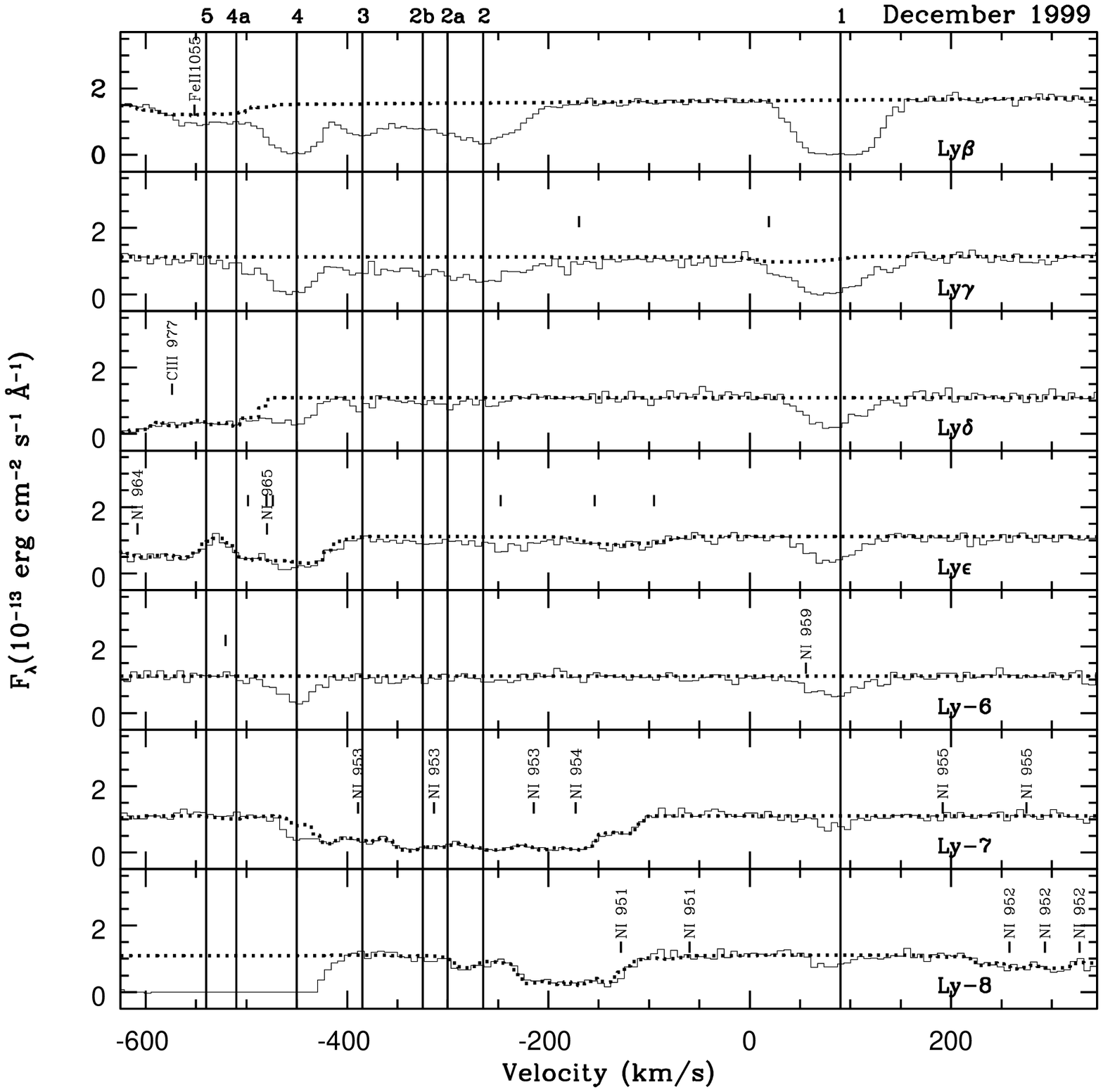}
\caption{\small 1999 December \fuse\ observations:
Velocity components of 
intrinsic Lyman series absorption with continuum, emission line, and ISM fits (dotted lines),
and component velocities defined by Ly$\beta$ (bold vertical lines). Tick marks with
labels are ISM metal lines, those without labels are ${\rm H}_{2}$.
\label{fig:ly99}}
\end{figure}

\begin{figure}
\plotone{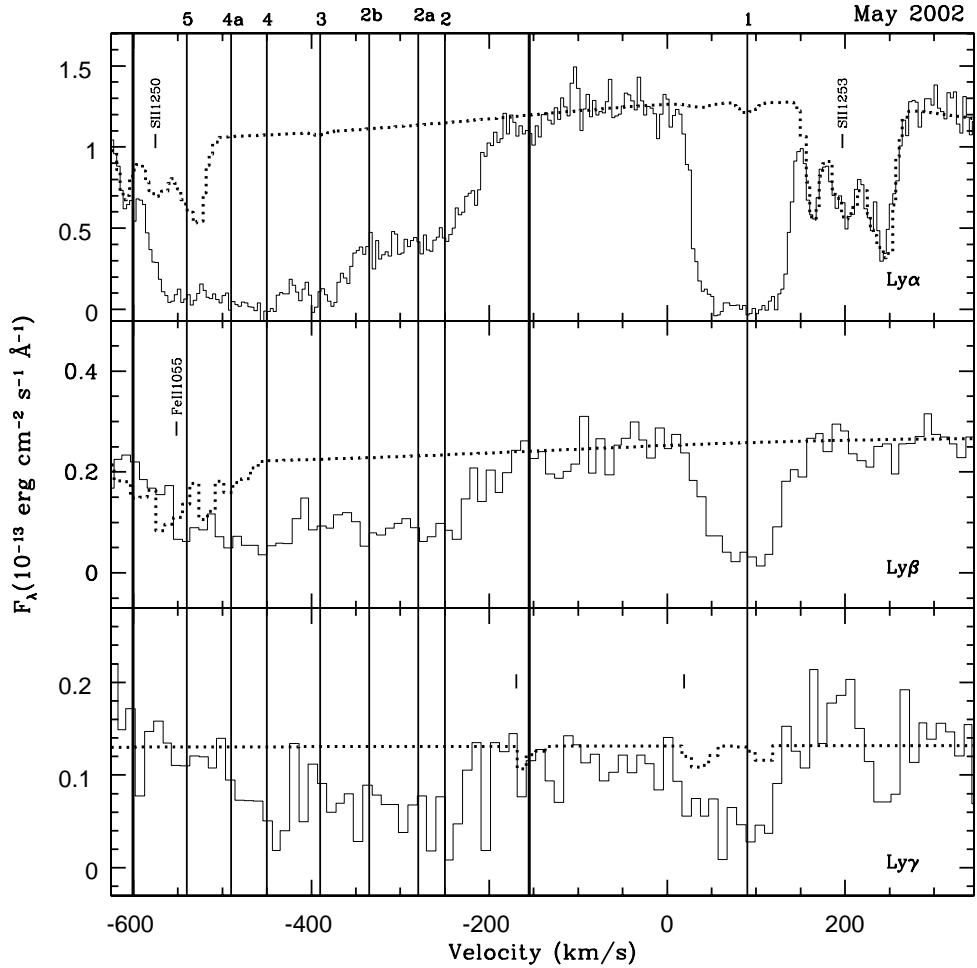}
\caption{\small 2002 May \stis\ and \fuse\ observations:
Velocity components of
intrinsic Lyman series absorption with continuum, emission line, and ISM fits (dotted lines), 
and component velocities defined by Ly$\beta$ (bold vertical lines).
Tick marks with
labels are ISM metal lines, those without labels are ${\rm H}_{2}$.
Bolder vertical lines denote Ly$\alpha$ velocity
components not visible in the \fuse\ data.  \label{fig:ly502}}
\end{figure}

The interstellar medium (ISM) in the direction of Mrk~279 is kinematically complex,
as this line of sight
intersects the high-velocity cloud Complex C (Wakker et al.\ 2001;
Collins, Shull, \& Giroux 2002).
We find eight components of
the ISM absorption in a prominent but unsaturated line such as O~{\sc i}~\lam936
at heliocentric velocities of
-145,  -109, -85, -57, -30, -9, 5, and 20~\kms.
We fit three different models
to the interstellar lines that contaminate the intrinsic absorption features
using {\it specfit}.
The first of these employs the published oscillator strengths of the
ISM lines (Morton 1991; Howk et al.\ 2000). The second model uses average
$f$ ratios from best fit
ISM models to six other \fuse\ spectra of AGNs
that show no intrinsic absorption: Mrk~335, PKS~0558-504, PG~0804+761,
3C~273, Mrk~1383, and Mrk~817.
The third ISM model uses empirically determined $f$ ratios
from the \fuse\ spectrum of Mrk~279
itself; the relative $f$-values that give the lowest $\chi^{2}$ are adopted.
This model is illustrated in Figures~\ref{fig:ly99}-\ref{fig:metstis}.
We use the third model for removing the ISM absorption
from the Mrk~279 spectra.  However, we
will evaluate the systematic errors in the removal of ISM contamination
by comparing the results using each of these ISM models.
Mrk~279 lies at a reasonably large Galactic latitude, $b= +46.86\arcdeg$, 
so its spectrum is uncomplicated
by strong molecular hydrogen absorption.
The ISM features contaminating the intrinsic Lyman series lines are 
shown in Figures~\ref{fig:ly99}, \ref{fig:ly00},
\ref{fig:ly02}, and \ref{fig:ly502}.
The ISM features contaminating the intrinsic metal lines are 
shown in Figures~\ref{fig:met99}, \ref{fig:met00},
\ref{fig:met02}, \ref{fig:met502}, and \ref{fig:metstis}.

The interstellar absorption contaminating the intrinsic absorption features
in the \stis\ spectrum are shown in Figures~\ref{fig:ly502}
and \ref{fig:metstis}. 
Component 1 of intrinsic Ly$\alpha$ absorption is blended with
interstellar S~{\sc ii}~\lam1254, which is visible in four prominent velocity components
just redward of the intrinsic absorption. Also, the 
N~{\sc v}~\lam1238 absorption from velocity Component 1 is contaminated
with interstellar C~{\sc i}~\lam1277.  The velocity components of the 
intrinsic absorption
are described in further detail in Section~\ref{sec-intabs} below.

Penton et al.\ (2000)  identify six intergalactic Ly$\alpha$  lines in their
spectrum of Mrk~279 taken with the Goddard High Resolution Spectrograph
on board {\it HST}.
We note that the features they find at 1243.75, 1241.80, and 1241.50~\AA\
coincide with the positions of intrinsic Si~{\sc iii}~\lam1206 for Components 1, 3, and 4,
low-ionization absorption systems, which, as we discuss below, may arise in the host
galaxy of Mrk~279.
Also, while we see no feature in the \stis\ spectrum
that corresponds to the feature they
identify at 1236.9~\AA, it lies at the expected wavelength of  
intrinsic N~{\sc i}~\lam1200 for
Component 1.

\subsection{Intrinsic Absorption Features}
\label{sec-intabs}

\begin{figure}
\plotone{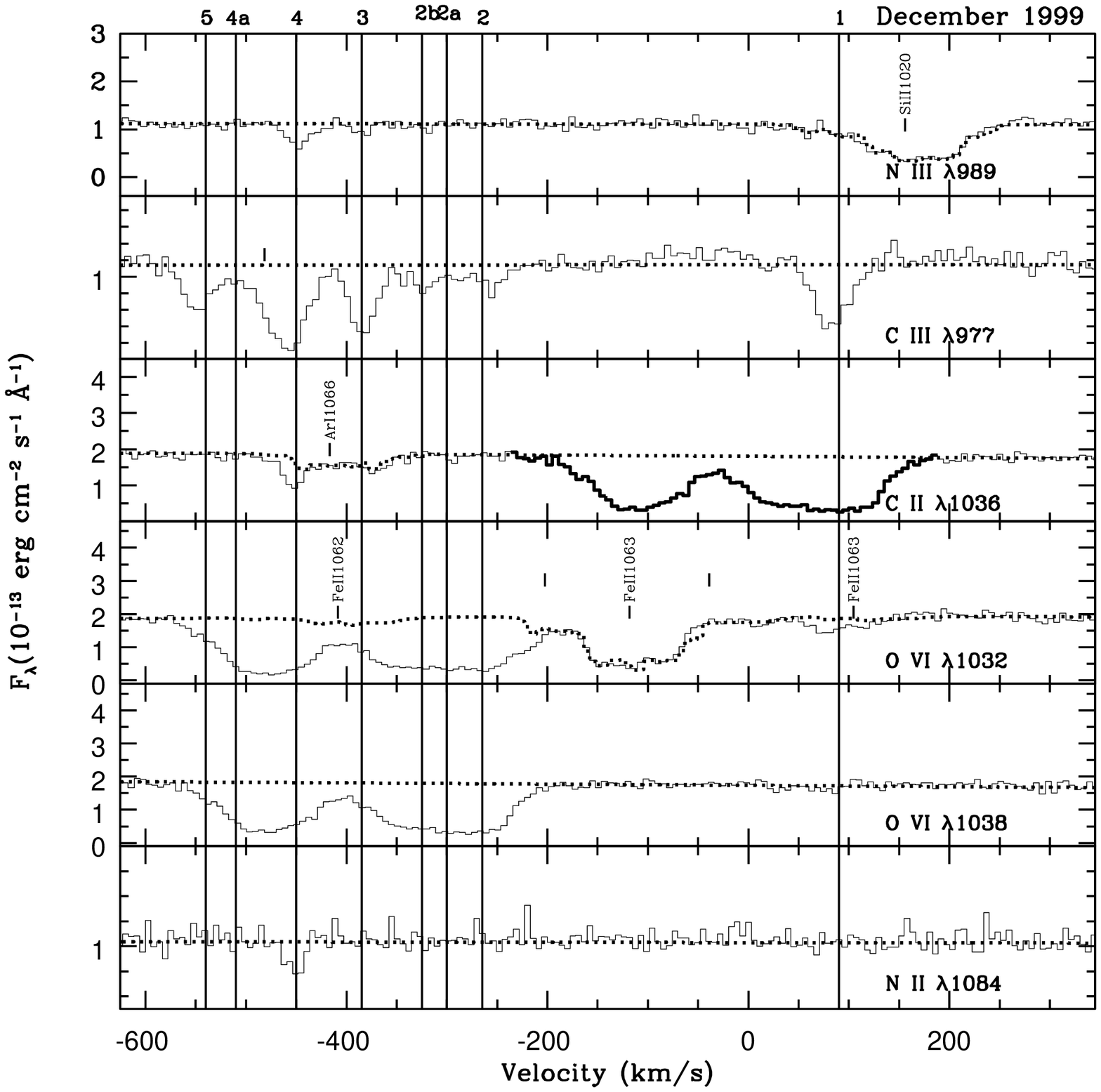}
\caption{\small 1999 December \fuse\ observations: 
Velocity components of 
intrinsic metal absorption with continuum, emission line, and ISM fits (dotted lines), 
and component velocities defined by Ly$\beta$ (bold vertical lines).
Tick marks with
labels are ISM metal lines, those without labels are ${\rm H}_{2}$. 
Bold segment on C~{\sc ii}~\lam1036 plot indicates region contaminated by
O~{\sc vi}~\lam1038 absorption.
\label{fig:met99}}
\end{figure}

\begin{figure}
\plotone{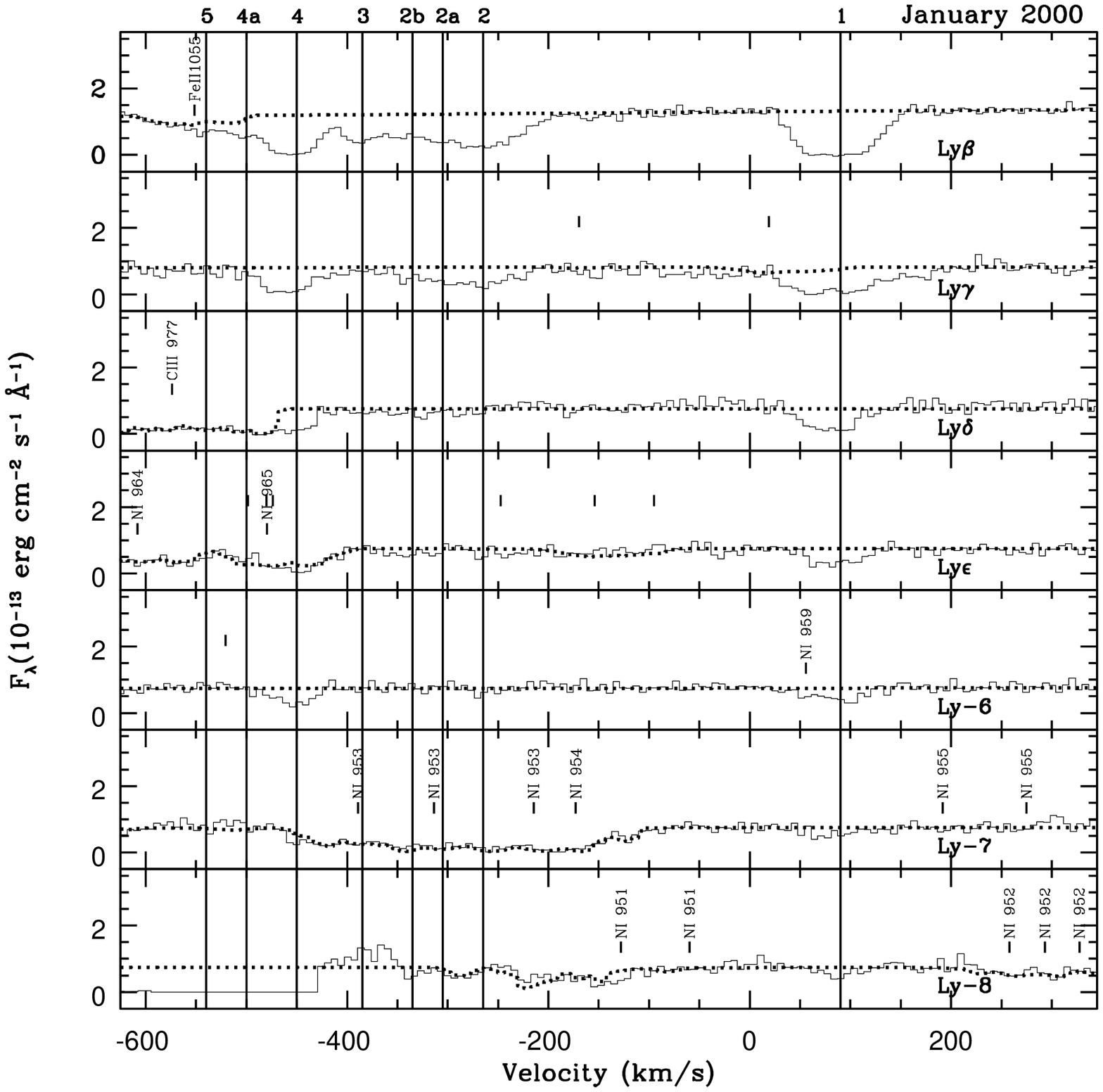}
\caption{\small 2000 January \fuse\ observations:
Velocity components of
intrinsic Lyman series absorption with continuum, emission line, and ISM fits (dotted lines),
and component velocities defined by Ly$\beta$ (bold vertical lines).
Tick marks with
labels are ISM metal lines, those without labels are ${\rm H}_{2}$.
\label{fig:ly00}}
\end{figure}

\begin{figure}
\plotone{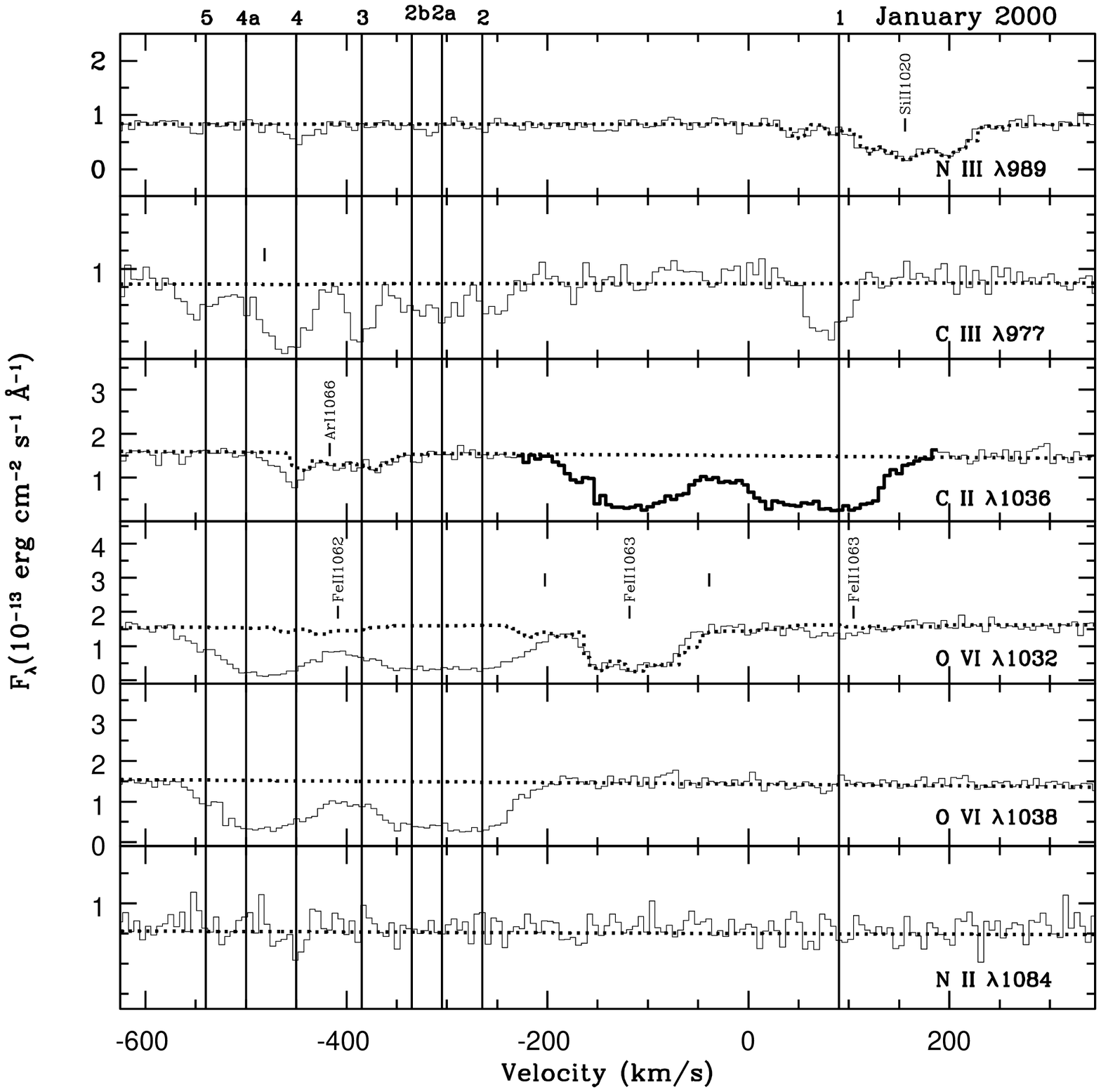}
\caption{\small 2000 January \fuse\ observations:
Velocity components of
intrinsic metal absorption with continuum, emission line, and ISM fits (dotted lines),
and component velocities defined by Ly$\beta$ (bold vertical lines)
Tick marks with
labels are ISM metal lines, those without labels are ${\rm H}_{2}$.
Bold segment on C~{\sc ii}~\lam1036 plot indicates region contaminated by
O~{\sc vi}~\lam1038 absorption.
\label{fig:met00}}
\end{figure}

\begin{figure}
\plotone{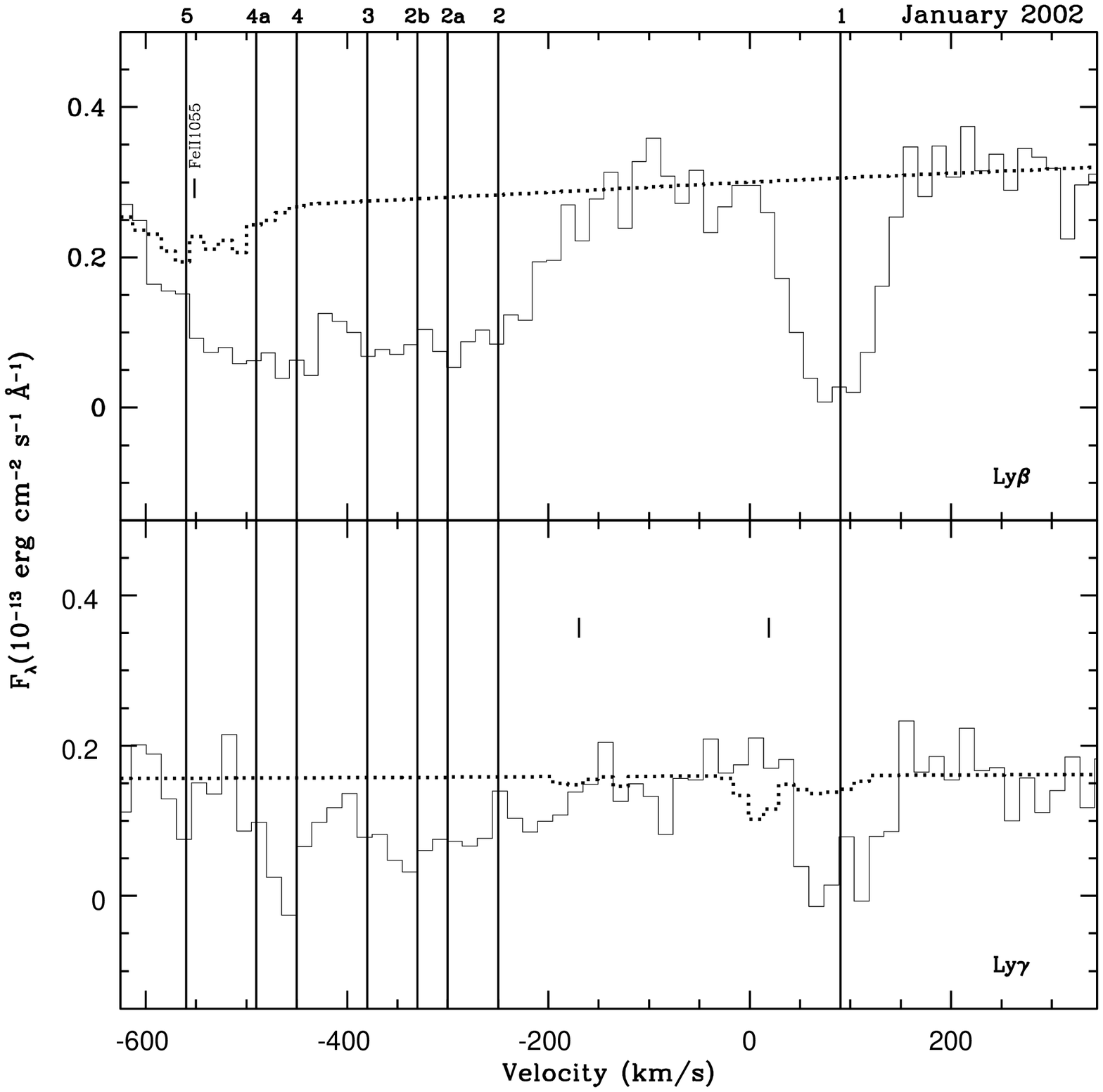}
\caption{\small 2002 January \fuse\ observations:
Velocity components of
intrinsic Lyman series absorption with continuum, emission line, and ISM fits (dotted lines),
and component velocities defined by Ly$\beta$ (bold vertical lines).
Tick marks with
labels are ISM metal lines, those without labels are ${\rm H}_{2}$.
\label{fig:ly02}}
\end{figure}

\begin{figure}
\plotone{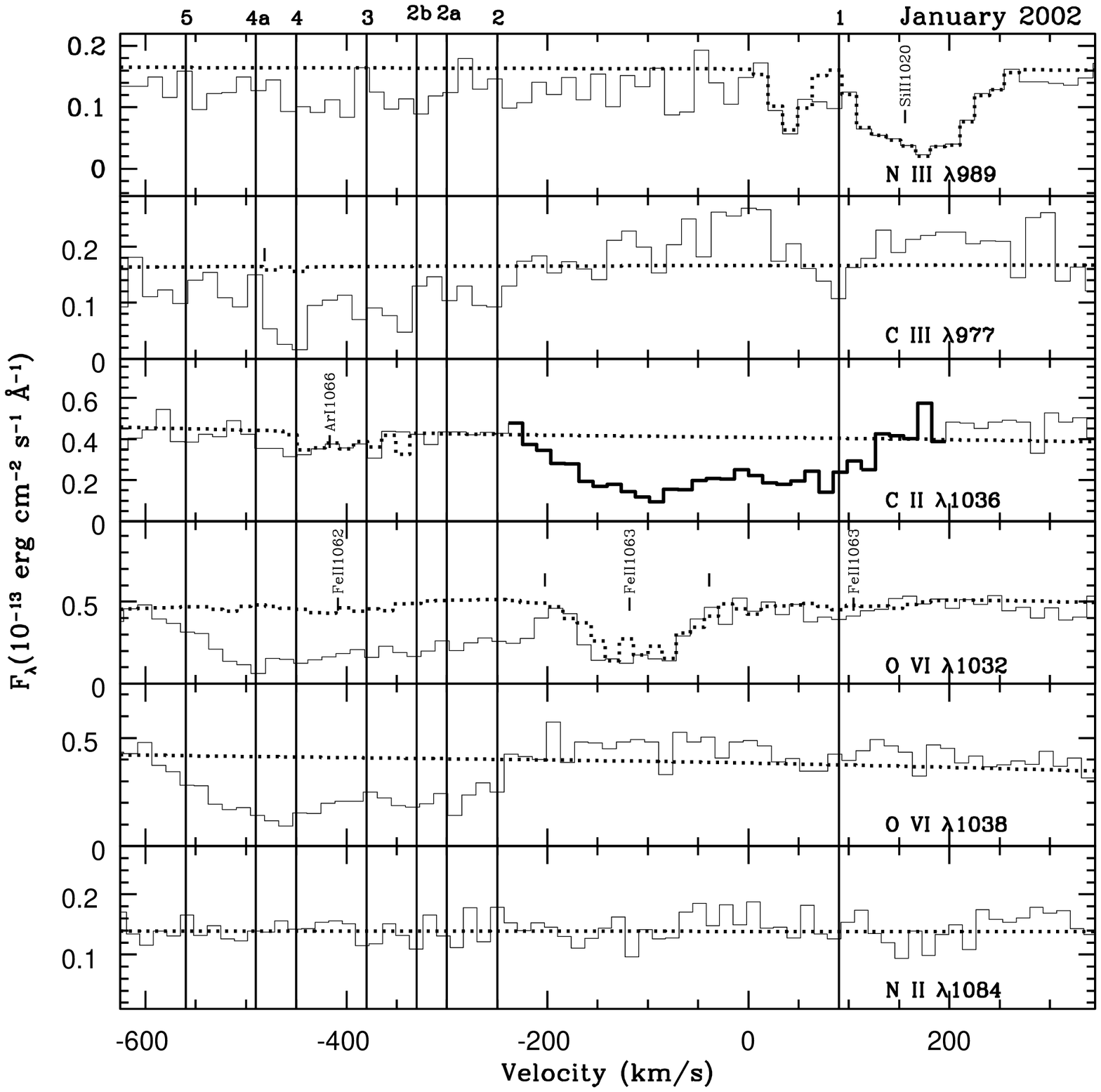}
\caption{\small 2002 January \fuse\ observations:
Velocity components of
intrinsic metal absorption with continuum, emission line, and ISM fits (dotted lines),
and component velocities defined by Ly$\beta$ (bold vertical lines).
Tick marks with
labels are ISM metal lines, those without labels are ${\rm H}_{2}$.
\label{fig:met02}}
\end{figure}

\begin{figure}
\plotone{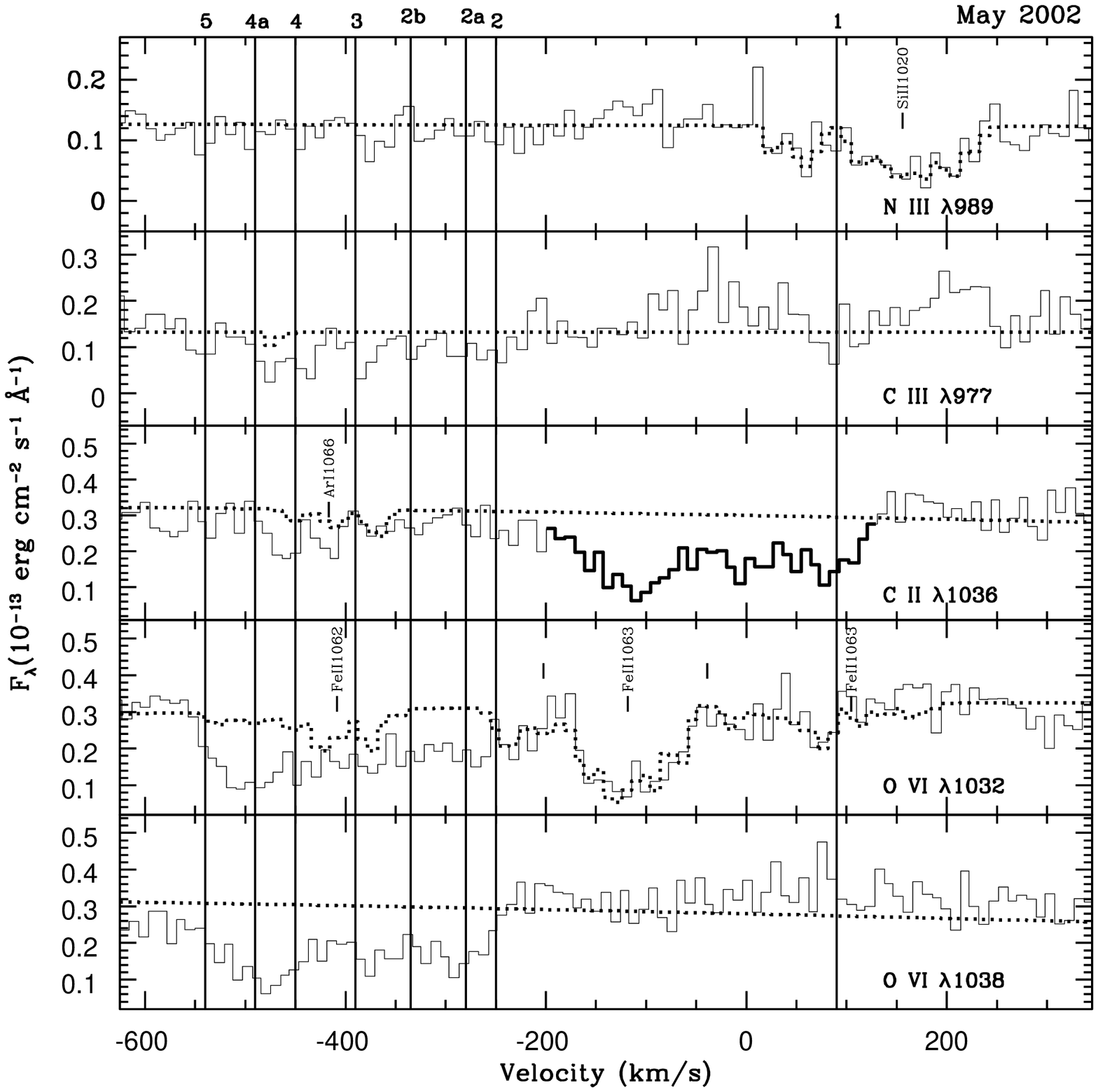}
\caption{\small 2002 May \fuse\ observations:
Velocity components of
intrinsic metal absorption with continuum, emission line, and ISM fits (dotted lines), 
and component velocities defined by Ly$\beta$ (bold vertical lines).
Tick marks with
labels are ISM metal lines, those without labels are ${\rm H}_{2}$.
Bold segment on C~{\sc ii}~\lam1036 plot indicates region contaminated by
O~{\sc vi}~\lam1038 absorption.
\label{fig:met502}}
\end{figure}

\begin{figure}
\plotone{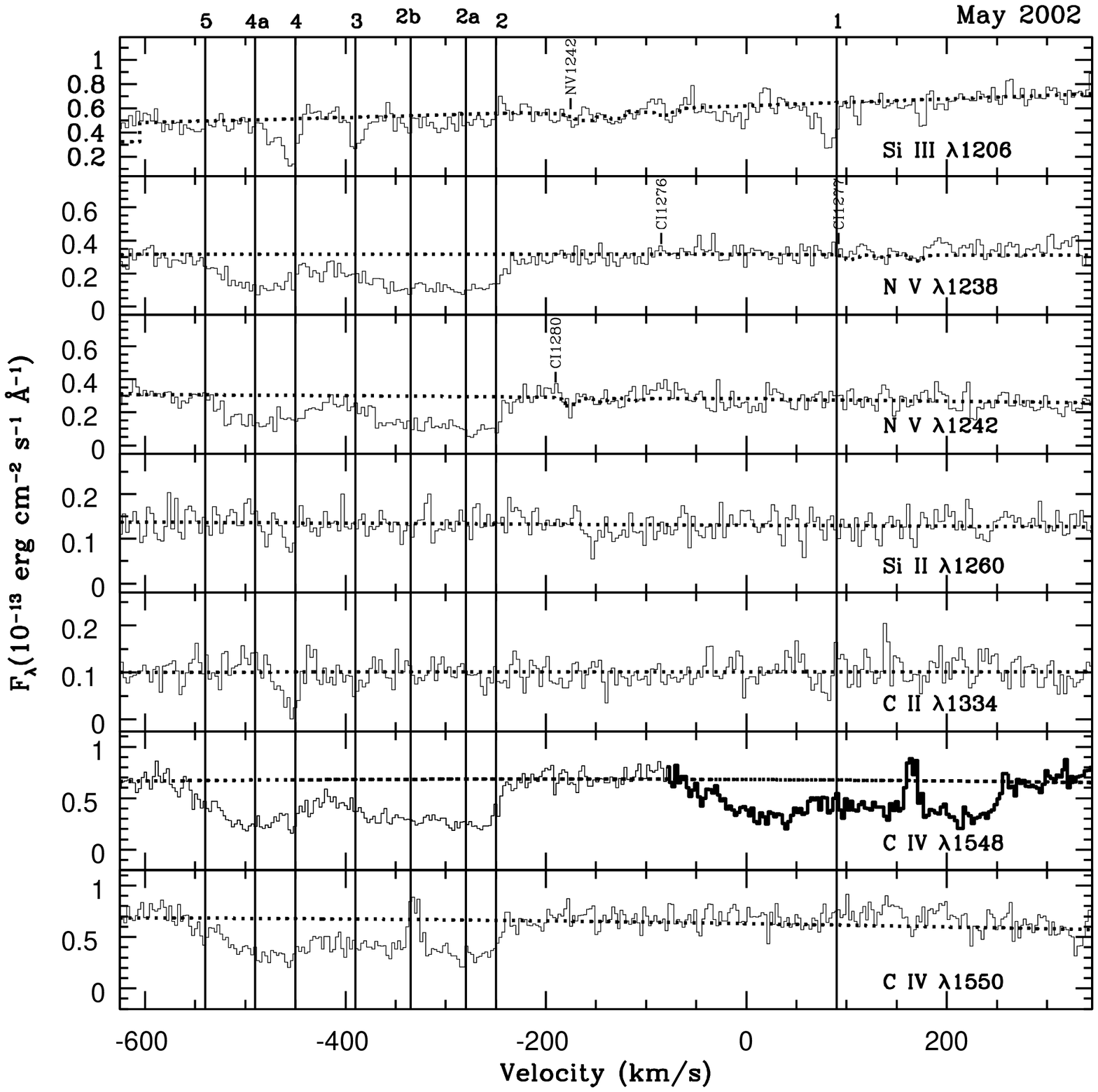}
\caption{\small 2002 May \stis\ observations:
Velocity components of
intrinsic metal absorption with continuum, emission line, and ISM fits (dotted lines), 
and component velocities defined by Ly$\beta$ (bold vertical lines).
Tick marks with
labels are ISM metal lines.
Bold segment on C~{\sc iv}~\lam1548 plot indicates region contaminated by
C~{\sc iv}~\lam1550 absorption.
\label{fig:metstis}}
\epsscale{1.0}
\end{figure}

Of our four \fuse\ spectra of Mrk~279, the
1999 December spectrum has the highest S/N.  We therefore
use this as a basis for comparison with the other three.
The intrinsic absorption features in the \fuse\ and \stis\ spectra of
Mrk~279 extend over $\sim$700 \kms.
We identify five distinct components we label Components 1-5, with
subcomponents associated with
Components 2 and 4.
The delineation of individual velocity
components of the intrinsic absorption is
problematic and somewhat arbitrary, particularly in cases where
low-ionization and high-ionization lines show different absorption
trough structures.
We will therefore attempt to describe the properties of the
intrinsic absorption as a continuous function of velocity,
using the individual velocity components for reference purposes, and
when speaking of the average properties of the absorption in the line cores.
The positions of the individual components in Ly$\beta$, C~{\sc iii}, and
O~{\sc vi}~\lam1032 are listed
in Table~\ref{table-vel}, and the Ly$\beta$ centroids 
are indicated by bold vertical lines in
Figures~\ref{fig:ly99}-\ref{fig:metstis}.

In our discussion of the intrinsic absorption features in  
the spectrum of Mrk~279, we use
the term $``$intrinsic" to refer to
absorption arising in gas associated either with an outflow
from the AGN itself, 
or with the interstellar medium in the host galaxy
of the AGN.   In fact we shall argue that 
several velocity components of the intrinsic absorption, namely, the low-ionization
species in Components 1, 3, 4, and perhaps 5,
are likely to arise in the host galaxy, either
in gas associated with the disk or halo, in material analogous to
the high-velocity clouds (HVCs) observed in the Milky Way,
or in gas participating in an interaction with a companion galaxy to Mrk~279,
MCG +12-13-024.
In these components,
we find \hi\ covering fractions consistent with unity,
prominent C~{\sc iii}~\lam977 in most \fuse\ epochs, 
and Si~{\sc iii}~\lam1206 lines in the \stis\ spectrum.  
Component 4 also shows absorption
from other low-ionization species such as
N~{\sc iii}, C~{\sc ii}, and N~{\sc ii}.
In general, the centroids of these 
features agree with the centroids of the
Ly$\beta$ lines, illustrated by the vertical lines in 
Figures~\ref{fig:ly99}-\ref{fig:metstis}.
Figures~\ref{fig:met99}, \ref{fig:met00}, \ref{fig:met02},
\ref{fig:met502}, and \ref{fig:metstis}
show that the low-ionization absorption line profiles
such as those of C~{\sc iii} and C~{\sc ii} differ from
the high-ionization profiles such as O~{\sc vi}, 
or C~{\sc iv} in the \stis\ data, in velocity centroids
and, to a greater extent, in overall profile shapes.
In Figure~\ref{fig:c3comp}, we show comparisons of 
the C~{\sc iii}~\lam977 and O~{\sc vi}~\lam1032 flux profiles
with Ly$\beta$ in the 1999 December \fuse\ spectrum.  
The comparison between C~{\sc iii}~\lam977 and Ly$\beta$ 
in the top panel shows that the
profile shapes are in general agreement, although
the C~{\sc iii}~\lam977
lines are narrower in velocity extent by $\sim$15 \kms\ on average, and 
the C~{\sc iii} line in Component 4 is blueshifted
with respect to Ly$\beta$ by $\sim$10 \kms.
There are distinct differences
between Ly$\beta$ and O~{\sc vi}~\lam1032, which we compare
in the bottom panel of Figure~\ref{fig:c3comp}. 
Component 1 shows little O~{\sc vi}~\lam1032
absorption and the component we have labeled Component 5 forms the blue
wing of the O~{\sc vi} profile of Component 4/4a. 
The Ly$\beta$ and O~{\sc vi} profiles of Components 2-4 are dramatically
different in their overall shapes. 
The O~{\sc vi}/\hi\ ratios in Components 2a and 2b
are substantially larger relative to the ratio in Component 2. 
Also, Component 3 is less well-defined in O~{\sc vi} than in Ly$\beta$, and
the profiles of Components 4 and 4a in these two species are highly offset with
respect to one another.

Before we continue with the detailed 
discussion of the intrinsic absorption in Mrk~279,
some words about the systemic redshift of this galaxy are in order.
The redshifts listed in the NASA Extragalactic Database
span nearly 700 \kms.  If we restrict ourselves to recent references, we find
$z=0.030608\pm0.000113$ in the 
Third Reference Catalog of Bright Galaxies  (RC3, de Vaucouleurs et al.\ 1991)
and $z=0.029937\pm0.000133$ in the
Updated Zwicky Catalog (UZC, Falco et al.\ 1999).
The RC3 redshift is a weighted mean of 
various measurements found in the literature, while the redshifts in the UZC are 
based on a more homogeneous set of spectra taken as part of the CfA Redshift Survey.
However, the UZC measurement is the result of a cross correlation with both
absorption line templates and narrow emission line templates, the latter of which may
bias the final redshift owing to bulk motions of material in relatively close
proximity to the central engine of the AGN.  Therefore, we choose a redshift
that is measured from the UZC spectral data, but based on absorption lines only.
For the systemic redshift of Mrk~279, we adopt
$z=0.0305 \pm 0.0003$ (Falco 2003, private communication).

We find intrinsic absorption in the 1999 December \fuse\ spectrum of
Mrk~279 
extending from -570 to +145~\kms.
In the strongest components, we identify up to 10 Lyman series absorption features,
although to avoid confusion with strong Galactic interstellar lines and
with overlapping Lyman series lines from adjacent velocity components,
the highest order Lyman series line we use in any one component is
Ly-8. 
The velocity component with the largest blueshift with respect to systemic, Component 5, 
shows little neutral hydrogen absorption. It is
seen only in Ly$\beta$, where it is blended with interstellar Fe~{\sc ii}~\lam1055, 
and possibly in Ly$\gamma$.  This component
is visible mainly in C~{\sc iii},  and as a blue wing on the 
broad O~{\sc vi} profile.  Conversely, Component 1 
shows strong Lyman series absorption and
strong C~{\sc iii}~\lam977, but a weak
O~{\sc vi} doublet.  We can identify the Lyman series in this component reliably
to Ly-8 at a rest wavelength of 923.15~\AA.  

At some velocities within the 
-570 to +145~\kms\ range, Ly$\beta$ and the
1032~\AA\ component of O~{\sc vi} are contaminated by
Galactic Fe~{\sc ii} absorption, though the 1038~\AA\ component is unaffected by
any ISM features.  Absorption due to C~{\sc iii}~\lam977 is present over the
entire velocity range of the intrinsic absorption; and
notable N~{\sc iii}~\lam989 features are visible at
$-385$ and $-450$~\kms, in Components 3 and 4.
Absorption features due to intrinsic C~{\sc ii}~\lam1036 and
N~{\sc ii}~\lam1084 are identified in Component 4 only.
If C~{\sc ii}~\lam1036 is present in Component 1, it 
is blended with the intrinsic O~{\sc vi}~\lam1038 lines of Component 2, 2a, and 2b.
We note, however, that we detect no C~{\sc ii}~\lam1334 absorption at the expected position for
Component 1 in the 2002 May \stis\ spectrum.
We show the intrinsic metal lines in Figure~\ref{fig:met99},
with the
velocity components defined by the Ly$\beta$ lines, listed
in Table~\ref{table-vel}.  As discussed above,
the centroids of the C~{\sc iii} features
agree with the \hi\ centroids, and
the broad O~{\sc vi} profile in Components 4 and 4a is
offset in velocity with respect to the
Ly$\beta$ profiles.  
The near-equal depths of the two O~{\sc vi} doublet lines in Components 2 and 5
indicate that the absorption is saturated.
Therefore, the nonzero flux levels in the centers of these 
O~{\sc vi} features 
imply partial coverage of the O~{\sc vi}-absorbing gas.

In the 2000 January \fuse\ spectrum, 
we find the intrinsic
absorption extends from -570 to +145~\kms.   Once again, Component 1
is visible down to Ly-8 with strong C~{\sc iii}~\lam977 and weak
O~{\sc vi} \lam\lam1032,1038 features; and
Component 5
shows relatively weak Ly$\beta$, blended with Fe~{\sc ii}~\lam1055, with
a prominent O~{\sc vi} doublet, and well-defined C~{\sc iii}~\lam977 absorption.
As in the 1999 December spectrum, N~{\sc iii}~\lam989, C~{\sc ii}~\lam1036, and
N~{\sc ii}~\lam1084 features are seen in Component 4.
The velocity components are
tabulated in Table~\ref{table-vel}, and they
are plotted in Figures~\ref{fig:ly00}
and \ref{fig:met00}.

In the 2002 January \fuse\ spectrum, the low signal-to-noise limits
the analysis of the \hi\ lines to Ly$\beta$ and Ly$\gamma$.
The intrinsic
absorption extends from -595 to +145~\kms.   The above description of Components
1 and 5 still holds at this epoch. 
We see no significant N~{\sc iii}~\lam989 or N~{\sc ii}~\lam1084 lines, but
we do identify C~{\sc ii}~\lam1036 in Component 4. 
The velocity components are
tabulated in Table~\ref{table-vel}, and the intrinsic features 
are plotted in Figures~\ref{fig:ly02}
and \ref{fig:met02}.  We show the expected positions of all the 
metal lines discussed above in the 1999 December and 2000 January
spectra for reference, even if we do not identify a significant 
absorption feature.

In the 2002 May \fuse\ spectrum, 
the intrinsic Ly$\beta$
absorption extends from -530 to +145~\kms. Since there is no SiC data
for $\lambda < 1000$ \AA\ for this observation, the analysis is 
limited to Ly$\beta$ and Ly$\gamma$.
The \stis\ spectrum,
taken simultaneously with the 2002 May \fuse\ spectrum, shows
absorption due to Ly$\alpha$ extending from -610 to +145~\kms,
as listed  in Table~\ref{table-vel}.  
We identify two velocity components in Ly$\alpha$ 
that are not visible in the higher-order transitions
in the \fuse\ data:  at -155 and -600~\kms.
These components are marked by bold vertical lines in 
Figure~\ref{fig:ly502}.  
The absorption complex at 1253.4-1253.8~\AA\ is
attributed to S~{\sc ii}~\lam1254, indicating the 
presence of the S~{\sc ii}~\lam1250 line at the position of the
-605~\kms component of the intrinsic Ly$\alpha$.  We take this
feature
into account in our Galactic ISM model and find that the -605~\kms\ feature 
cannot be accounted for by Milky Way ISM absorption alone.

We show the intrinsic metal lines in Figure~\ref{fig:met502}, with the
velocity components defined by the Ly$\beta$ lines, as before.
At this epoch, Component 1 still shows strong \hi\ absorption and a weak 
O~{\sc vi} doublet. Component 5 is also still relatively weak in Ly$\beta$,
but it is reasonably prominent in Ly$\alpha$.  This component is
coincident in velocity with the bluest edge of the broad O~{\sc vi} absorption profile.
We identify no significant N~{\sc iii}~\lam989 features in this spectrum, but
Component 4 does show  C~{\sc ii}~\lam1036 absorption at this epoch.
A possible component of
C~{\sc ii}~\lam1036 is seen at -420 \kms, though this is likely to be due
to Ar~{\sc i}~\lam1067 in the ISM, as in the spectra taken at other epochs
(see Figures~\ref{fig:met99} and \ref{fig:met00}).

The intrinsic metal absorption identified in the \stis\ spectrum
is shown in Figure~\ref{fig:metstis}.
We find Si~{\sc iii}~\lam1206 in Components 1, 3, and 4, and
Si~{\sc ii}~\lam1260 and C~{\sc ii}~\lam1334 in Component 4 only.
We identify the N~{\sc v}~\lam\lam1238,1242 and C~{\sc iv}~\lam\lam1558,1550 doublets
in all velocity components with the exception of Component 1.
In Component 1, the \lam1548 C~{\sc iv} line
is blended with the \lam1550 line of Component 3, but the
absence of the \lam1550 feature at the velocity of Component 1, combined with 
the lack of N~{\sc v}~\lam\lam1238,1242 and weak O~{\sc vi}~\lam\lam1032,1038, indicate
that there is no significant C~{\sc iv} absorption at this velocity.  
We find no detectable Si~{\sc iv} features in the \stis\ spectrum.

\subsection{Covering Fractions, Optical Depths and Column Densities}
\subsubsection{Lyman series}
\label{sec-ly}
The availability of many lines in the Lyman series of hydrogen allows
a direct solution for the line and continuum covering fractions of the
absorbing medium and its optical depth. 
The normalized flux at a position in the absorption profile, ${\cal I}$,
is a function of the optical depth and the covering fraction,
including both line and continuum components:
\begin{equation}
{\cal I}= R_{l}[C_{l} e^{-\tau_{l}} + (1-C_{l})] + R_{c}[C_{c} e^{-\tau_{c}}
 + (1-C_{c})],
\label{equ:cover}
\end{equation} 
where $R_{l,c}= F_{l,c}/(F_{l}+F_{c})$ is the relative
flux in the line or continuum and $C_{l,c}$ is the
covering fraction in the line or continuum, and
where in general it is assumed that $\tau_{c}=\tau_{l}$
(Ganguly et al.\ 1999.)

We determine the best-fit continuum and emission line covering fraction
and the optical depth using information from the entire Lyman series.
At each velocity position, $i$, we construct 
the $\chi^{2}$ function of merit by
summing over the Lyman series components, $x$, in the \fuse\ and \stis\ data:
\begin{equation}
\chi_{i}^{2} = \sum_{x=\alpha,\beta,\gamma,...} \left(
  \frac{{\cal I}_{i}^{x}-I_{i}^{x}}{\sigma^{x}_{i}} \right) ^{2},
\label{equ:minchi}
\end{equation}
where $I_{i}^{x}$ and $\sigma^{x}_{i}$ are the measured
normalized flux and error in Lyman series component $x$ at position $i$, and
the model ${\cal I}_{i}^{x}$ is defined by Equation~\ref{equ:cover}, where
\begin{equation}
\tau_{i}^{x} = \frac{f^{x} \lambda^{x}} {f^{\alpha} \lambda^{\alpha}} \tau_{i}^{\alpha}.
\label{equ:tau}
\end{equation}
The $\chi^{2}$ function is minimized using
a nonlinear least
squares technique to solve for $C_{c}$, $C_{l}$, and $\tau^{\alpha}$ (or $\tau^{\beta}$)
at each velocity.
We use the method described by Savage \& Sembach (1991) to compute
column density profiles from the optical depth profiles:
\begin{equation}
N(v) = \frac{m_{e} c }{\pi e^{2} f \lambda} \tau(v).
\label{equ:column}
\end{equation}

\begin{figure}
\epsscale{0.52}
\plotone{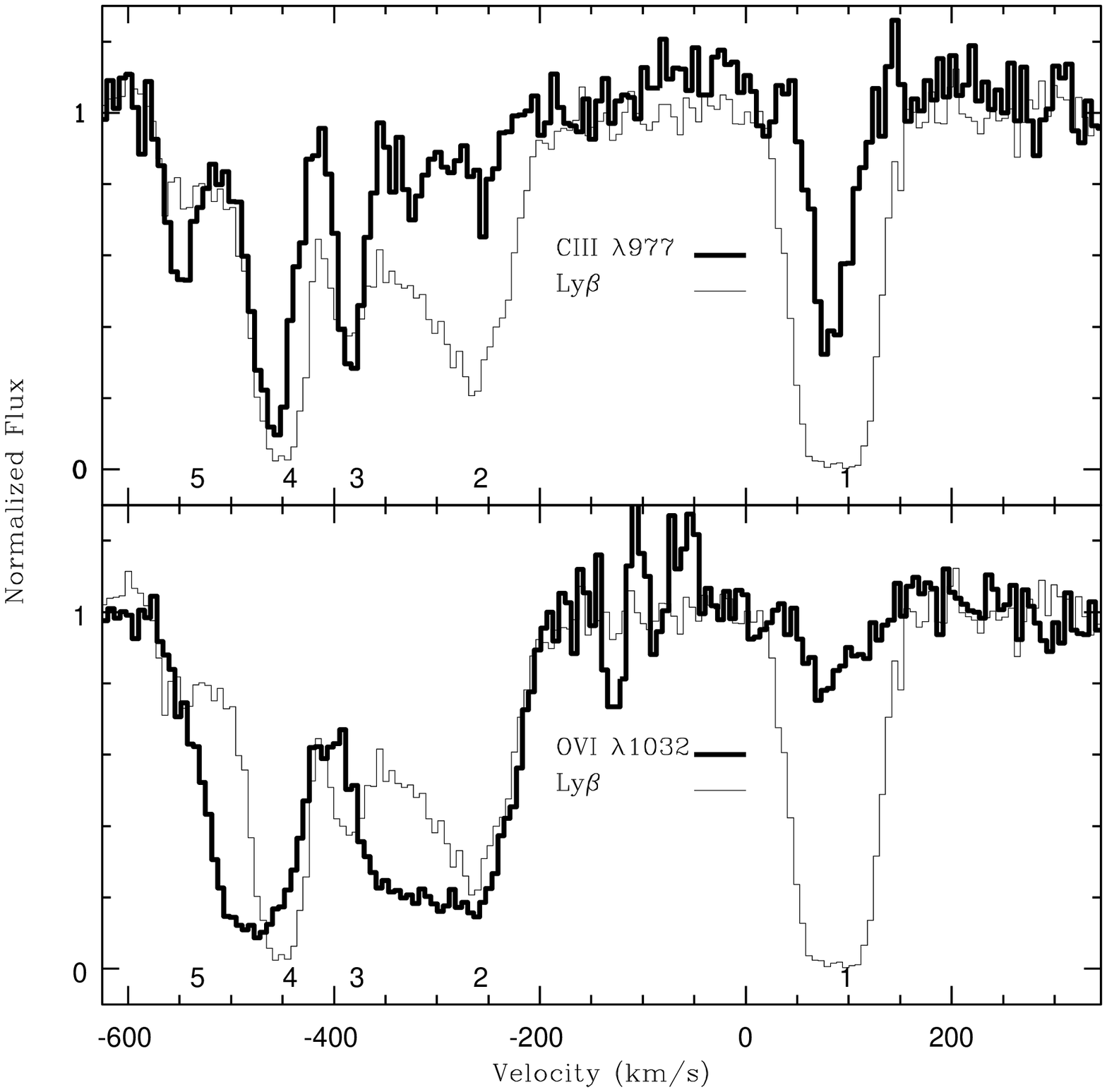}
\caption{\small
Profile comparisons of C~{\sc iii}~\lam~977 and Ly$\beta$ and
O~{\sc vi}~\lam~1032 and Ly$\beta$ in the 1999 December \fuse\ spectrum
\label{fig:c3comp}}
\plotone{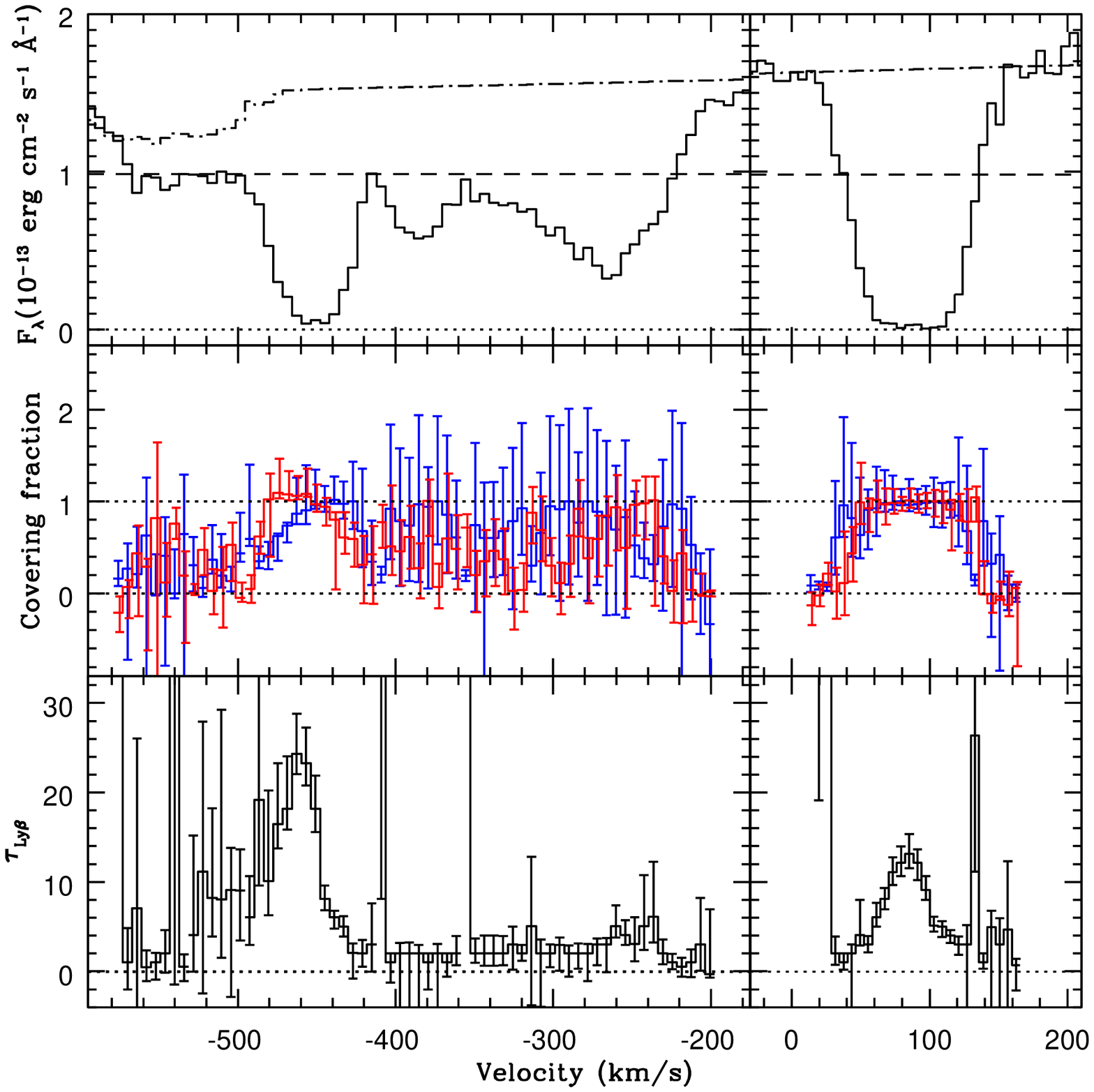}
\caption{\small 1999 December: 
{\it Top panel:} 
Ly~$\beta$ profile with continuum and emission line fits with ISM 
absorption (dot-dashed line) and
the power-law continuum fit alone (dashed line)
{\it Middle panel:} Histogram of solutions to the minimization of
Equation~\ref{equ:minchi} for
$C_{c}$ (blue lines), $C_{l}$ (red lines),
the continuum and line covering fractions vs.\ velocity with respect to systemic
{\it Bottom panel:}
$\tau^{\beta}$, the optical depth in the Ly~$\beta$ line vs.\ velocity.
\label{fig:hist99}}
\epsscale{1.0}
\end{figure}

In Components 1, 2, 3, and 4 of the 1999 December \fuse\ spectrum,
we use the Lyman series lines Ly$\beta$, $\gamma$, $\delta$, Ly-5, and Ly-6
and Equations~\ref{equ:cover}-\ref{equ:tau}
to derive $C_{c}$, $C_{l}$, and $\tau^{\beta}$ at each
resolution element in the intrinsic absorption profile. 
In Figure~\ref{fig:hist99}, we show the Ly$\beta$ absorption profile in the top
panel, with the power law continuum and the full continuum plus emission line fit
with the ISM absorption model.  In the middle and bottom panels on this figure,
we show the corresponding profiles in line and continuum covering fractions
and Ly$\beta$ optical depth from our solution.
The error bars on $C_{c}$, $C_{l}$, and $\tau^{\beta}$, shown in the histogram in
Figure~\ref{fig:hist99}, are 1$\sigma$ confidence limits from $\Delta\chi^{2}=1$.  
For the fit to Component 1, we add Ly-7 and Ly-8 to the Lyman series lines listed above;
and for Component 4, we fit the series down to Ly-7. 
Component 4a suffers from blending with interstellar features:  Ly$\beta$ is
blended with Fe~{\sc ii}~\lam1055, Ly$\delta$ with C~{\sc iii}~\lam977, and 
Ly$\epsilon$ with N~{\sc i}~\lam965.   
This blending is even more 
severe for Component 5, to the extent that it is plausible
to attribute the putative Ly$\beta$ component entirely to Fe~{\sc ii}~\lam1055.
We note, however, that this component is
seen clearly in C~{\sc iii}~\lam977 (Figure~\ref{fig:met99}) and 
it does appear
convincingly in Ly$\alpha$ in the 2002 May \stis\ spectrum (Figure~\ref{fig:ly502}).
Because of the particularly severe blending of
the Ly$\delta$ line of Components 4a and 5 with interstellar O~{\sc i}~\lam976 
and C~{\sc iii}~\lam977,
we fit Ly$\beta$ through Ly-7, excluding Ly$\delta$ for these components.

The average covering fraction and column density in the core of 
each velocity component
are listed in Table~\ref{table-cover2}.
The velocity over which the
column density is calculated, $\Delta v$,  is listed in the final column of the table. 
We determined $\Delta v$ for each component so that (1) the
$\Delta v$ is equal for each species for each component;
(2) no pixels are summed twice for two different adjacent
velocity components of a single species; and (3)
the optical depths
are summed over the core of each absorption feature, generally avoiding the
wings of the lines, particularly for features with large velocity widths.

We use the same Lyman series lines and the
procedure outlined above for solving for the covering fractions, optical depths,
and \hi\ column densities of the absorbers in the 2000 January \fuse\
spectrum.   
For the 2002 May data, the absence of data from the \fuse\ SiC channels
means that we have no coverage of intrinsic lines blueward of Ly$\gamma$.
However, the \stis\ spectrum at this epoch provides coverage of
intrinsic Ly$\alpha$ which we use for the covering fraction and optical
depth solutions.
The results of these calculations
for the 2000 January and the 2002 May spectra are also listed in Table~\ref{table-cover2}.
In some resolution elements, where the absorption is saturated for instance, the
optical depth is poorly constrained on the high end and
the $\chi^{2}$ function has no well-defined minimum.
In these cases, we quote a
lower limit on the column density summed only from velocity bins with derived optical depths
lower than the maximum value
that we can reliably measure.  We estimate this maximum from the S/N in the spectrum
at the location of the absorption component with the lowest optical depth used in the solution.
For the \hi\ solutions that
extend far down the Lyman series, the maximum Ly$\beta$ optical depth can be
as large as $\sim$70. 

\begin{figure}
\epsscale{0.5}
\plotone{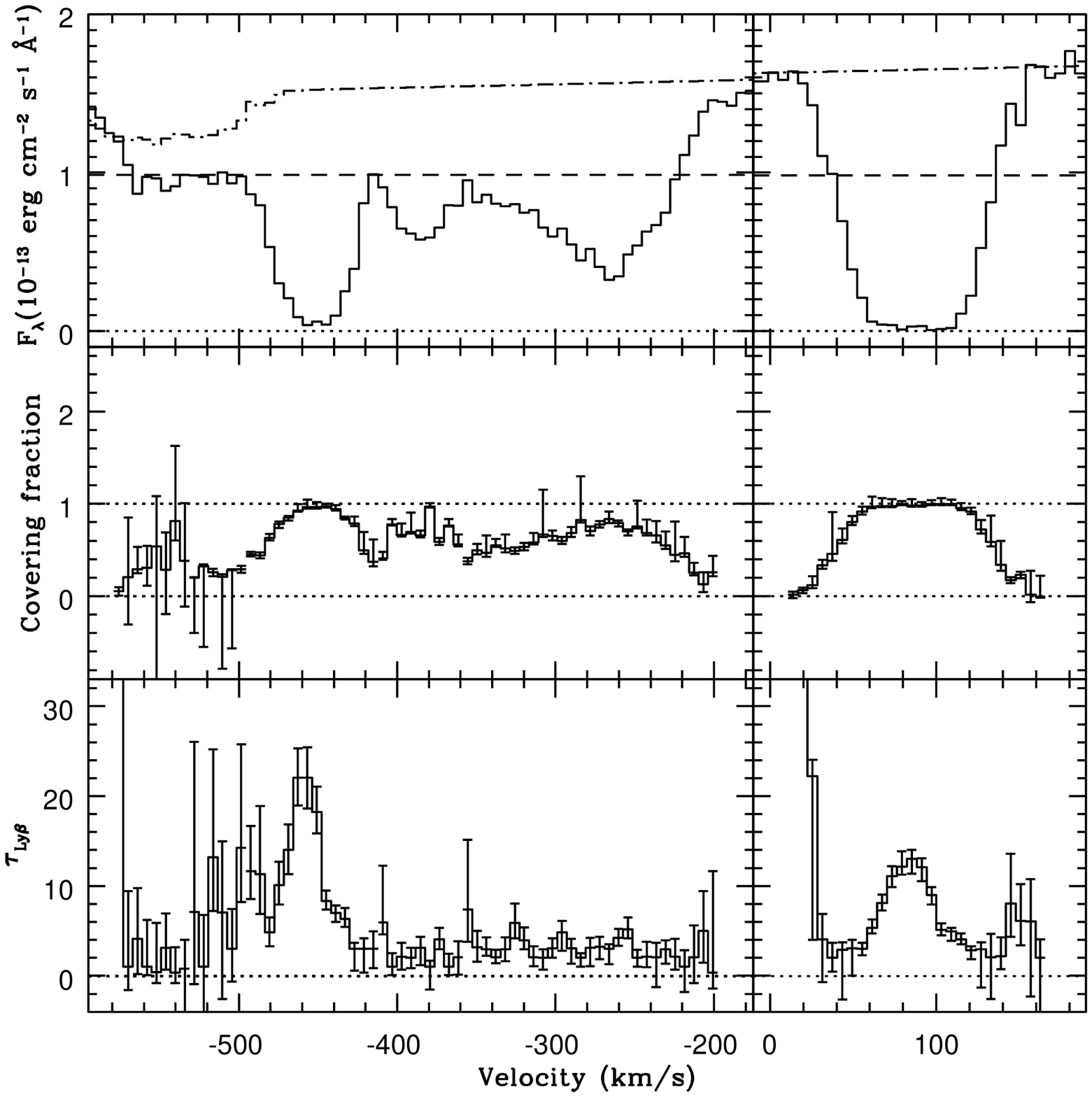}
\caption{\small 1999 December:
{\it Top panel:}
Ly~$\beta$ profile with continuum and emission line fits with ISM absorption (dot-dashed line) and
the power-law continuum fit alone (dashed line).
{\it Middle panel:} Histogram of solutions to the minimization of
Equation~\ref{equ:minchi} for $C_{c}=C_{l}=C_{f}$,
the effective covering fraction vs.\ velocity with respect to systemic.
{\it Bottom panel:}
$\tau^{\beta}$, the optical depth in the Ly~$\beta$ line vs.\ velocity.
\label{fig:hist99eff}}
\plotone{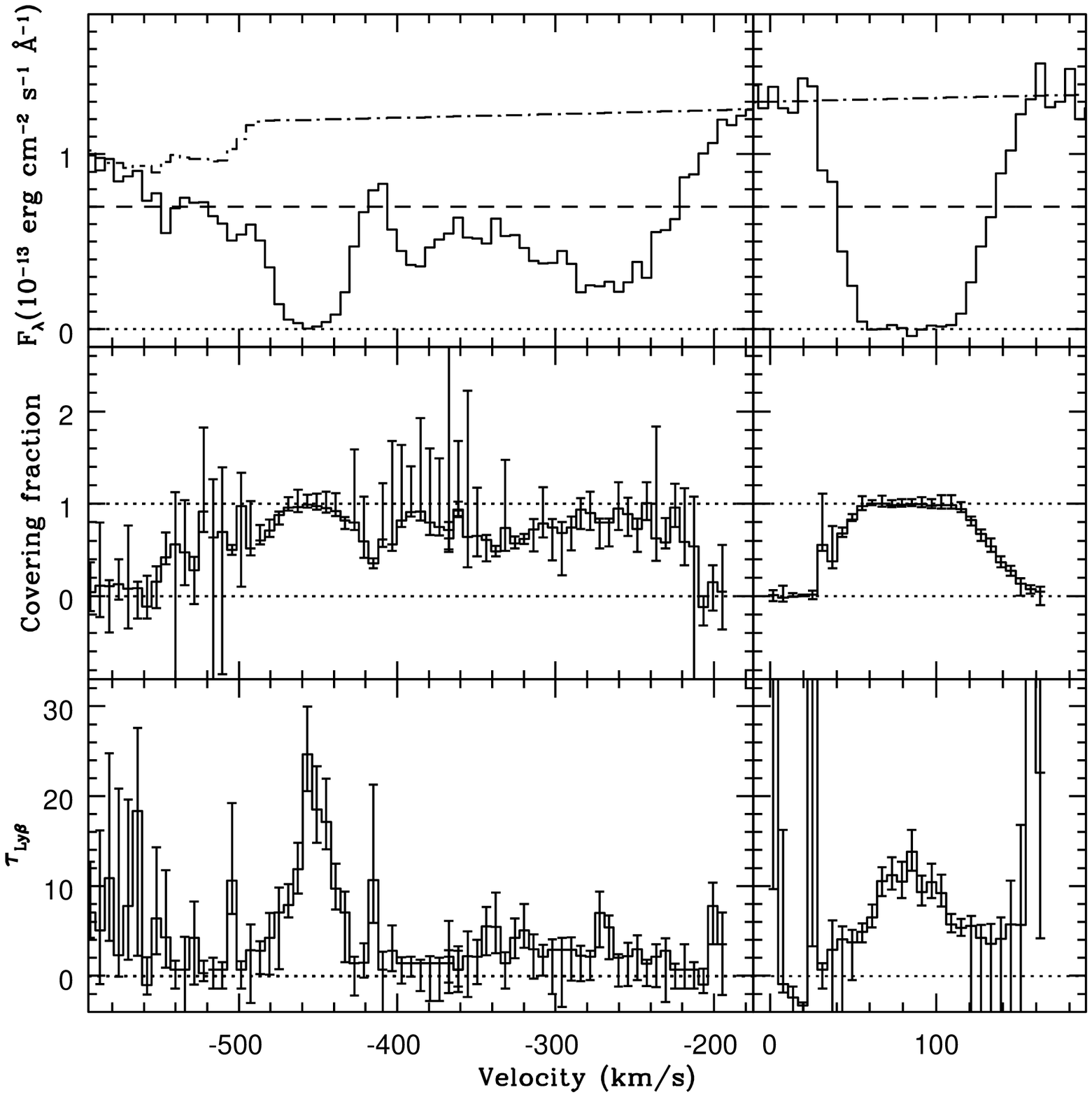}
\caption{\small 2000 January:
{\it Top panel:}
Ly~$\beta$ profile with continuum and emission line fits with ISM absorption (dot-dashed line) and
the power-law continuum fit alone (dashed line).
{\it Middle panel:} Histogram of solutions to the minimization of
Equation~\ref{equ:minchi} for $C_{c}=C_{l}=C_{f}$,
the effective covering fraction vs.\ velocity with respect to systemic.
{\it Bottom panel:}
$\tau^{\beta}$, the optical depth in the Ly~$\beta$ line vs.\ velocity.
\label{fig:hist00eff}}
\epsscale{1.0}
\end{figure}

\begin{figure}
\epsscale{0.5}
\plotone{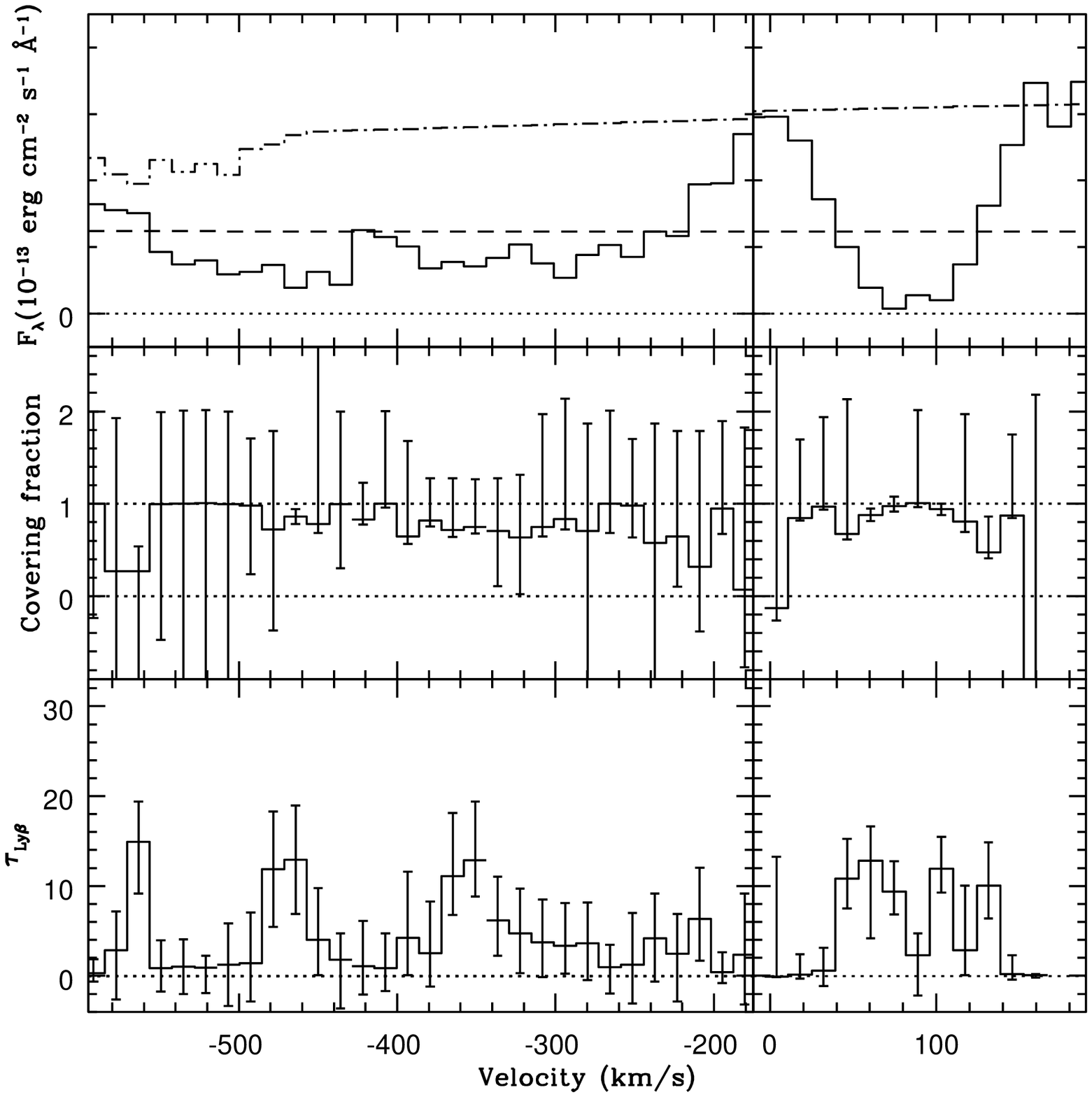}
\caption{\small 2002 January:
{\it Top panel:}
Ly~$\beta$ profile with continuum and 
emission line fits with ISM absorption (dot-dashed line) and
the power-law continuum fit alone (dashed line).
{\it Middle panel:} Histogram of solutions to the minimization of
Equation~\ref{equ:minchi} for $C_{c}=C_{l}=C_{f}$,
the effective covering fraction vs.\ velocity with respect to systemic.
{\it Bottom panel:}
$\tau^{\beta}$, the optical depth in the Ly~$\beta$ line vs.\ velocity
\label{fig:hist102eff}}.
\plotone{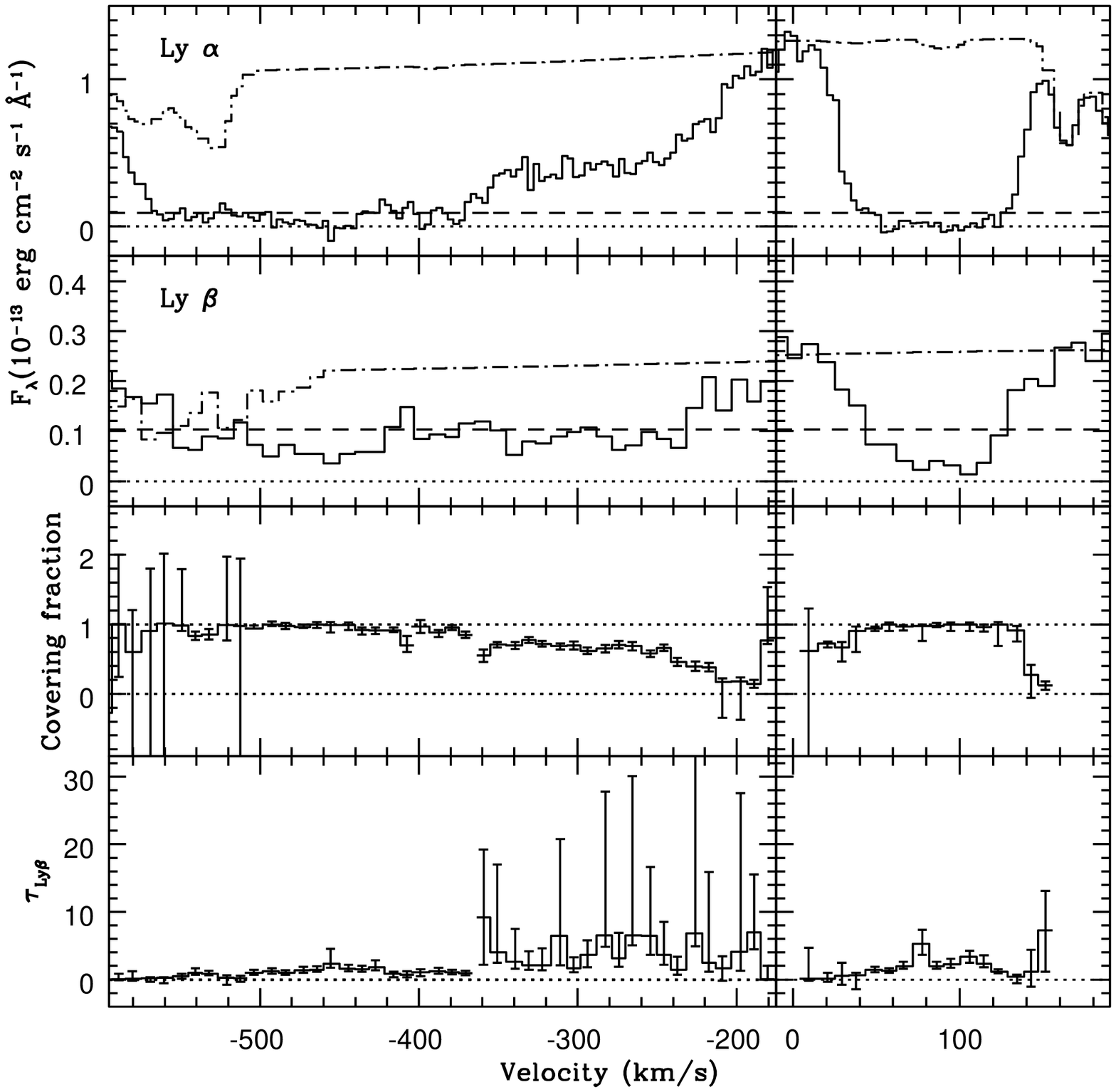}
\caption{\small 2002 May:
{\it Top two panels:} Ly~$\alpha$ and
Ly~$\beta$ profiles with continuum and emission line fits with ISM absorption 
(dot-dashed line) and
the power-law continuum fit alone (dashed line).
{\it Third panel:} Histogram of solutions to the minimization of
Equation~\ref{equ:minchi} for $C_{c}=C_{l}=C_{f}$,
the effective covering fraction vs.\ velocity with respect to systemic.
{\it Bottom panel:}
$\tau^{\beta}$, the optical depth in the Ly~$\beta$ line vs.\ velocity.
\label{fig:hist502eff}}
\epsscale{1.0}
\end{figure}

The quantities $C_{c}$ and $C_{l}$ are highly correlated, giving
rise to large uncertainties in the solutions.
With the exception of two velocity bins in the blue wing of Component 4 at
$\sim$470~\kms, visible in Figure~\ref{fig:hist99},
we find that 
the line and continuum covering fractions
do not differ significantly from one another.
Hence, we find no significant difference in the column densities derived
if we instead
assume $C_{c} = C_{l}= C_{f}$ and solve for an effective covering fraction, $C_{f}$, and
optical depth at each point using the $\chi^{2}$ minimization described above.
This procedure is necessary for the 
2002 January \fuse\ spectrum, because the S/N in the spectrum restricts our analysis to
the Ly$\beta$ and Ly$\gamma$ lines only.
The Ly$\beta$, $C_{f}$, and $\tau_{{\rm Ly}\beta}$ profiles for each
epoch are shown in Figures~\ref{fig:hist99eff}, \ref{fig:hist00eff}, \ref{fig:hist102eff},
and \ref{fig:hist502eff}.  In Figure~\ref{fig:hist502eff}, we show the Ly$\alpha$
profile from the 2002 May \stis\ spectrum in addition to Ly$\beta$. 
In comparing Figure~\ref{fig:hist99eff} with Figure~\ref{fig:hist99}, we see 
that the assumption of an effective covering fraction provides better constraints
on the covering fraction than the attempt to solve for the line and continuum covering 
fractions separately.
Because this assumption also provides better constraints 
on the column density solutions, we use the values of $N$(\hi) from this
procedure in the discussion below.  These results are tabulated in Table~\ref{table-cover}.
The histograms of $C_{f}$ and $\tau_{{\rm Ly}\beta}$ in 
Figures\ref{fig:hist99eff}-\ref{fig:hist502eff} demonstrate
that the shapes of the absorption profiles are dictated to a large
degree by the velocity-dependent covering fraction (Arav et al.\ 2002).
Note that the trend of decreasing \hi\ covering fraction in the wings of 
Component 1 in particular is likely due to the convolution of the
flux profile with the instrumental line spread function of the \fuse\
spectrograph.

\begin{figure}
\includegraphics[angle=-90,scale=0.46]{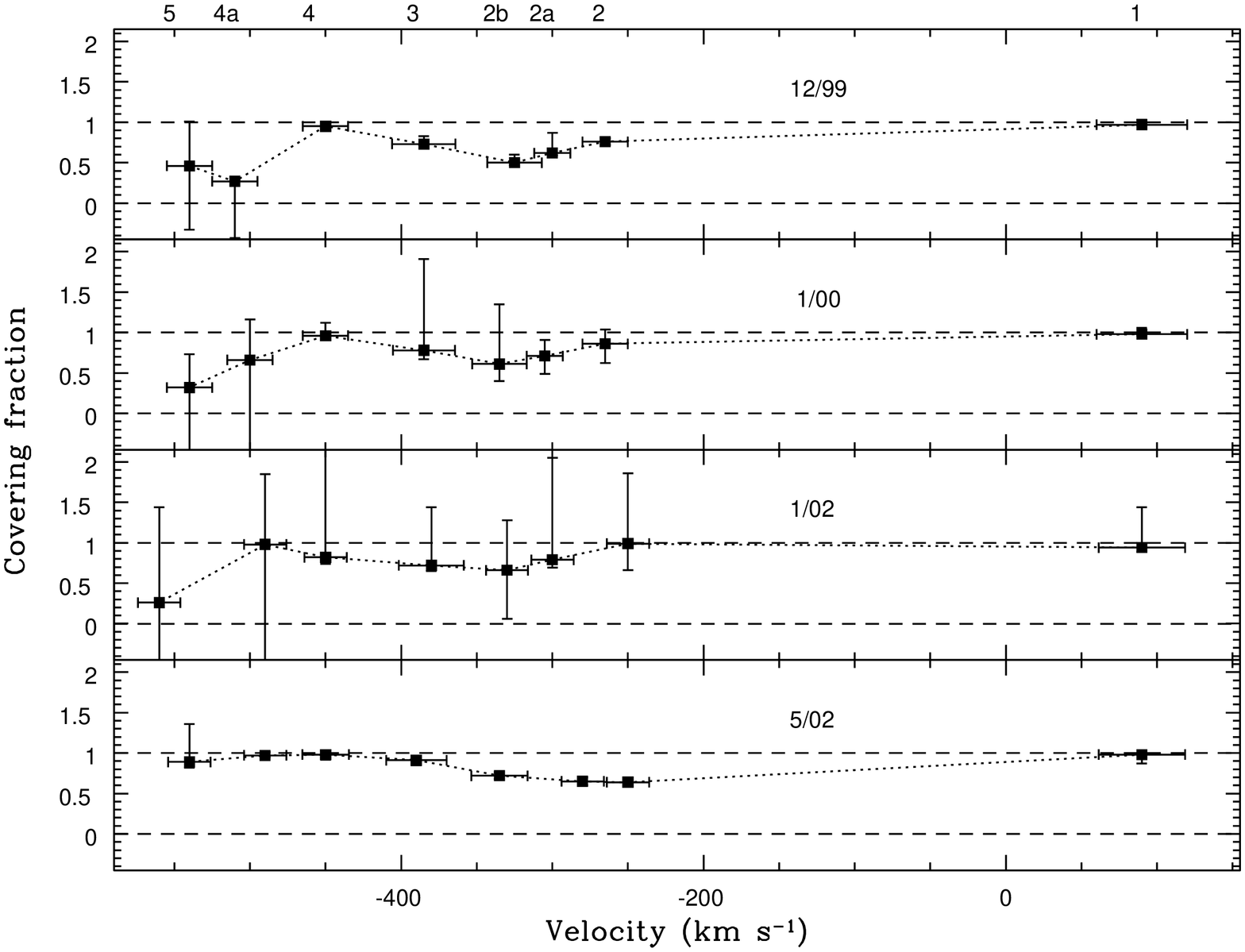}
\caption{\small Averaged solutions to
Equation~\ref{equ:minchi} for the effective covering fraction of \hi.
Dotted line connects the data points.
Dashed lines at $C_{f}=0$ and $C_{f}=1$ are plotted for reference.
\label{fig:avgc}}
\includegraphics[angle=-90,scale=0.46]{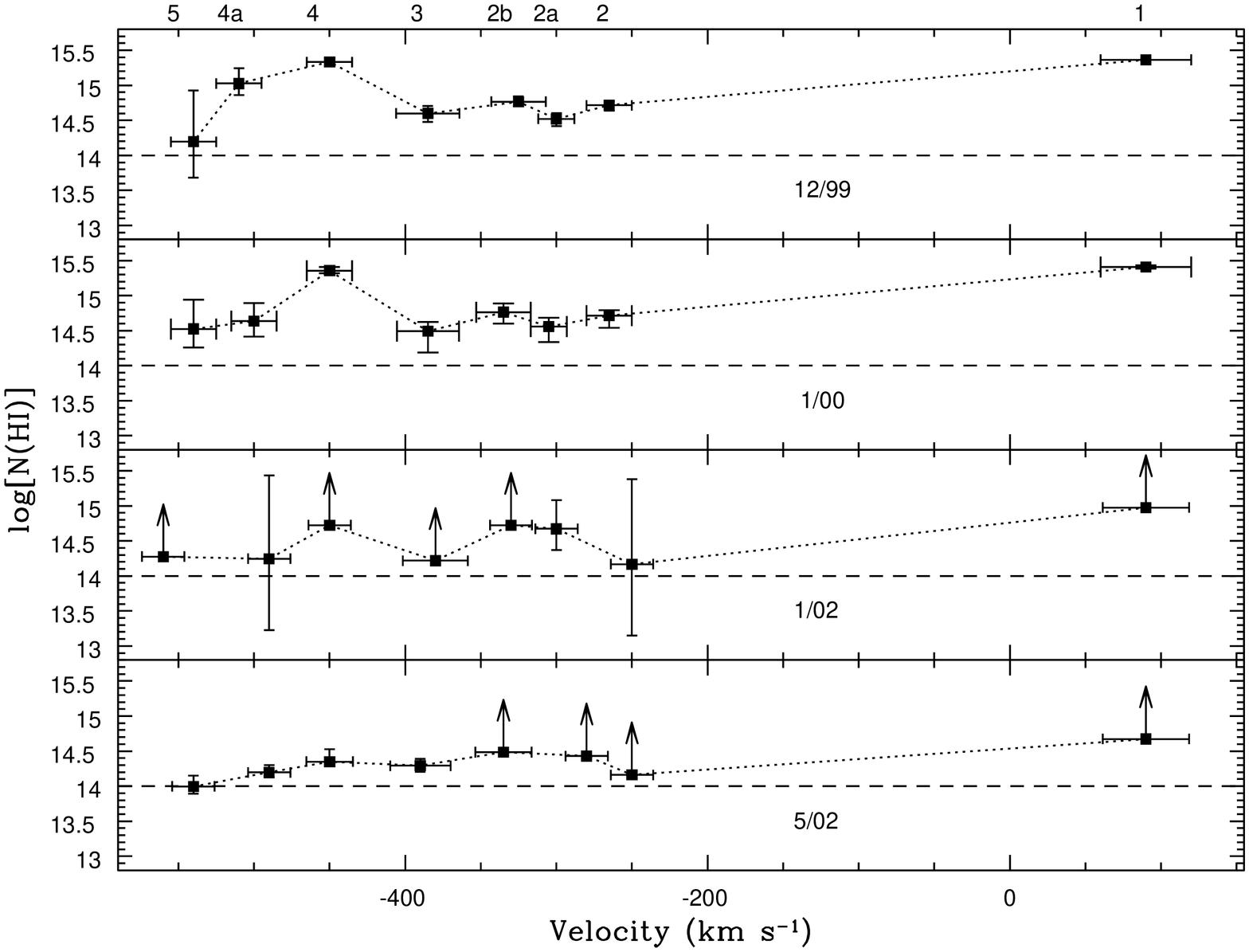}
\caption{\small Averaged solutions to
Equation~\ref{equ:minchi} for
\hi\ column densities. Dotted line connects the data points.
Dashed line at log[$N$(\hi)]=14 is plotted for reference.
\label{fig:avgnh}}
\end{figure}

We plot the \hi\ covering fractions and summed column densities tabulated in 
Table~\ref{table-cover} in 
Figures~\ref{fig:avgc} and \ref{fig:avgnh}.
In Figure~\ref{fig:avgc}, we show average values of the
effective covering fractions within the $\Delta v$ listed in Table~\ref{table-cover}
of the velocity centroids of each component,
and in Figure~\ref{fig:avgnh}, we plot
the \hi\ column density in each component, summed over $\Delta v$.
The intrinsic absorption in the first two epochs 
of observations is similar in terms of
both covering fractions and \hi\ column densities.  These
quantities are largest in Components 1 and 4, where $C_{f}$
is consistent with unity and, in the 1999/2000 epochs, 
$N$(\hi) $\sim (2.3-2.5) \times
10^{15}$~cm$^{-2}$ in Component 1 and $N$(\hi) $\sim (2.1-2.2) \times 
10^{15}$~cm$^{-2}$ in Component 4.
The velocity of Component 1, +90 \kms,
is consistent with the systemic redshift of Mrk~279.  This fact,
the large \hi\ column density, the unity covering
fraction, and the weak O~{\sc vi} absorption
all indicate that this absorption is due to the ISM of Mrk~279.
For all components, the \hi\ column densities in the first two epochs are
consistent with one another within the uncertainties.

The \hi\ column densities in  Components 2, 2a, and 2b do not
vary significantly from the 1999/2000 epochs to the 2002
observations.
By contrast, there is a slight drop in $N$(\hi) in Components 3 and 5
and $N$(\hi) in Components 4 and 4a
drops dramatically over this time period. In particular, $N$(\hi)
in Component 4 drops by a factor of $\sim$10 from 2000 January to 2002 May.
We will discuss the variability of the absorption components in
Section~\ref{sec-varabs}.
Like Component 1, Component 4 also shows a unity covering fraction, a high
column density of \hi, and absorption due to a number of
low-ionization species.   This suggests that at least some
of the gas that contributes to this component does not
arise in the AGN outflow.   We will discuss this component
further in Section~\ref{sec-disc}.

\subsubsection{Doublets}
\label{sec-doub} 

For doublets, we define an effective covering fraction, 
$C_{f}=R_{l}C_{l} + R_{c} C_{c}$ 
(Ganguly et al.\ 1999; Gabel et al.\ 2003a) as outlined in the discussion 
of the Lyman series above, and require 
$\tau_{b} \approx 2 \tau_{r}$, where $b$ and $r$ denote the blue and
red components of the doublet, respectively.
Though
applying Equation~\ref{equ:cover} to each doublet component allows
for a direct solution of the two unknowns, $C_{f}$ and $\tau$
(\cite{hamann1997}), we choose instead to use the $\chi^{2}$ minimization
technique described above by Equation~\ref{equ:minchi} so that
we are able to restrict the values of $C_{f}$ and $\tau$ to physically
meaningful ranges.  In most cases, however, this method gives
solutions indistinguishable from the direct solution of Equation~\ref{equ:cover}.

We use the optical depths from this method
and Equation~\ref{equ:column} to derive the column densities of 
O~{\sc vi} at all epochs, as well as N~{\sc v} and C~{\sc iv} in the 
2002 May \stis\ data. 
For Component 1, for which we found no obvious N~{\sc v} lines, we
place an upper limit of $1.5 \times 10^{14}$~cm$^{-2}$
on the N~{\sc v} column density.
To do this, we assume that the covering fraction
of the N~{\sc v} in this component is equal to the O~{\sc vi} covering
fraction.  The  N~{\sc v} and O~{\sc vi} covering fractions
are consistent with one another in the other components, where they
are determined independently.
For Component 1, the 1548~\AA\ member of the C~{\sc iv} doublet is blended with the
1550~\AA\ line of Component 3.
We set an upper limit of $7.5 \times 10^{13}$~cm$^{-2}$ 
for the C~{\sc iv} column density in Component 1
using the 1550~\AA\ line
and the O~{\sc vi} covering fraction.
The C~{\sc iv} and O~{\sc vi} covering fractions
are consistent with one another in the other components, with the exception of Component 5,
where the C~{\sc iv} covering fraction is $\sim$3.5$\sigma$ larger than the
O~{\sc vi} covering fraction.   For Component 3, we use the 1548 \AA\ line
and the O~{\sc vi} covering fraction to derive
$N$(C~{\sc iv})$ < 9.4 \times 10^{13}$~cm$^{-2}$.
We find no detectable Si~{\sc iv} features in the \stis\ spectrum, so we
do not attempt to solve directly for Si~{\sc iv} column densities using
the doublet method.  Instead, we use the O~{\sc vi} covering fractions
to set 3$\sigma$ upper limits on the Si~{\sc iv} column density in each component.
The results of these calculations are listed in Table~\ref{table-compdoub}.

\subsubsection{Other Metals}
\label{sec-metals}

For metals that do not give rise to doublets, we do not have enough  
information to solve for the covering fraction and column density
self-consistently. 
In principle, where metal features are coincident in velocity with
hydrogen lines, we could assume that the
metal absorption in question arises in the same location as the \hi\ and 
use the covering fractions derived from the hydrogen lines to solve for
optical depths in the cores of these metal lines.
However, because we cannot unambiguously determine what portion of the \hi\ 
absorption in any component 
is due to the host galaxy ISM and what portion is due to an AGN outflow,
we cannot make this assumption. 
However, because the low-ionization
lines in Components 1, 3, 4, and 5 
are likely to arise in regions far from the AGN in Mrk~279,
we simply assume that the covering fractions are unity and 
use Equation~\ref{equ:column} to calculate 
the column densities of low-ionization species using
C~{\sc iii}~\lam977, N~{\sc iii}~\lam989, C~{\sc ii}~\lam1036, and N~{\sc ii}~\lam1084
in the \fuse\ data, and Si~{\sc iii}~\lam1206, Si~{\sc ii}~\lam1260,
and C~{\sc ii}~\lam1334 in the \stis\ data.
In cases where we do not measure a
column density of greater than 3$\sigma$ significance,
we quote a 3$\sigma$ upper limit.
The results of these
measurements are listed in Tables~\ref{table-compmetfuse} and \ref{table-compmetstis}.

\subsubsection{Foreground Galactic Absorption}

As mentioned in Section~\ref{sec-ism}, we computed models of foreground Galactic interstellar
absorption using three different methods for estimating the $f$ ratios
of the relevant species.  We created the baseline models by using
$f$ ratios that give the lowest $\chi^{2}$ between the modeled interstellar
absorption features and the observed spectra.
We calculated the other two models,
which we now compare with the 1999 December baseline model results,
from fits to \fuse\ spectra of six AGNs
that show no intrinsic absorption and
from published $f$ ratios from Morton (1991) and Howk et al.\  (2000).
We calculated the covering fractions and optical depths
of \hi, O~{\sc vi}, and C~{\sc iii} for Component 4,
and for O~{\sc vi} in Components 1 and 3
using both of these other
ISM models.
In no case do the results differ by more than 1.5$\sigma$
from the values derived using our fiducial ISM model, and in most cases,
the difference is less than 1$\sigma$. We conclude from this that uncertainties
in our ISM model are not a significant source of systematic error in 
our covering fraction and column density calculations.

\subsection{Variability}
\label{sec-var}

\subsubsection{Continuum and Emission Lines} 
\begin{figure}
\plottwo{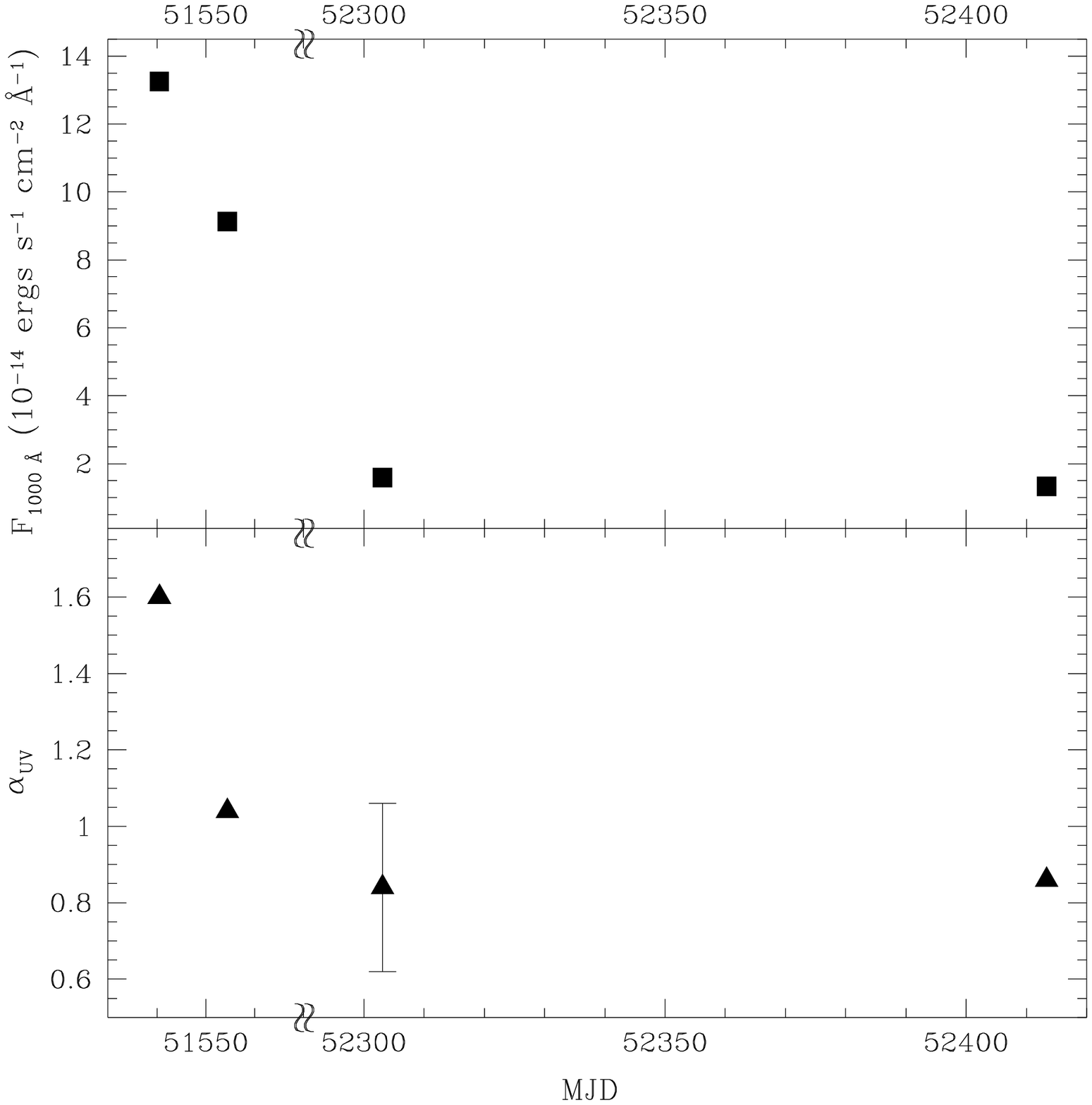}{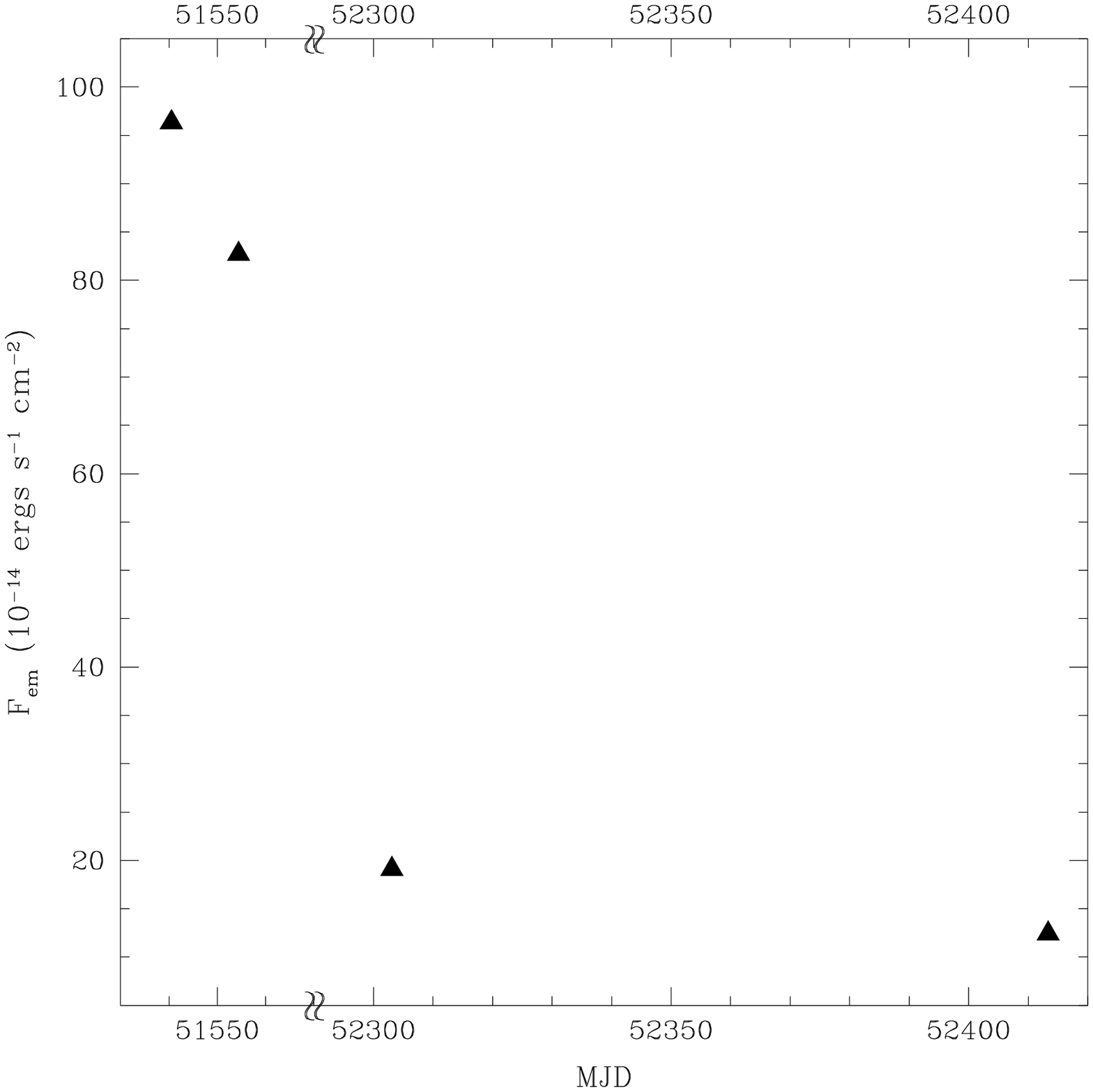}
\caption{\small {\it Left, Top panel:}
Flux at 1000~\AA\ from \fuse\ spectrum
vs.\ modified Julian date
(JD-2,400,000.5).
{\it Left, Bottom panel:} UV spectral index vs.\ modified Julian date. Note that the
most recent \fuse\ spectrum covers only 991-1180~\AA.
\label{fig:var}}
\caption{\small {\it Right:} 
Broad O~{\sc vi} emission line flux vs.\ modified Julian date
(JD-2,400,000.5).
\label{fig:var_em}}
\plotone{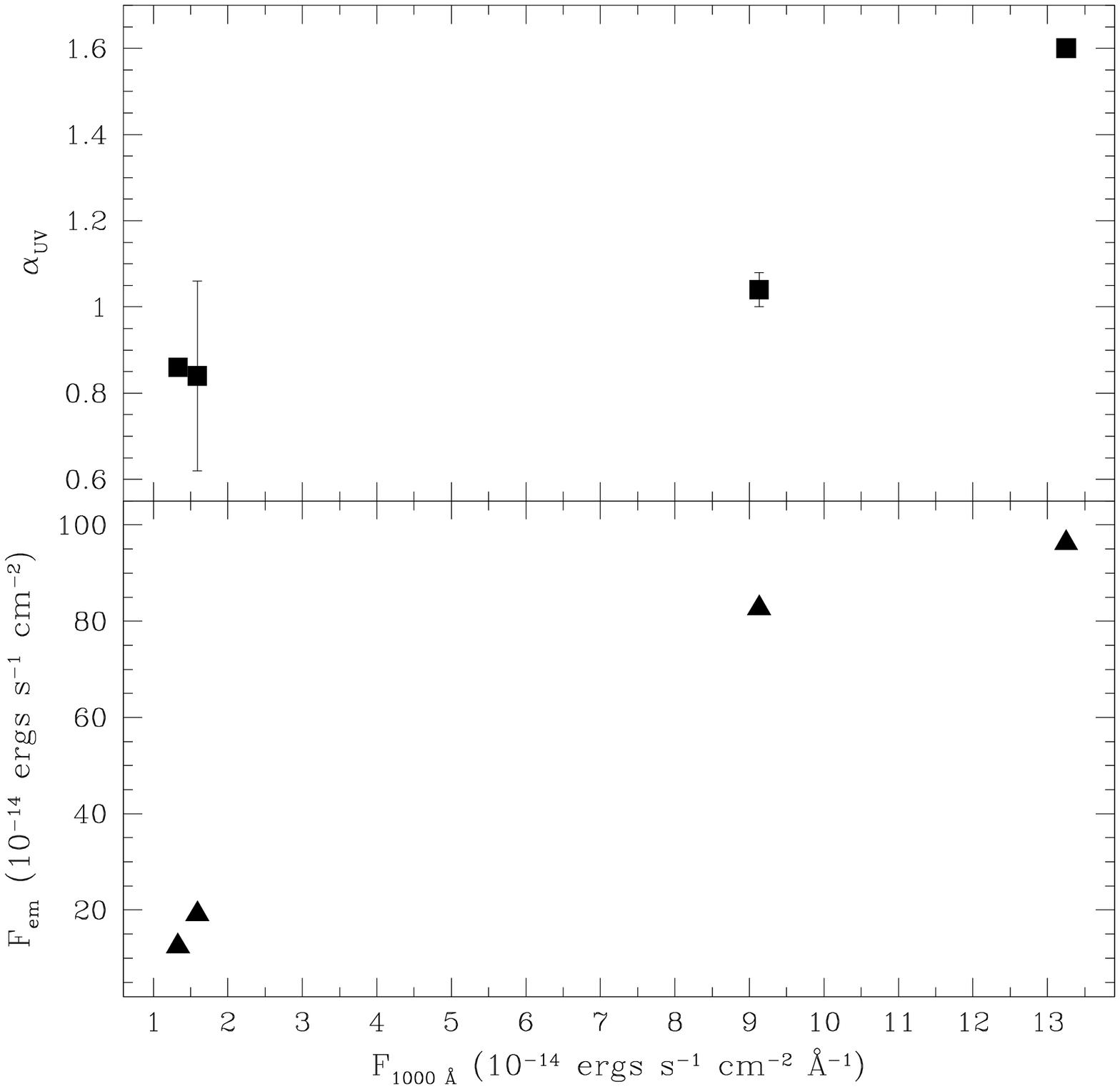}
\caption{\small {\it Top panel:}
UV spectral index vs.\
flux at 1000~\AA.
{\it Bottom panel:}
Broad O~{\sc vi}
emission line flux vs.\
flux at 1000~\AA.
\label{fig:fluxcor}}
\end{figure}

The variability of the 1000~\AA\ flux and the UV spectral index
of Mrk~279 with time are shown in Figure~\ref{fig:var}. 
The continuum flux drops by a factor of $\sim$7.5 from 1999 December 
to 2002 January and remains at roughly this level in 2002 May.
As discussed in Section~\ref{sec-xray}, the {\it Chandra} spectrum of 2002 May 
also shows Mrk~279 in a low state.
The brighter UV continua are fitted best by bluer power laws, as indicated 
in the bottom panel of Figure~\ref{fig:var}.
The variation of the flux in the broad O~{\sc vi} emission line
with time is shown in Figure~\ref{fig:var_em}.
The broad line flux varies dramatically in response to the changes in the
continuum flux, dropping by a factor of 
four between 2000 January and 2002 January.
The top panel of Figure~\ref{fig:fluxcor} illustrates the changes in the UV spectral
index with continuum flux.
As noted above, the continuum of Mrk~279 becomes systematically redder
as it becomes fainter.  

From studies of large AGN samples
(Laor et al.\ 1995; Zheng, Kriss, \& Davidsen 1995; Green 1996;
Green, Forster, \& Kuraszkiewicz 2001;
Dietrich et al.\ 2002; Kuraszkiewicz  et al.\ 2002)
the Baldwin effect has been found to be particularly 
pronounced for O~{\sc vi} emission lines. These studies have found Baldwin relation
slopes, $\beta$,
ranging from -0.50 to -0.15, where $W_{\lambda}$(O{\sc vi}) $\propto L^{\beta}$.
Kinney, Rivolo, \& Koratkar (1990) 
found that as individual AGNs vary in luminosity, 
they show an intrinsic Baldwin relation in the C~{\sc iv} and Ly~$\alpha$
emission
lines, with steeper slopes than the relation derived from their full AGN sample.
In the bottom panel of Figure~\ref{fig:fluxcor},
we show the O~{\sc vi} emission line flux
versus the monochromatic flux
at 1000~\AA, and we 
use the four observations of Mrk~279 to examine the
intrinsic Baldwin effect in the broad O~{\sc vi} emission line.
We derive a Baldwin relation slope 
for O~{\sc vi} of $\beta=-0.06 \pm 0.14$.  This result
runs counter to the expectation as it is flatter than the
ensemble Baldwin relation slopes for O~{\sc vi} quoted above.  
We note, however, that this slope is based on four observations
of Mrk~279, while Kinney et al.\ (1990) drew their conclusions
from seven AGNs with 16 or more observations each.
Also, we have not corrected for the response time of the BELR
to changes in the continuum, which may
introduce scatter into the Baldwin relation (Pogge \& Peterson 1992). 
Balmer line time-delay measurements 
suggest a BELR size of $6-17$ lt-days for Mrk~279
(Maoz et al.\ 1990; Stirpe \& de Bruyn 1991; Stirpe et al.\ 1994;
Santos-Lle\'{o} et al.\ 2001), but
we do not correct for the light travel time between the continuum source and
the BELR as we have not sampled the light curve
of Mrk~279 finely enough for such an analysis.

\subsubsection{Intrinsic Absorption}
\label{sec-varabs}

\begin{figure}
\epsscale{0.55}
\plotone{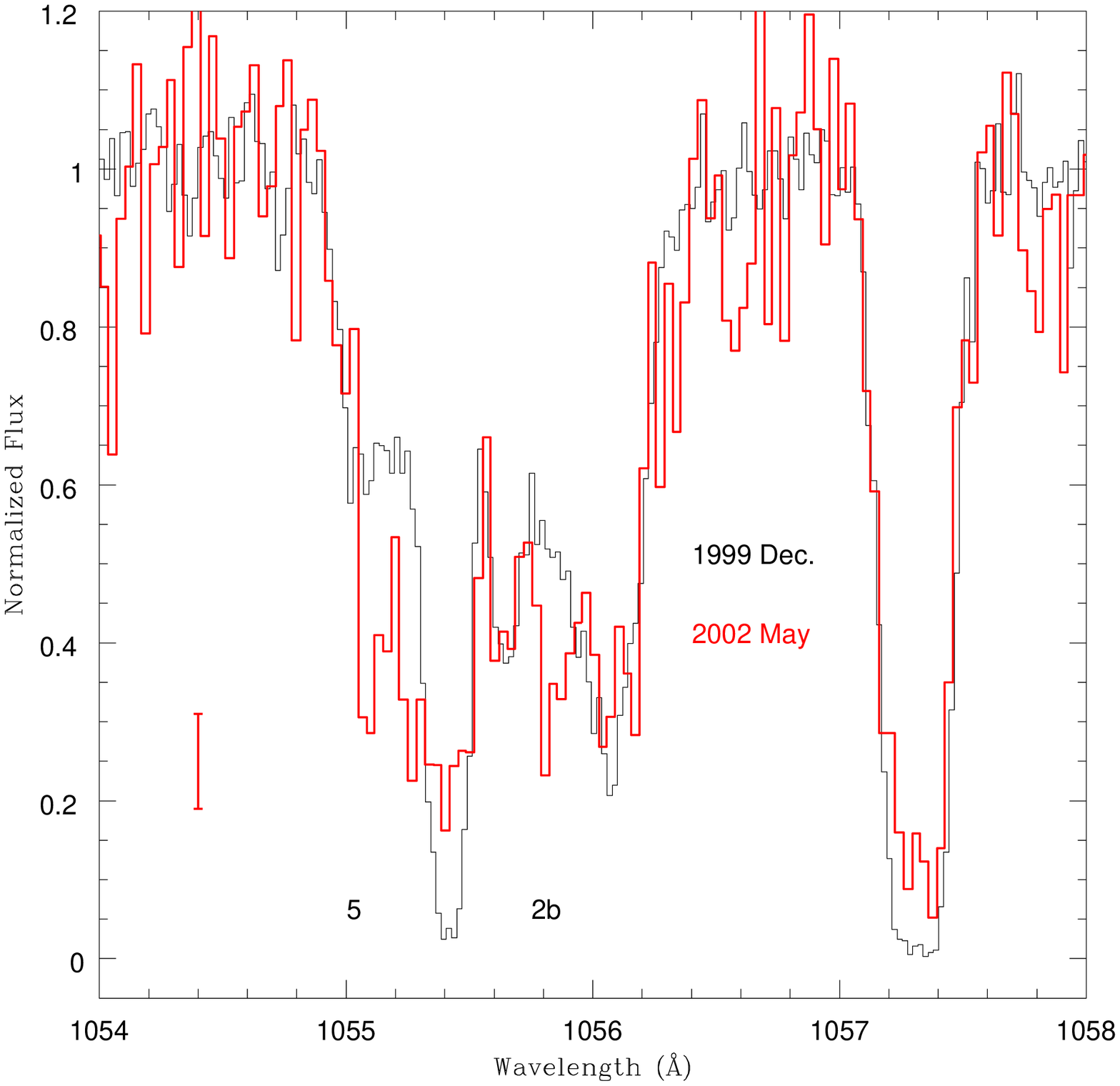}
\caption{\small 
Comparison of the normalized Ly$\beta$ profiles in the 1999 December and
2002 May spectra. Components discussed in Section~\ref{sec-varabs}
are marked.  Representative errorbar for 2002 May spectrum in line
troughs is shown in red at left. 
\label{fig:lyb}}
\plotone{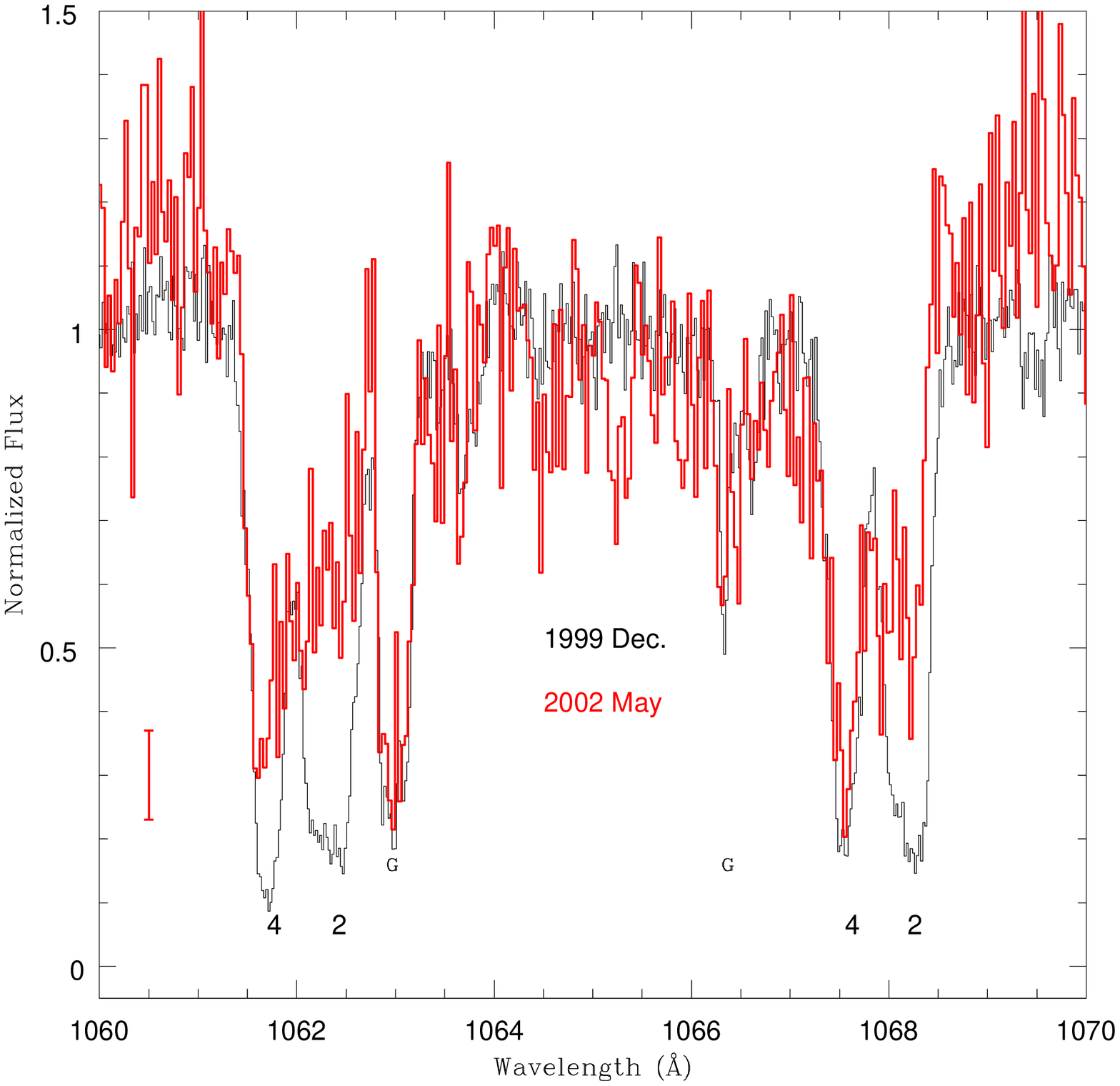}
\caption{\small Comparison of the normalized O~{\sc vi} profiles in the 1999 December and
2002 May spectra. Components discussed in Section~\ref{sec-var}
are marked.  Representative errorbar for 2002 May spectrum in line
troughs is shown in red at left.
\label{fig:o6}}
\epsscale{1.0}
\end{figure}

We see variability in the flux profiles of the 
intrinsic absorption over the course of the
observations presented here.
To investigate this, we use
the 1999 December \fuse\ spectrum as representative of the 2000 epoch
and the 2002 May \fuse\ spectrum as representative of the 2002 epoch.
The normalized Ly$\beta$ and 
O~{\sc vi} profiles in the 1999 December and 2002 May spectra are shown
in Figures~\ref{fig:lyb} and \ref{fig:o6}. 
The red errorbars at the left of each figure illustrate representative
flux errors in the troughs of the Ly$\beta$ and O~{\sc vi}
absorption lines in the 2002 May spectrum.
In Ly$\beta$, Components 2b and 5 appear to
be more pronounced in the 2002 May spectrum than in 1999 December.  
Comparing the region from -320 to -360 \kms\ in the covering fraction histograms
shown in 
Figures~\ref{fig:hist99eff} and \ref{fig:hist502eff} 
and examining the average covering fractions in each component shown
in Figure~\ref{fig:avgc}, we see that
the covering fraction in Component 2b is correspondingly larger in the
2002 solution, although not by a significant factor.
Also, Figure~\ref{fig:avgnh} demonstrates that 
there are no significant differences in the column densities
for this component in these two epochs, although
because of the S/N in the 2002 epoch data, we are only able
to place lower limits on the column density in Components 2, 2a, and 2b.
A similar situation holds for Component 5.

Figure~\ref{fig:o6} shows that the O~{\sc vi} absorption profiles
changed between the 2000 and 2002 epochs as well.  Specifically,
Component 2 appears to be weaker in the 2002 May spectrum.  
Because the O~{\sc vi} absorption is saturated in both epochs, 
we are only able to place lower limits on the column density in this
component.
Component 4 in the 2002 May spectrum is saturated as well, 
and it even appears that the red doublet line is
stronger than the blue line, perhaps indicating a problem with the
removal of the interstellar Fe~{\sc ii} in the blue component.
The limits placed on the O~{\sc vi} column density for Component 4 
in 2002 May are consistent
with no variation since the 1999 epoch.
Therefore, even though the absorption profiles of Ly$\beta$
and O~{\sc vi} appear to vary in some components, it is not clear whether
this is due to a change in the column density or the covering
fraction of the absorbers.

The overall $C_{f}$ and $\tau_{{\rm Ly}\beta}$ profiles,
shown in Figures~\ref{fig:hist99eff} and \ref{fig:hist502eff},
are similar in the
1999 December and 2000 January spectra, with the exception of Components 4a and 5.
Component 4a shows a smaller covering fraction and larger optical depth
in the 1999 December spectrum compared with 2000 January, while the
reverse is true for Component 5.  The constraints are poor enough, however,
that the uncertainties on $C_{f}$ and $\tau_{{\rm Ly}\beta}$ are consistent with no 
change in these components over this short time period.
In the 2002 May spectrum, however, the profiles are substantially different.
The broad Ly$\alpha$ absorption in Component 1
is reflected in the covering fraction profile in the third panel of 
Figure~\ref{fig:hist502eff}.  Similarly, the broad Ly$\alpha$ absorption trough
in Components 2-5 and the extension of the Ly$\alpha$
absorption blueward of Component 5 is reflected in the covering fraction solutions.
The bottom panels of Figures~\ref{fig:hist99eff} and \ref{fig:hist502eff}
illustrate the significantly lower optical depths derived in 
Components 4 and 4a 
in 2002 May than in 1999 December, and the corresponding summed column densities
in these components are shown in the top and bottom panels of Figure~\ref{fig:avgnh}.

\begin{figure}
\epsscale{0.57}
\plotone{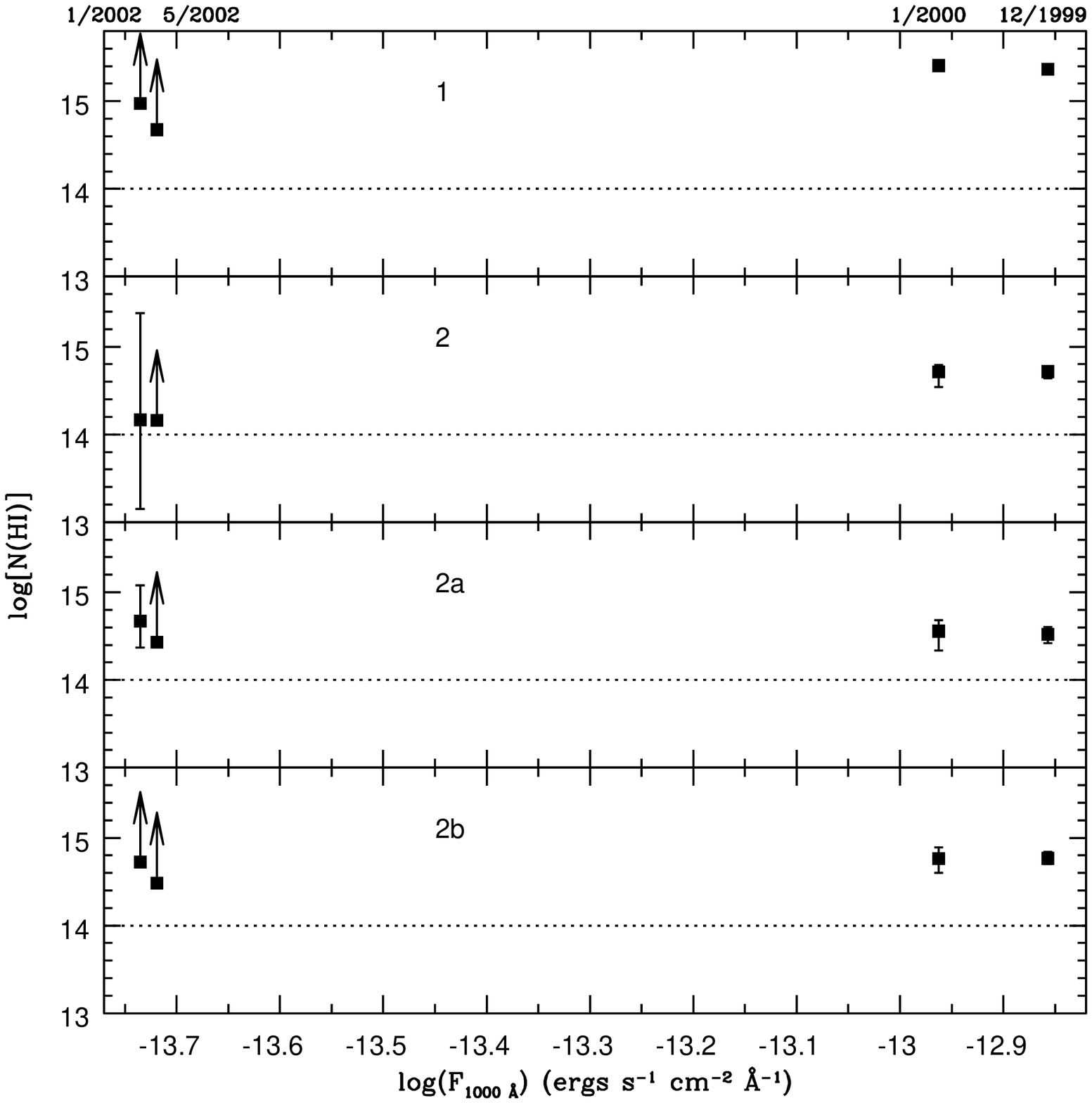}
\caption{\small 
Logarithm of the integrated column density 
of \hi\ vs.\ the logarithm of the
flux at 1000~\AA\ for Components 1, 2, 2a, and 2b.
Dotted line at log[$N$(\hi)]=14 is plotted for reference.
Month and year of observation are shown at top. 
\label{fig:nhflux1}}
\plotone{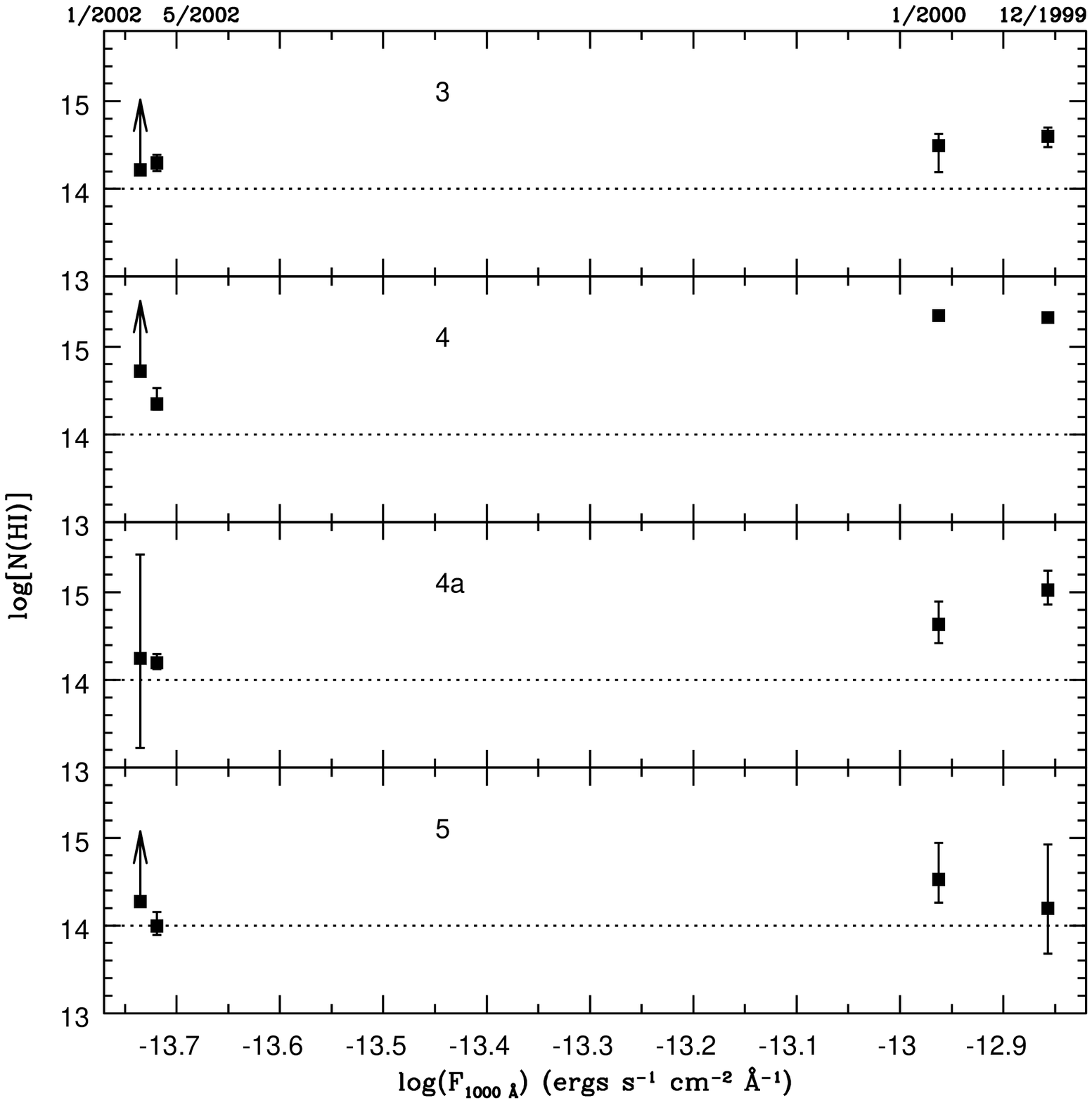}
\caption{\small
Same as Figure~\ref{fig:nhflux1},
for Components 3, 4, 4a, and 5
\label{fig:nhflux2}}
\epsscale{1.0}
\end{figure}

To investigate whether the intrinsic absorption has responded
to the dramatic changes in the ionizing continuum of Mrk~279, we
show the variation in $N$(\hi) with UV continuum flux
Figures~\ref{fig:nhflux1} and \ref{fig:nhflux2}.
The measurements and limits on $N$(\hi) in Component 1,
attributed to the AGN host galaxy, are 
consistent with no variation. 
The \hi\ column densities in Components 2, 2a, and 2b also do not appear to vary.
In Components 3 and 5, we find a marginal decrease in $N$(\hi), by a factor of $\sim$2,
between the 1999/2000 and 2002 epochs as the 1000 \AA\ flux decreased.
For Components 4 and 4a, however, we find that the \hi\ column density
decreased more significantly over this period.  The \hi\ column density in Component 4a
decreased by a factor of $\sim$7, and that in Component 4 decreased by
a factor of $\sim$10 between 1999 December and 2002 May, as the continuum
flux dropped by a factor of $\sim$7. 
This runs counter to expectation for a scenario in which the ionization state of the
absorbers responds directly to changes in the ionizing continuum.

\section{Discussion}
\label{sec-disc}

Figures~\ref{fig:met99}, \ref{fig:met00}, \ref{fig:met02}, and \ref{fig:met502}
clearly illustrate the 
strong, saturated, intrinsic O~{\sc vi} absorption present in the Mrk~279 spectrum.  
However, absorption due to ions 
such as C~{\sc iii}, N~{\sc iii}, Si~{\sc ii}, C~{\sc ii}, and N~{\sc ii}
indicates the presence of low-ionization gas along the line of sight.
The low S/N of the {\it Chandra} spectrum limits our analysis, but
we find no discernible O~{\sc vii} or O~{\sc viii} absorption features 
and few other high-ionization
X-ray absorption lines in those data.

We derive limits on the column densities of O~{\sc vii} and O~{\sc viii}, 
$N$(O~{\sc vii}, O~{\sc viii}) $\lesssim 5 \times 10^{16}$~cm$^{-2}$.
This $N$(O~{\sc vii}) limit is approximately one order of magnitude
lower than the O~{\sc vii} column densities measured in the X-ray spectra
of NGC~3783 ($1 \times 10^{18}$~cm$^{-2}$,
Kaspi et al.\ 2002, Behar et al.\ 2003) and 
NGC~5548 ($4 \times 10^{17}$~cm$^{-2}$,
Kaastra et al.\ 2002) and that inferred from the model for the warm absorber in 
Mrk~509 ($3.5 \times 10^{17}$~cm$^{-2}$, Yaqoob et al.\ 2003).
The $N$(O~{\sc viii}) limit is 1-2 orders of magnitude lower
than the values measured or inferred for these AGN (NGC~3783: $4 \times 10^{18}$~cm$^{-2}$,
Kaspi et al.\ 2002, Behar et al.\ 2003;  NGC~5548:  $2 \times 10^{18}$~cm$^{-2}$,
Kaastra et al.\ 2002; Mrk~509:  $8 \times 10^{17}$~cm$^{-2}$, Yaqoob et al.\ 2003).
Assuming solar abundances (Anders \& Grevesse 1989) and that all oxygen in
the absorber is in the form of either O~{\sc vii} or O~{\sc viii},
these column density limits imply $N_{H} < 6 \times 10^{19}$~cm$^{-2}$.
This total hydrogen column is 1-2 orders of magnitude less than 
values typically derived for warm absorbers (Kaspi et al.\ 2002, Behar et al.\ 2003,
Blustin et al.\ 2002, 2003, Kaastra et al.\ 2002,
Krongold et al.\ 2003, Netzer et al.\ 2003, Yaqoob et al.\ 2003).
However, a strong caveat to the $N$(O~{\sc vii}) and $N$(O~{\sc viii})
limits is that they assume the lines lie on the linear part of the 
curve-of-growth.  If the line widths of the X-ray absorbers
are comparable to those derived in the UV, $\sim$100 \kms,
the column densities in O~{\sc vii} and O~{\sc viii} can be
as large as $6 \times 10^{19}$ and $3 \times 10^{19}$~cm$^{-2}$,
respectively, and the equivalent
total hydrogen column density is totally consistent with
other X-ray absorbers.
Although it is difficult to draw strong conclusions from the low S/N
in the X-ray spectrum of Mrk~279 presented here, 
Mrk~279 may be qualitatively similar to Ton~S180 (Turner et al.\ 2001, 2002)
in the sense that it shows intrinsic absorption in O~{\sc vi}, but no evidence
for a warm absorber in the X-ray regime.
The weak \hi\ absorption and the limit on the 
$N$(O{\sc vii})/$N$(O{\sc vi}) ratio inferred from the
X-ray and UV spectra of Ton~S180 indicate that the absorbing gas
must be in a high ionization state.  
By contrast, the spectra of Mrk~279 show strong features due to both
\hi\ and O~{\sc vi}.

In Components 2, 2a, and 2b of the UV absorption, we 
find strong O~{\sc vi} and \hi\ absorption, although
the Ly$\beta$ and Ly$\alpha$ profiles are somewhat shallower
than in adjacent components.   The \hi\ covering fractions
of these components measured from the 1999 December and 2002 May spectra 
are significantly less than 1, and the results from 2000 January and 2002 January 
are consistent with $C_{f} < 1$.
The absorption due to low-ionization
species is less prominent in these components than in Components 1 and 4.
The O~{\sc vi} and
Ly$\beta$ profiles of Components 2, 2a, and 2b change over the course of the observations,
owing to variations in the covering fractions and/or the column densities, but
the uncertainties in our measurements do not allow us to determine
which.  In either case, this variability, the weak low-ionization
absorption, and the non-unity covering fractions 
imply that these components do arise in an
outflow from the AGN in Mrk~279. 

In Components 1 and 4, 
we find significant absorption due
to low-ionization species such as
C~{\sc iii}, N~{\sc iii}, and Si~{\sc iii}, and even 
C~{\sc ii} and N~{\sc ii} in the case of Component 4. We 
suggest
that these lines may not
in fact arise in the AGN outflow but
rather in gas associated with the host galaxy in either its ISM,
or in material analogous to the HVCs observed in the Milky Way,
or in material participating in an interaction with a physical companion.
There are five pieces of evidence that support these interpretations:
(1) the velocity of Component 1 is consistent with the systemic
redshift of Mrk~279; 
(2) the covering fractions of the \hi\ absorption in 
these components
are consistent with unity;
(3) the absorption profiles of the low- and  high-ionization species
in Component 4 are markedly different; 
(4) the density inferred for Component 4 is significantly less than that
typically found for intrinsic absorbers in AGN outflows;
(5) the \hi\ column density  of Component 1 does not vary.

In Components 1 and 4, the
\hi\ covering fractions are consistent
with unity in all the observations.   Component 4 shows absorption due
to low-ionization species such as C~{\sc ii} and N~{\sc ii}.
Intrinsic absorption due to low-ionization species 
has been observed in the spectrum of NGC~4151,
but unlike Component 4 here, it shows 
partial covering (Kriss et al.\ 1995, 
Espey et al.\ 1998, Weymann et al.\ 1997). 
We do observe absorption due to highly ionized species such as O~{\sc vi},
N~{\sc v}, and C~{\sc iv} at the velocities of these components
and find that the velocity widths of the
absorption in the high-ionization lines are larger those that of
\hi\ and the low-ionization lines (Figure~\ref{fig:c3comp}).
This phenomenon was observed in one of the intrinsic absorption components of
NGC~3783 (Gabel et al.\ 2003a).
However, highly ionized species are also
observed in the Galactic halo and in  HVCs
(Savage, Sembach, \& Lu 1997; Sembach, Savage, \& Hurwitz 1999a;
Savage, Meade, \& Sembach 2001;
Wakker et al.\ 2003; Savage et al.\ 2003; Sembach et al.\ 2003)
with some HVCs showing strong absorption due to highly-ionized
species and little or no absorption due to low ions (Sembach et al.\ 1999b).
Moreover, Howk, Sembach, \& Savage (2003) have observed that the absorption profile of
O~{\sc vi} along a sight line through the 
warm-hot thick disk ISM of the Milky Way 
shows a larger velocity width than those of the
low and intermediate ions while coinciding with them in
velocity centroid and overall extent.
This is interpreted to be a result of the
O~{\sc vi} arising at the interfaces surrounding warm interstellar clouds.
Therefore, although the
broader profile shapes of the high-ionization lines suggest a different origin 
from that of the low-ionization gas seen in C~{\sc iii}, for example, some portion of
the high-ionization absorption may also be attributable to the host galaxy
of Mrk~279.
We note, however, that the O~{\sc vi} and C~{\sc iv}
covering fractions of Components 1 and 4
do tend to be smaller than the \hi\ covering fractions.

For Component 4, we use the 2.3$\sigma$ detection of C~{\sc ii}~\lam1334 and
the non-detection of C~{\sc ii}$^{*}$~\lam1335 in the \stis\ spectrum
to place a limit on the density.
We find $N$(C~{\sc ii}$^{*}$)/$N$(C~{\sc ii})~$<$~1.3. 
For pure collisional excitation and 
$\log T < 5.5$, this implies $n_{e} \lesssim$~750~cm$^{-3}$.
(Srianand \& Petitjean 2000).  
This is lower than the broad range of densities 
typically inferred for AGN outflows from a variety of methods,
$10^{6} < n_{e} < 10^{10}$~cm$^{-3}$ (see, however, Behar et al.\ 2003). 
We tentatively interpret this as another piece of evidence that the low-ionization
species observed in Component 4 in the 
Mrk~279 spectrum do not arise in the AGN outflow, but in
gas of lower density associated with the host galaxy.  

As we noted above, the \hi\ column density
in Component 4 decreases significantly
as the UV continuum flux of Mrk~279 decreases
over the time period of our observations.
However, we stress that in the 2002 May observations, we lack coverage
of the Lyman series blueward of Ly$\gamma$.  In the 1999/2000 epoch,
it is the absorption in the high-order Lyman lines that drives the solution for
$N$(\hi) to large values.  Also, the deep Ly$\alpha$ trough of Components 3-5 
in the 2002 May STIS spectrum drives the covering fraction solution to values
near unity.  This would require smaller $N$(\hi) for the 2002 May solution
even if the depth of the Ly$\beta$ trough was similar to that observed
in the 1999/2000 epoch.  Figure~\ref{fig:lyb} shows that
the Ly$\beta$ absorption in Component 4 is in fact marginally shallower than observed in 
1999 December.  Solving for $C_{f}$ and $\tau_{{\rm Ly}\beta}$ in Component 4 using only
the lines in common between the 1999 December and  2002 May observations,
Ly$\beta$ and Ly$\gamma$, we find the same result, namely that 
$N$(\hi) is significantly lower in 2002 May than in 1999 December.

A variation in column density
that cannot be explained by photoionization may indicate the influence
of bulk transverse 
motion in an absorbing outflow, as observed in the spectrum
of NGC~3783 (Kraemer et al.\ 2001).   
However, as discussed above, 
the low density of Component 4 suggests that it does 
not arise in an AGN outflow. 
A large contribution to the absorption
from the host galaxy may also help resolve this puzzle.
If low-density gas lies at several
kiloparsecs from the continuum source 
one would not expect to
observe changes in the ionization structure of the gas in
direct response to the variability of the AGN continuum on short timescales.
To illustrate with a simple calculation, we choose 
a total column density typical of Galactic HVCs, $\log N = 19$ (Sembach et al.\ 1999b), and
an ionization parameter near the peak of the C~{\sc ii} abundance curve,
$\log U$=-3, since Component 4 shows substantial C~{\sc ii} absorption.
Given the upper limit on the density set by the absence 
of C~{\sc ii}$^{*}$ absorption,
$\sim$750 cm$^{-3}$, and the luminosity of Mrk~279, this places
the absorbing gas at $\sim$1 kpc from the ionizing source, well
outside the nuclear region.  For a neutral fraction of $\sim$$7 \times 10^{-3}$,
the recombination timescale is $\sim$1.5 yr.
This period of time is short enough for the ionization state of the 
absorber
to vary between the 1999/2000 and 2002 epochs but long enough that the gas may
not have responded to changes in the ionizing continuum that occurred
prior to the 2002 observations.

The origin of Components 3 and 5 of the intrinsic absorption is ambiguous.
These components show low-ionization absorption with profiles that
differ from those seen in O~{\sc vi}, as discussed in Section~\ref{sec-intabs}.
These components show a
marginal decrease in $N$(\hi) with decreasing UV continuum flux, in the same
sense observed for Component 4.
The \hi\ covering fractions of Components 3 and 5 are consistent with unity within the uncertainties
with the exception of $C_{f}=0.73^{+0.10}_{-0.01}$ found for Component 3
in the 1999 December spectrum. 
However, the \hi\ covering fractions of Components 3
and 5 are uniformly less than unity, and lower than those measured in Components 1 and 4,
suggesting an origin in a nuclear outflow.

With the exception of
Component 1, all velocity components are viewed with large blueshifts
with respect to the systemic redshift of the galaxy, itself 
defined by galactic absorption lines, as discussed in Section~\ref{sec-intabs}.
This implies that the low-ionization gas 
and perhaps some fraction of the high-ionization gas
seen in Component 4
is analogous to the Milky Way HVCs
and is involved in some kind of large-scale galactic outflow, 
such as a galactic fountain. In a fountain flow, we may
expect to see both blueshifted and redshifted absorption as the material
falls back onto the disk of the galaxy. In light of this, we note
that the host galaxy of Mrk~279 is an S0 galaxy with an off-center nucleus, indicating
an interaction with a neighboring galaxy, MCG +12-13-024, from which it is
separated by $\sim$20~$h^{-1}$~kpc
and 360~\kms\ in line-of-sight
velocity (Keel et al.\ 1996).  High-resolution imaging of Mrk~279 
reveals the presence
of a nuclear dust spiral and a large-scale bar (Pogge \& Martini 1998;
Peletier et al.\ 1999), and  its inclination is $\sim50-60\arcdeg$.  It
is therefore plausible that our line of sight traces a substantial path
through the disk and halo of the early-type host galaxy of Mrk~279, 
that the dynamical state of its ISM and immediate environment is complex,
and that this is reflected in the absorption patterns we observe.

\section{Summary}
\label{sec-summary}

We have presented far-UV observations of Mrk~279 from \fuse\ obtained over a 
2.4-yr period.  We have also presented UV and X-ray spectra 
from \stis\ and {\it Chandra}
taken simultaneously with the final \fuse\ spectrum.  From these data, we conclude the
following:

\begin{enumerate}

\item {Over the 2.4-yr epoch of observations, the UV continuum flux of Mrk~279
decreased by a factor of $\sim$7.5.  The brighter spectra
show bluer power-law continuum slopes.
The flux in the broad O~{\sc vi} emission
lines decreased by a factor of $\sim$8 over this period, 
in tandem with changes in the continuum flux.}

\item{We do not detect absorption due to O~{\sc vii} or
O~{\sc viii} in the merged MEG + HEG {\it Chandra} spectrum.
We use this non-detection to set limits of
$N$(O~{\sc vii},O~{\sc viii}) $< 5 \times 10^{16}$~cm$^{-2}$,
under the assumption of unsaturated absorption.  
However, for a reasonable
assumption about the Doppler parameter of the X-ray absorber,
the column densities in O~{\sc vii} and O~{\sc viii} can
be large enough to be consistent with X-ray observations of other
AGNs that also show intrinsic absorption in the UV.}

\item{We detect significant, narrow (FWHM 4200$^{+3350}_{-2950}$ \kms) Fe K$\alpha$
emission in both the HEG and  MEG {\it Chandra} spectra.  The
line is marginally blueshifted with respect to the systemic
redshift of Mrk~279 and is reduced in strength in direct proportion
to changes in the continuum flux as compared with previous observations
with {\it ASCA}.}

\item{Components 2, 2a, and 2b show strong O~{\sc vi} and
prominent \hi\ absorption with little absorption due to other low-ionization
species.  The \hi\ covering fractions are consistently less than unity.
The Ly$\beta$ and O~{\sc vi} flux profiles in these components
vary, but it is not clear whether this is due to a change in the
covering fractions or column densities of the absorbers.
These characteristics lead us to conclude that these components
arise in an outflow from the AGN in Mrk~279.}

\item{The large \hi\ column density of UV absorption Component 1, constant over the
four observation epochs, 
its low ionization state, high covering fraction, and its
velocity consistent with the systemic redshift of Mrk~279 indicate that
it likely arises in the  
host galaxy of this AGN.}

\item{Component 4 displays
absorption due to a wide
range of ionization states including a strong Lyman series from neutral
hydrogen, 
prominent C~{\sc iii}, N~{\sc iii}, and Si~{\sc iii} and detectable 
C~{\sc ii}, N~{\sc ii}, and Si~{\sc ii} as well as features due to
highly ionized species such as O~{\sc vi}, C~{\sc iv}, and N~{\sc v}.
The absorption profiles of the high-ionization lines are distinctly
different from those of the low-ionization lines, showing velocity offsets and 
larger velocity widths.
The \hi\ column density in this component decreased by a factor
of $\sim$10 as the UV continuum flux of Mrk~279 decreased by a factor
of $\sim$7.  
This component also shows a neutral hydrogen covering fraction consistent
with unity at all epochs. 
These characteristics lead us
to conclude that the ISM of the host galaxy contributes to the
total absorption in this component.}

\item{The low density inferred from \linebreak
\mbox{$N$(C{\sc ii}$^{*}$)/$N$(C~{\sc ii})}
in Component 4, $n_{e} \lesssim$~750~cm$^{-3}$, 
also indicates that the low-ionization gas
arises in the host galaxy rather than the AGN outflow.}

\item{The origin of Components 3 and 5 is uncertain.
They show some similarities to Component 4 that suggest an
association with the AGN host galaxy, namely
low-ionization absorption with profiles that differ from the 
high-ionization absorption profiles and
marginal variability in $N$(\hi) with UV continuum flux in the same sense
as Component 4. 
However, although the \hi\ covering fractions of these components are
generally consistent with unity, they are uniformly less than unity,
implying an origin in an AGN outflow.
}

\item{
The large blueshifts of Components 3, 4, and 5 with respect to the 
systemic redshift
of the host of Mrk~279 imply that if some part of the absorption in
these components arises in the host galaxy,
some type of galactic outflow may be present. Alternatively, the absorption may
arise in a physical interaction between Mrk~279 and a companion galaxy,
MCG~+12-13-024.}

\end{enumerate}

Another set of simultaneous UV and X-ray observations of
Mrk~279 obtained in 2003 March with \fuse, \stis, and {\it Chandra}, 
when the AGN was once again in a
high state, are forthcoming in a future paper (Arav et al.\ 2004, in preparation). 
Because of the complicated, multiphase nature of the intrinsic absorption
in the spectra we have presented, we
defer detailed photoionization modeling for this future paper
in order to incorporate these new data into that work. 

\acknowledgements
J.\ E.\ S.\ acknowledges helpful discussions with K.\ Sembach, R.\ Ganguly, and J.\ Gabel and
thanks the anonymous referee for constructive comments that improved the paper.
J.\ C.\ L.\  thanks the  Chandra fellowship 
grant PF2-30023 for financial support. This grant is issued by the
Chandra X-ray Observatory Center which is operated by the Smithsonian Astrophysical
Observatory for NASA under contract NAS8--39073.
This research has made use of the NASA/IPAC Extragalactic Database (NED), 
which is operated by the Jet
Propulsion Laboratory, Caltech, under contract 
with the National Aeronautics and Space Administration.

\onecolumn
\begin{deluxetable}{llccll}
\tablecolumns{5}
\tablewidth{27pc}
\tablecaption{Observations of MRK 279 \label{table-obs}}
\tablehead{
\colhead{Instrument} &\colhead{ID} &\colhead{Start Date} &\colhead{UT} &\colhead{Exp. (s)} }
\startdata
{\it Chandra} &700501/3062 &2002-05-18   &06:28:37       &114200 \\
{\it FUSE} &P1080303     &1999-12-28     &00:37:05       &61139   \\
{\it FUSE} &P1080304     &2000-01-11     &01:28:19       &30288   \\
{\it FUSE} &S6010501     &2002-01-28     &22:52:53       &18115   \\
{\it FUSE} &S6010502     &2002-01-29     &07:10:53       &19082   \\
{\it FUSE} &C0900201     &2002-05-18     &18:13:44       &47414  \\
STIS &O6JM01       &2002-05-18     &12:53:58       &13193   \\
\enddata
\end{deluxetable}

\begin{deluxetable}{lccr}
\tablecolumns{4}
\tablewidth{25pc}
\tablecaption{Continuum Fits to {\it FUSE} and STIS Spectra of Mrk~279 
\tablenotemark{1}
\label{table-contspec}}
\tablehead{\colhead{Obs. Date}
&\colhead{$f_{\lambda}$(1000 \AA)\tablenotemark{2}} 
&\colhead{$\alpha$} &\colhead{$\Delta\lambda$ (\AA)} }
\startdata
1999Dec28    &$1.325\pm0.004$  &$1.60\pm0.02$ &927-1150 \\
2000Jan11    &$0.913\pm0.004$  &$1.04\pm0.04$ &927-1150 \\
2002Jan28-29 &$0.159\pm0.003$  &$0.84\pm0.22$ &1000-1182 \\
2002May18    &$0.133\pm0.001$  &$0.86\pm0.02$ &994-1682\tablenotemark{3}\\
\enddata
\tablenotetext{1}{Fits to power law of the form $f_{\lambda} \propto \lambda^{-\alpha}$}
\tablenotetext{2}{Flux at 1000 \AA\ in units of 10$^{-13}$ ergs s$^{-1}$ cm$^{-2}$ \AA$^{-1}$}
\tablenotetext{3}{{\it FUSE} spectrum from LiF channel only, combined with STIS spectrum}
\end{deluxetable}

\begin{deluxetable}{llccc}
\tablecolumns{5}
\tablewidth{29pc}
\tablecaption{Emission Line Fits to {\it FUSE} and STIS Spectra of Mrk~279
\label{table-emspec}}
\tablehead{
\colhead{Line} &\colhead{$\lambda_{\rm vac}$} 
&\colhead{Flux\tablenotemark{1}}
&\colhead{Velocity\tablenotemark{2}}
&\colhead{FWHM} \\
\colhead{} &\colhead{(\AA)} 
&\colhead{}
&\colhead{(\kms)}
&\colhead{(\kms)} }
\startdata
\multicolumn{5}{c}{1999 December}\\
\hline
S{\sc vi}   &933.37  &6.0  $\pm$ 2.2  &929 $\pm$ 1091 &7504 $\pm$ 34 \\
S{\sc vi}   &944.52  &3.0  $\pm$ 1.1  &929 $\pm$ 1091 &7504 $\pm$ 34 \\
C{\sc iii}  &977.02  &30.0 $\pm$ 6.8  &491 $\pm$ 208  &8852 $\pm$ 770 \\
N{\sc iii}  &989.79  &7.3  $\pm$ 1.7  &599 $\pm$ 2637 &8852 $\pm$ 770 \\
Ly$\beta$ broad  &1025.72 &29.7 $\pm$ 1.0 &-317 $\pm$ 110 &8555 $\pm$ 869 \\
Ly$\beta$ int.   &1025.72 &8.9  $\pm$ 2.2 &-252 $\pm$ 48  &3308 $\pm$ 428 \\
Ly$\beta$ narrow\tablenotemark{3} &1025.72 &0.3 $\pm$ 0.6  &315 $\pm$ 17 &697 $\pm$ 34 \\ 
O{\sc vi} broad  &1031.93 &96.3 $\pm$ 1.9 &595 $\pm$ 115  &7504 $\pm$ 34 \\
O{\sc vi} broad  &1037.62 &48.1 $\pm$ 0.9 &595 $\pm$ 115  &7504 $\pm$ 34 \\
O{\sc vi} int.   &1031.93 &31.4 $\pm$ 0.5 &-67 $\pm$ 5    &2695 $\pm$ 187 \\
O{\sc vi} int.   &1037.62 &15.7 $\pm$ 0.2 &-67 $\pm$ 5    &2695 $\pm$ 187 \\
S{\sc iv}   &1062.66 &12.8 $\pm$ 0.4 &-222 $\pm$ 104 &3213 $\pm$ 345 \\
S{\sc iv}   &1072.97 &12.8 $\pm$ 0.4 &-222 $\pm$ 104 &3213 $\pm$ 345 \\
He{\sc ii}  &1085.15 &30.8 $\pm$ 0.7 &-63  $\pm$ 45  &7504 $\pm$ 34 \\
\hline
\multicolumn{5}{c}{2000 January} \\
\hline
S{\sc vi}   &933.37  &5.0 $\pm$ 2.1  &1138 $\pm$ 932 &7023 $\pm$ 9 \\
S{\sc vi}   &944.52  &2.5 $\pm$ 1.0  &1138 $\pm$ 932 &7023 $\pm$ 9 \\
C{\sc iii}  &977.02  &24.7 $\pm$ 5.7 &220  $\pm$ 986 &8576 $\pm$ 314 \\
N{\sc iii}  &989.79  &21.0 $\pm$ 5.3 &-147 $\pm$ 658 &8576 $\pm$ 314 \\
Ly$\beta$ broad  &1025.72 &34.1 $\pm$ 2.3 &150 $\pm$ 42 &7858 $\pm$ 305 \\
Ly$\beta$ int.   &1025.72 &7.6  $\pm$ 1.8 &102 $\pm$ 70 &2395 $\pm$ 361 \\
Ly$\beta$ narrow\tablenotemark{3} &1025.72 &0.7 $\pm$ 0.4  &315 $\pm$ 17  &697 $\pm$ 34 \\
O{\sc vi} broad  &1031.93 &82.7 $\pm$ 0.4 &372 $\pm$ 59  &7023 $\pm$ 9  \\
O{\sc vi} broad  &1037.62 &41.4 $\pm$ 0.2 &372 $\pm$ 59  &7023 $\pm$ 9 \\
O{\sc vi} int.   &1031.93 &27.7 $\pm$ 0.4 &144 $\pm$ 31  &2164 $\pm$ 58 \\
O{\sc vi} int.   &1037.62 &13.8 $\pm$ 0.2 &144 $\pm$ 31  &2164 $\pm$ 58 \\
S{\sc iv}   &1062.66 &17.9 $\pm$ 0.5 &278 $\pm$ 59   &3500 $\pm$ 21 \\
S{\sc iv}   &1072.97 &17.9 $\pm$ 0.5 &278 $\pm$ 59   &3500 $\pm$ 21 \\
He{\sc ii}  &1085.15 &30.4 $\pm$ 0.9 &-125 $\pm$ 40 &7023 $\pm$ 9 \\
\hline
\multicolumn{5}{c}{2002 January} \\
\hline
C{\sc iii}  &977.02  &7.0 $\pm$ 1.2 &-348 $\pm$ 628  &8315 $\pm$ 333 \\
N{\sc iii}  &989.79  &7.9 $\pm$ 0.7 &-482 $\pm$ 426  &8315 $\pm$ 333 \\
Ly$\beta$ broad    &1025.72 &8.0 $\pm$ 0.9 &150 $\pm$ 363  &9488 $\pm$ 61 \\
Ly$\beta$ int.     &1025.72 &3.8 $\pm$ 1.2 &-130$\pm$ 275  &1702 $\pm$ 277 \\
Ly$\beta$ narrow\tablenotemark{3}  &1025.72 &0.3 $\pm$ 0.5 &315 $\pm$ 17  &697 $\pm$ 34 \\
O{\sc vi} broad    &1031.93 &19.1 $\pm$ 1.0 &615 $\pm$ 166  &7081 $\pm$ 56 \\
O{\sc vi} broad    &1037.62 & 9.5 $\pm$ 0.5 &615 $\pm$ 166  &7081 $\pm$ 56 \\
O{\sc vi} int.     &1031.93 &17.0 $\pm$ 0.5 &45 $\pm$ 25    &1674 $\pm$ 123 \\
O{\sc vi} int.     &1037.62 & 8.5 $\pm$ 0.2 &45 $\pm$ 25    &1674 $\pm$ 123 \\
S{\sc iv}   &1062.66 &1.6 $\pm$ 0.2 &-198 $\pm$ 95 &3438 $\pm$ 335 \\
S{\sc iv}   &1072.97 &1.6 $\pm$ 0.2 &-198 $\pm$ 95 &3438 $\pm$ 335 \\
He{\sc ii}  &1085.15 &4.8 $\pm$ 0.4 &-79 $\pm$ 367 &7081 $\pm$ 57 \\
\hline
\multicolumn{5}{c}{2002 May\tablenotemark{4}} \\
\hline
S{\sc vi}           &933.37  &4.8 $\pm$ 3.0  &607 $\pm$ 893 &9495 $\pm$ 58 \\
S{\sc vi}           &944.52  &2.4 $\pm$ 1.5  &607 $\pm$ 893 &9495 $\pm$ 58 \\
C{\sc iii}          &977.02  &9.4 $\pm$ 1.2  &394 $\pm$ 664 &9849 $\pm$ 2133 \\
N{\sc iii}          &989.79  &0.4 $\pm$ 0.4  &665 $\pm$ 1141 &9849 $\pm$ 2133 \\
Ly$\beta$ broad     &1025.72 &4.3 $\pm$ 2.7  &-396 $\pm$ 36 &7862 $\pm$ 119 \\
Ly$\beta$ int.      &1025.72 &5.2 $\pm$ 0.7  &-243 $\pm$ 18 &2990 $\pm$ 39 \\
Ly$\beta$ narrow    &1025.72 &0.4 $\pm$ 0.2  &315 $\pm$ 17  &697 $\pm$ 34  \\
O{\sc vi} broad     &1031.93 &12.4 $\pm$ 2.0 &403 $\pm$ 491 &9495 $\pm$ 58 \\
O{\sc vi} broad     &1037.62 &6.2 $\pm$ 1.0  &403 $\pm$ 491 &9495 $\pm$ 58 \\
O{\sc vi} int.      &1031.93 &13.2 $\pm$ 0.5 &327 $\pm$ 78  &2758 $\pm$ 43 \\
O{\sc vi} int.      &1037.62 &6.6 $\pm$ 0.2  &327 $\pm$ 78  &2758 $\pm$ 43 \\
S{\sc iv}\tablenotemark{5}   &1072.97 &2.2 $\pm$ 0.4 &-124$\pm$96 &1104$\pm$166 \\
He{\sc ii}          &1085.15 &0.2 $\pm$ 0.008&360 $\pm$ 170 &9495 $\pm$ 58 \\
Si{\sc ii}          &1192.33 &1.3 $\pm$ 0.3  &545 $\pm$ 234 &9195 $\pm$ 534 \\
Ly$\alpha$ broad    &1215.67 &130. $\pm$ 1.8 &-396 $\pm$ 36 &7862 $\pm$ 119 \\
Ly$\alpha$ int.     &1215.67 &72.8 $\pm$ 1.6 &-243 $\pm$ 18 &2990 $\pm$ 39 \\
Ly$\alpha$ narrow   &1215.67 &9.0 $\pm$ 0.5  &315 $\pm$ 17  &697 $\pm$ 34 \\
N{\sc v}   broad    &1240.15 &29.6 $\pm$ 0.9 &119 $\pm$ 171 &9495 $\pm$ 58 \\
N{\sc v}   int.     &1240.15 &11.5 $\pm$ 0.5 &226 $\pm$ 47  &2758 $\pm$ 43 \\
Si{\sc ii}          &1260.45 &2.0 $\pm$ 0.6  &544 $\pm$ 233 &9195 $\pm$ 534 \\
O{\sc i}+Si{\sc ii}  &1304.35 &6.8 $\pm$ 0.3  &143 $\pm$ 89  &4043 $\pm$ 354 \\
C{\sc ii}           &1335.30 &1.5 $\pm$ 0.1  &-57 $\pm$ 145 &2460 $\pm$ 147 \\
Si{\sc iv}          &1393.76 &9.8 $\pm$ 0.2  &-283 $\pm$ 270 &9495 $\pm$ 58 \\
Si{\sc iv}          &1402.77 &4.9 $\pm$ 0.1  &-283 $\pm$ 270 &9495 $\pm$ 58  \\
O{\sc iv}]          &1402.06 &9.4 $\pm$ 0.4  &-405 $\pm$ 89  &3609 $\pm$ 138  \\ 
C{\sc iv} broad     &1549.05 &153. $\pm$ 1.1 &-177 $\pm$ 8   &9495 $\pm$ 58 \\
C{\sc iv} int.      &1549.05 &50.1 $\pm$ 0.9 &-196 $\pm$ 18  &2758 $\pm$ 43 \\
He{\sc ii}          &1640.50 &15.4 $\pm$ 0.9 &184 $\pm$ 185  &9495 $\pm$ 58 \\ 
\enddata
\tablenotetext{1}{Flux in units of 10$^{-14}$ ergs cm$^{-2}$ s$^{-1}$}
\tablenotetext{2}{Velocity relative to systemic redshift, $z=0.0305$}
\tablenotetext{3}{Velocity and FWHM fixed at 2002 May values}
\tablenotetext{4}{{\it FUSE} spectrum from LiF1 channel only, combined 
with STIS spectrum}
\tablenotetext{5}{Lack of LiF2 and SiC channels prevents coverage
of 1083-1093~\AA\ and fit to S{\sc iv}\lam1062}
\end{deluxetable}

\begin{deluxetable}{lrrrr}
\tablecolumns{5}
\tablewidth{27pc}
\tablecaption{Velocity Components of Intrinsic Absorption in Mrk~279
\tablenotemark{1} \label{table-vel}}
\tablehead{ \colhead{Component}
&\colhead{1999 Dec.} &\colhead{2000 Jan.} &\colhead{2002 Jan.} &\colhead{2002 May} }
\startdata
\multicolumn{5}{c}{Ly$\beta$} \\
\hline 
1  &90     &90      &90      &90   \\
\nodata &\nodata &\nodata &\nodata  &-155\tablenotemark{2} \\
2  &-265   &-265    &-250    &-250   \\
2a &-300   &-305    &-300    &-280  \\
2b &-325   &-335    &-330    &-335  \\
3  &-385   &-385    &-380    &-390  \\
4  &-450   &-450    &-450    &-450  \\
4a &-510   &-500    &-490    &-490  \\
5  &-540   &-540    &-560    &-540  \\
\nodata &\nodata &\nodata &\nodata  &-605\tablenotemark{2} \\
\hline
\multicolumn{5}{c}{C{\sc iii}\lam977}\\
\hline
1  &85     &85      &80      &85  \\
2  &-255   &-260    &-260    &-260  \\
2a &-290   &-300    &-290    &-300  \\
2b &-320   &-330    &-345    &-335  \\
3  &-385   &-385    &-380    &-385  \\
4  &-460   &-460    &-460    &-445  \\
4a &-505   &-495    &-510    &-485  \\
5  &-550   &-550    &-570    &-550  \\
\hline
\multicolumn{5}{c}{O{\sc vi}\lam1032} \\
\hline
1  &85     &90     &85      &70 \\
2  &-270   &-270   &-240    &-275  \\
2a &-300   &-300   &-285    &-310 \\
2b &-330   &-335   &-335    &-345 \\
3  &-385   &-380   &-390    &-380 \\
4  &-470   &-475   &-440    &-450 \\
4a &-490   &-500   &-500    &-500 \\
5  &-535 &-540  &-560  &-530 \\
\enddata
\tablenotetext{1}{Velocity centroids in \kms\ with respect to the systemic velocity of Mrk~279,
$z=0.0305$.}
\tablenotetext{2}{Velocity components seen in Ly$\alpha$ with no counterparts in 
the {\it FUSE} Ly$\beta$ profile, 
marked with bold lines in Figure~\ref{fig:ly502}.}
\end{deluxetable}

\begin{deluxetable}{llllc}
\tablecolumns{5}
\tablewidth{30pc}
\tablecaption{Covering Fractions and Column Densities 
of Intrinsic Hydrogen Absorption
\label{table-cover2}}
\tablehead{
 \colhead{Component}
 &\colhead{$C_{c}$} &\colhead{$C_{l}$} &\colhead{$N$ (10$^{14}$ cm$^{-2}$)}
 &\colhead{$\Delta v$ (\kms)} }
\startdata
\multicolumn{5}{c}{1999 December} \\
\hline
1  &$0.96^{+0.27}_{-0.21}$  &$0.98^{+0.12}_{-0.16}$     &$23.4^{+1.3}_{-1.1}$ &60  \\
2  &$0.84^{+0.85}_{-0.99}$  &$0.65^{+0.29}_{-0.52}$ &$4.7^{+0.8}_{-1.4}$ &30 \\
2a &$0.70^{+0.77}_{-0.66}$  &$0.49^{+0.29}_{-0.33}$ &$3.3^{+2.2}_{-3.1}$ &24 \\
2b &$0.72^{+0.76}_{-0.94}$  &$0.24^{+0.30}_{-0.37}$ &$3.6^{+1.2}_{-1.0}$ &36 \\
3  &$0.88^{+0.82}_{-0.68}$  &$0.50^{+0.44}_{-0.33}$ &$3.0^{+0.7}_{-7.6}$ &42  \\
4  &$0.94^{+0.35}_{-0.09}$  &$0.94^{+0.19}_{-0.35}$ &$22.1^{+2.0}_{-1.2}$   &30  \\
4a &$0.23^{+0.34}_{-0.11}$  &$0.25^{+0.22}_{-0.37}$ &$12.6^{+8.1}_{-5.6}$  &30  \\
5  &$0.33^{+0.53}_{-1.32}$  &$0.43^{+0.43}_{-1.23}$ &$10.7^{+1.7}_{-10.1}$   &30  \\
\hline
\multicolumn{5}{c}{2000 January} \\
\hline
1  &$0.97^{+0.41}_{-0.24}$  &$0.98^{+0.15}_{-0.33}$ &$25.2^{+3.3}_{-1.5}$   &60  \\
2  &$0.86^{+0.82}_{-0.36}$  &$0.86^{+0.56}_{-0.69}$ &$5.8^{+3.4}_{-3.7}$ &30  \\
2a &$0.79^{+0.84}_{-1.22}$  &$0.67^{+0.30}_{-0.37}$ &$4.8^{+4.9}_{-2.5}$ &24  \\
2b &$0.63^{+0.77}_{-0.25}$  &$0.56^{+0.33}_{-0.50}$ &$>4.7$                &36  \\
3  &$0.61^{+0.83}_{-0.91}$  &$0.89^{+0.86}_{-0.43}$ &$4.9^{+2.4}_{-4.0}$ &41  \\
4  &$0.94^{+0.45}_{-0.27}$  &$0.97^{+0.15}_{-0.25}$ &$23.0^{+5.0}_{-1.9}$   &30  \\
4a &$0.20^{+0.36}_{-0.30}$  &$0.84^{+0.81}_{-2.42}$ &$6.9^{+9.3}_{-4.0}$  &30  \\ 
5  &$0.20^{+0.31}_{-0.88}$  &$0.61^{+0.69}_{-2.03}$ &$>1.9$                &30  \\
\hline
\multicolumn{5}{c}{2002 May} \\
\hline
1  &$0.91^{+0.63}_{-0.20}$  &$0.99^{+0.04}_{-0.18}$ &$>3.3$                &57  \\
2  &$0.86^{+0.60}_{-0.26}$  &$0.60^{+0.06}_{-0.07}$ &$>0.8$                &28 \\
2a &$0.91^{+0.78}_{-0.31}$  &$0.61^{+0.07}_{-0.08}$ &$>1.6$                &28 \\
2b &$0.74^{+0.88}_{-0.31}$  &$0.71^{+0.06}_{-0.08}$ &$>1.8$                &37 \\ 
3  &$1.00^{+1.00}_{-0.48}$  &$0.90\pm0.09$          &$1.8^{+1.1}_{-0.6}$ &40 \\
4  &$0.91^{+0.81}_{-0.28}$  &$0.98^{+0.06}_{-0.05}$ &$2.8^{+1.4}_{-0.7}$ &28 \\
4a &$0.95^{+1.00}_{-0.49}$  &$0.97^{+0.09}_{-0.05}$ &$1.6^{+1.3}_{-0.5}$ &31 \\
5  &$0.16^{+0.65}_{-0.52}$  &$0.94^{+0.09}_{-0.10}$ &$3.0^{+0.4}_{-1.5}$ &28  \\
\enddata
\end{deluxetable}

\begin{deluxetable}{lllc}
\tablecolumns{4}
\tablewidth{25pc}
\tablecaption{Covering Fractions and Column Densities
of Intrinsic Hydrogen Absorption
\label{table-cover}}
\tablehead{
 \colhead{Component}
 &\colhead{$C_{f}$} &\colhead{$N$ (10$^{14}$ cm$^{-2}$)}
 &\colhead{$\Delta v$ (\kms)} }
\startdata
\multicolumn{4}{c}{1999 December} \\
\hline
1   &$0.97^{+0.06}_{-0.01}$ &$23.1^{+0.8}_{-0.9}$   &60 \\
2   &$0.76\pm0.04$          &$5.1^{+0.6}_{-0.9}$ &30 \\ 
2a  &$0.62^{+0.25}_{-0.03}$ &$3.3^{+0.6}_{-0.7}$ &24 \\
2b  &$0.50^{+0.10}_{-0.03}$ &$5.8^{+1.0}_{-0.8}$ &36 \\
3   &$0.73^{+0.10}_{-0.01}$ &$3.9^{+0.9}_{-1.1}$ &42 \\
4   &$0.95^{+0.05}_{-0.01}$ &$21.5^{+1.5}_{-1.4}$   &30 \\
4a       &$0.27^{+0.02}_{-0.70}$ &$10.6^{+5.4}_{-4.0}$   &30 \\
5        &$0.46^{+0.55}_{-0.79}$ &$1.5^{+2.6}_{-1.8}$ &30 \\
\hline
\multicolumn{4}{c}{2000 January} \\
\hline
1    &$0.98^{+0.08}_{-0.02}$ &$25.5^{+1.5}_{-1.3}$  &60 \\
2    &$0.86^{+0.18}_{-0.24}$ &$5.1^{+0.8}_{-2.0}$ &30 \\ 
2a   &$0.71^{+0.20}_{-0.22}$ &$3.6^{+1.0}_{-1.8}$ &24 \\ 
2b   &$0.61^{+0.74}_{-0.21}$ &$5.8^{+1.6}_{-2.1}$ &36 \\ 
3    &$0.78^{+1.13}_{-0.11}$ &$3.1^{+0.9}_{-2.1}$ &41 \\ 
4    &$0.96^{+0.16}_{-0.04}$ &$22.7^{+2.6}_{-2.0}$   &30 \\ 
4a   &$0.66^{+0.50}_{-1.16}$ &$4.3^{+2.5}_{-2.1}$ &30 \\ 
5    &$0.32^{+0.41}_{-0.78}$ &$3.3^{+3.2}_{-2.0}$ &30 \\ 

\hline
\multicolumn{4}{c}{2002 January} \\
\hline
1  &$0.94^{+0.50}_{-0.05}$  &$>9.4$                  &57  \\
2  &$0.99^{+0.87}_{-0.33}$  &$1.4^{+4.1}_{-3.4}$   &28  \\
2a &$0.79^{+1.26}_{-0.10}$  &$4.7^{+4.4}_{-3.2}$   &28  \\
2b &$0.66^{+0.62}_{-0.60}$  &$>5.2$                  &28  \\
3  &$0.72^{+0.72}_{-0.07}$  &$>1.6$                  &43  \\
4  &$0.82^{+1.46}_{-0.08}$  &$>5.2$                  &28  \\
4a &$0.98^{+0.87}_{-2.18}$  &$1.7^{+4.8}_{-4.1}$   &28  \\
5  &$0.26^{+1.18}_{-3.51}$  &$>1.8$                  &28  \\
\hline
\multicolumn{4}{c}{2002 May} \\
\hline
1    &$0.98^{+0.02}_{-0.11}$  &$>4.7$ &57\\ 
2    &$0.64\pm0.05$           &$>1.4$ &28\\
2a   &$0.65\pm0.05$           &$>2.6$ &28\\
2b   &$0.72\pm0.04$           &$>3.0$ &37\\
3    &$0.91^{+0.05}_{-0.06}$  &$1.9\pm0.4$          &40\\
4    &$0.98^{+0.03}_{-0.06}$  &$2.2^{+0.9}_{-0.3}$  &28\\
4a   &$0.97^{+0.04}_{-0.02}$  &$1.5^{+0.3}_{-0.2}$  &31\\
5    &$0.89^{+0.47}_{-0.07}$  &$0.9^{+0.3}_{-0.2}$  &28\\
\enddata
\end{deluxetable}

\begin{deluxetable}{lllc}
\tablecolumns{4}
\tablewidth{25pc}
\tablecaption{Properties of Intrinsic Metal Absorption:  Doublets
\label{table-compdoub}}
\tablehead{
 \colhead{Component} &\colhead{$C_{f}$} &\colhead{$N$ (10$^{14}$ cm$^{-2}$)}
 &\colhead{$\Delta v$ (\kms)} \\
\hline  \\
\multicolumn{2}{c}{Species} &\multicolumn{2}{c}{Date} } 
\startdata
\multicolumn{2}{c}{O{\sc vi}\lam\lam1032,1038} &\multicolumn{2}{c}{1999 December} \\
\hline
1  &$0.68^{+0.60}_{-1.33}$       &$1.1\pm0.3$ &59  \\
2   &$0.83^{+0.09}_{-0.13}$  &$>0.8$                &30  \\
2a  &$0.82^{+0.09}_{-0.14}$  &$>3.9$                &24  \\
2b  &$0.79^{+0.09}_{-0.30}$  &$>3.4$                &36  \\
3   &$0.51^{+0.14}_{-0.19}$  &$2.8^{+0.7}_{-0.6}$ &41  \\
4       &$0.90^{+0.04}_{-0.02}$  &$3.5\pm0.2$         &30  \\
4a      &$0.86^{+0.04}_{-0.02}$  &$3.8^{+0.5}_{-0.2}$ &30  \\
5       &$0.40^{+0.05}_{-0.20}$  &$>1.2$  &30  \\
\hline
\multicolumn{2}{c}{O{\sc vi}\lam\lam1032,1038} &\multicolumn{2}{c}{2000 January} \\
\hline
1  &$0.76^{+0.61}_{-0.25}$  &$>1.2$                &59  \\
2  &$0.81^{+0.03}_{-0.05}$  &$>0.9$                &30  \\
2a &$0.79^{+0.03}_{-0.05}$  &$>1.4$                &24  \\
2b &$0.79^{+0.04}_{-0.05}$  &$>2.9$                &36  \\
3  &$0.62^{+0.22}_{-0.06}$  &$2.9^{+1.1}_{-0.6}$   &41  \\
4  &$0.92^{+0.08}_{-0.03}$  &$2.9\pm0.3$           &30  \\
4a &$0.81\pm0.04$           &$>4.0$                &30  \\
5  &$0.35^{+0.11}_{-0.06}$  &$>0.9$                &30  \\
\hline
\multicolumn{2}{c}{O{\sc vi}\lam\lam1032,1038} &\multicolumn{2}{c}{2002 January} \\
\hline
1  &$0.72^{+1.13}_{-0.01}$  &$0.4^{+0.4}_{-0.5}$ &56  \\
2  &$1.00^{+1.01}_{-0.03}$  &$0.3^{+0.1}_{-0.8}$ &28  \\
2a &$0.56^{+0.20}_{-0.07}$  &$>1.1$                &28  \\
2b &$0.63^{+0.13}_{-0.07}$  &$4.4^{+2.0}_{-2.6}$ &42  \\
3  &$0.60^{+0.16}_{-0.08}$  &$4.4^{+2.6}_{-1.1}$ &42  \\
4  &$0.70\pm0.04$           &$>3.8$                &28  \\
4a &$0.77^{+0.21}_{-0.09}$  &$2.2^{+1.9}_{-0.7}$ &28  \\
5  &$0.27^{+0.10}_{-0.05}$  &$>1.4$                &28  \\
\hline
\multicolumn{2}{c}{O{\sc vi}\lam\lam1032,1038} &\multicolumn{2}{c}{2002 May} \\
\hline
1  &$0.16^{+0.50}_{-0.16}$  &$>0.03$ &59  \\
2  &$0.52^{+0.39}_{-0.09}$  &$>0.6$                &28  \\
2a &$0.40^{+0.26}_{-0.08}$  &$>0.7$                &28  \\ 
2b &$0.45^{+0.62}_{-0.22}$  &$>0.9$                &40  \\
3  &$0.36^{+0.38}_{-0.08}$  &$>1.1$                &39  \\
4  &$0.46^{+0.20}_{-0.11}$  &$>1.4$                &25  \\
4a &$0.71^{+0.58}_{-0.05}$  &$>0.3$                &28  \\
5  &$0.30\pm0.08$           &$>0.8$                &31  \\
\hline
\multicolumn{2}{c}{N{\sc v}\lam\lam1238,1242} &\multicolumn{2}{c}{2002 May} \\
\hline
1  &\nodata                 &$<1.5$\tablenotemark{1}    &60 \\
2  &$0.65^{+0.11}_{-0.07}$  &$>0.2$    &28  \\
2a &$0.72^{+0.69}_{-0.11}$  &$>0.8$    &25  \\
2b &$0.73^{+0.44}_{-0.95}$  &$>0.9$    &35  \\
3  &$0.73^{+1.15}_{-2.11}$  &$>0.7$    &39  \\
4  &$0.59^{+1.01}_{-1.98}$  &$>0.4$    &28  \\
4a &$0.76^{+0.52}_{-1.07}$  &$>0.4$    &28 \\
5  &$0.77^{+0.94}_{-2.14}$  &$>0.2$    &32  \\
\hline
\multicolumn{2}{c}{C{\sc iv}\lam\lam1548,1550} &\multicolumn{2}{c}{2002 May} \\
\hline
1  &\nodata                 &$<0.7$\tablenotemark{1}     &59  \\
2  &$0.70^{+0.68}_{-0.15}$  &$>0.7$           &31  \\
2a &$0.71^{+0.55}_{-0.72}$  &$>0.5$           &25  \\
2b &$0.80^{+0.80}_{-0.19}$  &$>0.5$           &37  \\
3  &\nodata                 &$<0.9$\tablenotemark{1}           &39 \\
4  &\nodata                 &\nodata           &\nodata \\
4a &$0.63^{+0.27}_{-0.11}$  &$>0.2$                &28  \\
5  &$0.59^{+0.89}_{-0.06}$  &$>0.2$                &31  \\
\hline
\multicolumn{2}{c}{Si{\sc iv}\lam\lam1393,1402} &\multicolumn{2}{c}{2002 May} \\
\hline
1  &\nodata  &$<0.2$\tablenotemark{1}  &59  \\
2  &\nodata  &$<0.08$\tablenotemark{1}  &28  \\
2a &\nodata  &$<0.1$\tablenotemark{1}  &28  \\
2b &\nodata  &$<0.9$\tablenotemark{1}  &40  \\
3  &\nodata  &$<0.09$\tablenotemark{1}  &39 \\
4  &\nodata  &$<0.7$\tablenotemark{1}  &25 \\
4a &\nodata  &$<0.5$\tablenotemark{1}  &28  \\
5  &\nodata  &$<0.1$\tablenotemark{1}  &31  \\
\enddata
\tablenotetext{1}{Limit calculated assuming $C_{f}=C_{f}$(O{\sc vi})}
\end{deluxetable}

\begin{deluxetable}{lllll}
\tablecolumns{5}
\tablewidth{32pc}
\tablecaption{Properties of Intrinsic Metal Absorption in {\it FUSE} Spectra 
\label{table-compmetfuse}}
\tablehead{
 \colhead{Component} &\colhead{$N$(C{\sc iii}\lam977)\tablenotemark{1}}
 &\colhead{$N$(N{\sc iii}\lam989)} &\colhead{$N$(C{\sc ii}\lam1036)}
 &\colhead{$N$(N{\sc ii}\lam1084)}}
\startdata
\multicolumn{5}{c}{1999 Dec.}\\
\hline
1   &$0.18\pm0.01$   &$0.10\pm0.02$  &\nodata       &$<0.04$ \\
3   &$0.16\pm0.01$   &$0.12\pm0.02$  &$<0.04$       &$<0.04$ \\
4   &$0.24\pm0.01$   &$0.37\pm0.02$  &$0.31\pm0.02$ &$0.17\pm0.01$ \\
5   &$0.08\pm0.01$    &$<0.06$        &$<0.04$       &$<0.03$ \\ 
\hline
\multicolumn{5}{c}{2000 Jan.}\\
\hline
1   &$0.22\pm0.01$    &$0.23\pm0.06$ &\nodata       &$<0.09$ \\
3   &$0.17\pm0.01$    &$<0.10$       &$<0.08$       &$<0.06$ \\
4   &$0.28\pm0.02$    &$0.34\pm0.04$ &$0.30\pm0.03$ &$0.11\pm0.03$ \\
5   &$0.06\pm0.01$    &$0.10\pm0.03$ &$<0.04$       &$<0.05$ \\
\hline
\multicolumn{5}{c}{2002 Jan.}\\
\hline
1  &$<0.10$   &$<0.47$  &\nodata       &$<0.34$ \\
3  &$<0.14$   &$<0.50$  &$<0.16$       &$<0.21$ \\
4  &$<0.47$   &$<0.49$  &$0.21\pm0.06$ &$<0.13$\\
5  &$<0.10$   &$<0.39$  &$<0.11$       &$<0.19$ \\
\hline
\multicolumn{5}{c}{2002 May}\\
\hline
1  &$<0.085$  &$<0.46$ &\nodata  &$<0.45$ \\
3  &$<0.13$   &$<0.39$ &$<0.19$  &$<0.42$ \\
4  &$<0.12$   &$<0.43$ &$<0.38$  &$<0.21$ \\
5  &$<0.057$  &$<0.32$ &$<0.14$  &$<0.31$ \\
\enddata
\tablenotetext{1}{Column densities in units of 10$^{14}$ cm$^{-2}$}
\end{deluxetable}

\begin{deluxetable}{llll}
\tablecolumns{4}
\tablewidth{26pc}
\tablecaption{Properties of Intrinsic Metal Absorption in STIS Spectrum
\label{table-compmetstis}}
\tablehead{
 \colhead{Component} &\colhead{$N$(Si{\sc iii}\lam1206)\tablenotemark{1}}
 &\colhead{$N$(Si{\sc ii}\lam1260)} &\colhead{$N$(C{\sc ii}\lam1334)}}
\startdata
1  &$0.03\pm0.01$  &$<0.038$ &$<0.45$\\
3  &$0.01\pm0.01$  &$<0.009$ &$<0.29$\\
4  &$0.04\pm0.01$  &$<0.024$ &$<0.27$\\
5  &$<0.011$       &$<0.018$ &$<0.17$\\
\enddata
\tablenotetext{1}{Column densities in units of 10$^{14}$ cm$^{-2}$}
\end{deluxetable}
\samepage

\clearpage

\clearpage

\clearpage

\clearpage

\clearpage

\clearpage

\clearpage

\clearpage

\clearpage

\clearpage

\clearpage

\clearpage

\end{document}